\documentclass[twocolumn]{aastex631}
\usepackage{xcolor}
\usepackage{amsmath}
\usepackage{booktabs}
\usepackage{array}
\usepackage{graphicx}
\usepackage{placeins}

\newif\ifarxiv
\arxivtrue        

\ifarxiv
  \usepackage{fancyhdr}
  \fancypagestyle{firstpage}{%
    \fancyhf{}%
    \fancyfoot[C]{\footnotesize \textcopyright\ 2026. All rights reserved.}%
  }

  \makeatletter
  \let\origmaketitle\maketitle
  \def\maketitle{\origmaketitle\thispagestyle{firstpage}}
  \makeatother
\fi

\makeatletter

\renewenvironment{acknowledgments}
{%
  \par
  \nolinenumbers
  \vskip 5.8mm plus 1mm minus 1mm
  \noindent\ignorespaces
}
{%
  \par
  \vskip 6pt
  \linenumbers
}

\makeatother

\definecolor{custom_blue}{rgb}{0.0588,0.1451,0.5804}
\hypersetup{
  colorlinks=true,
  linkcolor=custom_blue,
  citecolor=custom_blue,
  urlcolor=custom_blue
}

\newcommand{\hwc}[1]{(HWC; \textcolor{custom_blue}{\citeauthor{#1} \citeyear{#1}})}

\begin{document}

\shorttitle{Biosignature Detectability with ELT/ANDES}
\shortauthors{Kurzawa-Ferrandez et al.}
\title{Biosignature detectability on transiting habitable worlds with ELT/ANDES}

\author[0009-0006-1858-8166]{Evann Kurzawa-Ferrandez}
\affiliation{Jet Propulsion Laboratory, California Institute of Technology, Pasadena, CA 91109, USA}

\author[0000-0003-3355-1223]{Aaron Bello-Arufe}
\affiliation{Jet Propulsion Laboratory, California Institute of Technology, Pasadena, CA 91109, USA}

\author[0000-0003-2215-8485]{Renyu Hu}
\affiliation{Jet Propulsion Laboratory, California Institute of Technology, Pasadena, CA 91109, USA}
\affiliation{Department of Astronomy \& Astrophysics, The Pennsylvania State University, University Park,
PA 16802, USA}
\affiliation{Center for Exoplanets and Habitable Worlds, The Pennsylvania State University, University Park,
PA 16802, USA}
\affiliation{Institute for Computational and Data Science, The Pennsylvania State University, University Park,
PA 16802, USA}

\correspondingauthor{Evann Kurzawa-Ferrandez}
\email{contact@evannkurzawa.com}

\begin{abstract}
The search for life beyond the Solar System is at a turning point, transitioning from theoretical predictions to observations enabled by next-generation observatories. The Extremely Large Telescope (ELT) will host the ArmazoNes high Dispersion Echelle Spectrograph (ANDES), optimized for visible-to-near-infrared high-resolution spectroscopy. We present a simulation--detection pipeline and evaluate the detectability of CO$_2$, H$_2$O, and the biosignature gases O$_2$ and CH$_4$ in high-resolution transmission spectroscopy of transiting habitable-zone rocky planets with ANDES. Assuming cloud-free, modern Earth-like atmospheres, we model transmission spectra using noise estimates from the ANDES Exposure Time Calculator, based on the latest preliminary instrument design in seeing-limited mode. We introduce a novel Bayesian cross-correlation function (CCF) framework that incorporates molecule-specific kernels and a new autoregressive model to account for correlations in the CCF. We apply our framework to 18 known potentially habitable transiting exoplanets and estimate the number of transits required for a decisive detection ($\log_{10} B \ge 2.0$).
We find that H$_2$O is the most accessible species, with potential detections in 10--19 transits for the TRAPPIST-1 planets and 30 transits for LHS 1140\,b. CO$_2$, CH$_4$, and O$_2$ are more difficult to detect, requiring approximately 1.5, 3, and 4 times as many transits as H$_2$O. These estimates are lower limits that assume favorable observing conditions, perfect detrending, and the absence of systematics, yet still imply large observing campaigns. Alternative approaches, such as reflected-light high-dispersion coronagraphy of nearby nontransiting planets, may offer a promising complementary route for biosignature searches.
\end{abstract}

\keywords{Bayesian statistics (1900); Biosignatures (2018); Exoplanet atmospheres (487); Extrasolar rocky planets (511); Ground telescopes (687); Habitable planets (695); High resolution spectroscopy (2096); Near infrared astronomy (1093); Nested sampling (1894)}

\section{Introduction}\label{sec:one}

With a 39-meter diameter aperture, the ELT will be the largest optical and near-infrared telescope in the world, equipped with a wide range of instruments. These include ANDES \citep{Maiolino2013,Marconi2022,Marconi2024,Roederer2024,Martins2024,DOdorico2024,Palle2025}, a fiber-fed high-resolution spectrograph designed to provide $R\sim100,000$ spectroscopy across the optical and near-infrared.

One of the main goals of ANDES is to search for potential biosignature gases in exoplanet atmospheres through high-resolution transmission spectroscopy \citep{Palle2025}. Among the gases potentially linked to planetary habitability, O$_2$, CO$_2$, CH$_4$, and H$_2$O are of primary importance. Specifically, CO$_2$ and H$_2$O govern greenhouse warming and water cycling for climate stability, while O$_2$ and CH$_4$ are short-lived and may signal active biological replenishment \citep{Seager2012, Schwieterman2018, Ramirez2018, Meadows2018, Thompson2022}. Each of these four gases has distinct absorption features in the visible and/or near-infrared, making them prime candidates for searching for potentially habitable environments with ANDES. The instrument's high resolving power ($R\sim100,000$), combined with the ELT's extremely large aperture, will provide unprecedented data quality for potentially habitable exoplanets \citep{Kawahara2012,Snellen2013}, far exceeding that of JWST spectroscopic modes ($R\sim100$--$3000$) \citep[e.g.,][]{Morley2017,Lustig-Yaeger2019,Pidhorodetska2020}.

Over the past decade, significant theoretical and observational advancements have been made, particularly in the use of high-resolution spectroscopy and transit observations to probe the atmospheres of rocky planets orbiting M dwarfs. Early studies, such as \citet{Snellen2013}, proposed employing high-resolution spectroscopy ($R\sim100,000$) and cross-correlation techniques to search for biosignature gases with extremely large telescopes such as the ELT. They estimated that detecting O$_2$ in the transmission spectrum of an Earth-analog orbiting an M5V star at 12\,pc could be feasible after observing 30 transits. However, challenges such as the limited knowledge of the occurrence rate of habitable-zone Earth-sized planets around M dwarfs and the need to discover bright, nearby targets highlighted the difficulty of such endeavors.

\citet{Rodler2014} revisited the estimates from \citet{Snellen2013} and found far less favorable results, with O$_2$ requiring up to 175\,hr ($\sim$74 transits) to be detected on a target at 10\,pc. The authors concluded that nearby M4V stars are the most viable candidates, but even for these stars, achieving a significant detection remains challenging. \citet{Serindag2019} reinforced these findings by simulating O$_2$ detection in real observational data of Proxima Centauri, estimating the need for 20--50 transits for successful detection in optimal scenarios. Later, \citet{Currie2023} examined both inhabited and abiotic scenarios around M dwarfs and found that CO$_2$ and CH$_4$ are generally more accessible than O$_2$, which requires substantially more observational effort, particularly for late-type host stars.

In parallel, large-scale surveys and theoretical simulations have advanced our understanding of the detectability of biosignature gases. \citet{Hardegree-Ullman2023} expanded on earlier work by creating a catalog of over 286,000 main-sequence stars within 120\,pc and simulating a survey of M dwarfs within 20\,pc. They concluded that detecting Earth-like O$_2$ levels within 50\,yr is feasible for up to 21\% of nearby M dwarf systems with suitable transiting planets, particularly if high-resolution spectrographs (R $\sim$ 500,000) are utilized. They also noted that Earth-sized habitable-zone planets within 20\,pc are far more numerous in nontransiting configurations than in transiting ones, highlighting the importance of observational techniques beyond the transit method. In response, several studies have explored the potential of direct imaging with next-generation ground-based observatories, focusing on nearby exoplanets. \citet{Zhang2024} demonstrated that instruments such as ELT/METIS and ELT/HARMONI could detect key atmospheric species, such as CH$_4$, CO$_2$, H$_2$O and O$_2$, in nearby systems. Complementarily, \citet{Hardegree-Ullman2025} investigated the detectability of Earth-like O$_2$ levels in hypothetical habitable-zone exo-Earth candidates within 20\,pc. Under optimistic assumptions and a 10\,yr survey baseline, they found that O$_2$ at Earth-like abundances could be probed for up to $\sim$7 and $\sim$19 candidates using the GMT and ELT, respectively.

A complementary approach is to combine high-resolution spectroscopy with high-contrast imaging, often referred to as high-dispersion coronagraphy (HDC). In this technique, an adaptive-optics coronagraph spatially suppresses stellar light, and a high-resolution spectrograph disentangles the planet's Doppler-shifted molecular lines from the stellar spectrum \citep{Snellen2015,Wang2017,Mawet2017}. HDC is being developed on the ground (for example, with Keck/KPIC) and investigated for future space missions such as the Habitable Worlds Observatory \citep{Wang2021,Wang2018,Pueyo2019}. This approach may provide an alternative pathway for searching for biosignature gases in the reflected light of nearby habitable-zone planets. A major advantage of reflected-light HDC is that it does not require the planet to transit, thereby opening access to the nearest habitable worlds.

Although ground-based observations are affected by telluric contamination, they provide access to much larger collecting areas and spectral resolutions that remain unmatched by space telescopes. Two strategies for a comprehensive characterization of exoplanetary atmospheres are (i) combining ground- and space-based observations, as demonstrated by \citet{Brogi2017} and followed in other studies \citep[e.g.,][]{Kasper2023, Smith2024}, and (ii) increasing spectral resolution to facilitate the disentanglement of the planetary signal from the telluric and host-star spectra. \citet{Lopez-Morales2019} demonstrated that raising the resolution from $R \sim 100{,}000$ to $R\sim300,000-400,000$ can double the depth of O$_2$ lines in Earth-like atmospheres, reducing the required number of transits by over 30\%. Similarly, \citet{Rukdee2024} showed that even under the most challenging haze and cloud conditions, increasing the resolution from $R\sim100,000$ to $R\sim300,000$ enables robust detections with exposure times reduced by a factor of four, thereby requiring significantly fewer transits.

ANDES represents the most advanced and promising ground-based instrument currently planned for the search for biosignature gases \citep{Palle2025}. Yet, a dedicated study of the number of transits that could be required to detect key atmospheric molecules on potentially habitable rocky exoplanets with ANDES has not yet been published. To address this gap, we present a framework that estimates transit requirements and target rankings under a homogeneous set of simplifying assumptions.

Our paper is structured as follows. In Section~\ref{sec:two}, we present the assumptions and models used to simulate the atmospheres of known transiting habitable-zone terrestrial planets. In Section~\ref{sec:three}, we generate the transmission spectra using the \texttt{petitRADTRANS} library \citep{Molliere2019}. Section~\ref{sec:four} describes the ANDES observing setup, the Exposure Time Calculator-based S/N estimates, and the noise injection procedure. In Section~\ref{sec:five}, we define the CCF statistic adopted in this work and explain why a simple peak-based frequentist S/N estimate is not sufficiently robust for our sample. This motivates a more robust approach, introduced in Section~\ref{sec:six}, which details a custom Bayesian framework with \texttt{MultiNest} \citep{Feroz2009}, featuring interpolated kernels that mimic molecule-specific autocorrelation patterns and a first-order autoregressive AR(1) noise model that explicitly accounts for correlations between neighboring velocity bins in the CCF likelihood. Evidence estimation and Bayes factors are then used to quantify support for atmospheric-signal detection. In Section~\ref{sec:seven}, we apply this framework to the planets and gases of interest. The results are then interpreted in Section~\ref{sec:eight}, with emphasis on detectability, comparisons with prior work, and limitations of the adopted assumptions. Finally, in Section~\ref{sec:nine} we summarize the implications of these results for future ELT/ANDES observations.

\section{Target sample and simulation framework}\label{sec:two}

The Habitable Worlds Catalog\footnote{\url{http://phl.upr.edu/hwc}} \hwc{PHL}, last updated in March 2024, identifies 70 potentially habitable worlds among the more than five thousand exoplanets known at that time.\footnote{The NASA Exoplanet Archive (\url{https://exoplanetarchive.ipac.caltech.edu/}) currently lists over 6,000 confirmed exoplanets.} Our analysis focuses on the 41 HWC planets with confirmed transits, out of a total of 70, and excludes 11 targets that are disfavored by our simplified ELT visibility screening, based on a sampled 20$^\circ$ altitude criterion (see Section~\ref{subsec:two_three}). We further exclude 11 planets with radii larger than $2\,R_\oplus$, as planets above this threshold are unlikely to be rocky, and K2-288~B\,b, which falls within the radius valley, and whose lack of a measured mass means that this planet might not be rocky \citep{Fulton2017}.

\subsection{Global setup\label{subsec:two_one}}

The equilibrium temperature $T_{\mathrm{eq}}$ of each planet is estimated from blackbody theory, assuming a Bond albedo of 0.3 \citep[e.g.,][]{Suissa2020, Siffert2024}. Bulk planetary properties are listed in Table~\ref{tab:inputs} and are taken from the NASA Exoplanet Archive \citep{Christiansen2025}. Planetary masses are estimated using the Bayesian radius-density-mass relation of \citet{Parviainen2024} via the \texttt{SPRIGHT} code. We use the maximum a posteriori estimate of the predicted mass distribution, which together with the radius defines the surface gravity $g$. The planetary reference radius $R_p$ is defined at a reference pressure of 1\,bar. The vertical grid spans pressures $\log_{10}(p/\mathrm{bar}) \in [-7,0]$ using 100 layers, from 1\,bar down to $10^{-7}$\,bar ($\sim$120\,km altitude for Earth-like conditions). In practice, we restrict radiative transfer to a single absorber at a time in order to isolate molecule-by-molecule accessibility within a simplified setup.

\begin{deluxetable*}{lcccccccccccccc}
\tabletypesize{\footnotesize}
\tablewidth{0pt}
\tablecaption{Input parameters for \texttt{petitRADTRANS} and the ETC.\label{tab:inputs}}
\tablehead{
  \colhead{Planet} & \colhead{$a$} & \colhead{$d_\ast$} & \colhead{$P$} & \colhead{$T_{14}$} & \colhead{$T_{\rm eq}$} & \colhead{$R_p$} & \colhead{$R_\ast$} & \colhead{$M_p$} & \colhead{$T_\ast$} & \colhead{$v_{\rm sys}$} & \colhead{$v_{\rm bary}$} & \colhead{Sp. T} & \colhead{$m_J$} & \colhead{References} \\
  & \colhead{(au)} & \colhead{(pc)} & \colhead{(days)} & \colhead{(hr)} & \colhead{(K)} & \colhead{($R_\oplus$)} & \colhead{($R_\odot$)} & \colhead{($M_\oplus$)} & \colhead{(K)} & \colhead{($\mathrm{km\,s^{-1}}$)} & \colhead{($\mathrm{km\,s^{-1}}$)} & & \colhead{(Vega)} &
}
\startdata
K2-3\,d          & 0.210 & 44.07 & 44.556  & 4.170  & 286$^{\dagger}$ & 1.620 & 0.600 & 5.217$^{\dagger}$ & 3835 & 30.24  & 29.39$^{\dagger}$ & M0V   & 9.42  & 1,2,3,4,5 \\
K2-72\,e         & 0.106 & 66.43 & 24.159  & 2.250  & 272$^{\dagger}$ & 1.290 & 0.329 & 2.211$^{\dagger}$ & 3498 & -42.92 & 0.69$^{\dagger}$  & M2V   & 11.69 & 1,2,6,7,8 \\
Kepler-1649\,c   & 0.083 & 92.19 & 19.535  & 1.070  & 239$^{\dagger}$ & 1.060 & 0.232 & 1.101$^{\dagger}$ & 3240 & 0.00$^{\dagger}$     & -13.61$^{\dagger}$ & M5V   & 13.38 & 1,9,10 \\
Kepler-186\,f    & 0.432 & 177.59 & 129.943 & 5.448  & 193$^{\dagger}$ & 1.438 & 0.548 & 3.306$^{\dagger}$ & 3876 & -60.96 & -7.58$^{\dagger}$ & M1    & 12.47 & 1,2,7,11,12,13,14,15 \\
Kepler-440\,b    & 0.242 & 301.03 & 101.111 & 8.130  & 284$^{\dagger}$ & 1.899 & 0.663 & 4.140$^{\dagger}$ & 3891 & -20.61 & -7.23$^{\dagger}$ & K8V$^{\dagger}$   & 12.96 & 1,2,7,11,12,13,14 \\
Kepler-442\,b    & 0.409 & 365.97 & 112.303 & 5.630  & 248$^{\dagger}$ & 1.395 & 0.619 & 2.958$^{\dagger}$ & 4563 & -76.91 & -6.72$^{\dagger}$ & K4V$^{\dagger}$   & 13.23 & 1,2,7,11,12,13,14 \\
Kepler-452\,b    & 1.046 & 551.73 & 384.843 & 10.367 & 261$^{\dagger}$ & 1.511 & 1.119 & 3.983$^{\dagger}$ & 5728 & -25.49 & -7.80$^{\dagger}$ & G2    & 12.26 & 1,2,7,13,14,16 \\
Kepler-62\,e     & 0.428 & 300.87 & 122.386 & 7.196  & 279$^{\dagger}$ & 1.873 & 0.731 & 3.971$^{\dagger}$ & 4842 & 17.51  & -9.65$^{\dagger}$ & K2V   & 12.26 & 1,2,7,13,17,18 \\
Kepler-62\,f     & 0.720 & 300.87 & 267.283 & 7.460  & 215$^{\dagger}$ & 1.540 & 0.731 & 4.291$^{\dagger}$ & 4842 & 17.51  & -9.65$^{\dagger}$ & K2V   & 12.26 & 1,2,7,17,18,19 \\
LHS 1140\,b      & 0.095 & 14.99 & 24.737  & 2.150  & 206$^{\dagger}$ & 1.730 & 0.216 & 6.764$^{\dagger}$ & 3096 & -13.74 & -28.80$^{\dagger}$ & M4.5V & 9.61  & 1,2,20 \\
LP 890-9\,c      & 0.040 & 32.43 & 8.457   & 0.959  & 248$^{\dagger}$ & 1.367 & 0.156 & 2.740$^{\dagger}$ & 2850 & 28.84 & 19.99$^{\dagger}$  & M6V   & 12.26 & 1,2,21 \\
TOI-700\,d       & 0.163 & 31.13 & 37.424  & 3.314  & 245$^{\dagger}$ & 1.073 & 0.421 & 1.146$^{\dagger}$ & 3459 & -4.78 & 1.78$^{\dagger}$   & M2.5V & 9.47 & 1,22,23 \\
TOI-700\,e       & 0.134 & 31.13 & 27.810  & 2.777  & 270$^{\dagger}$ & 0.953 & 0.421 & 0.757$^{\dagger}$ & 3459 & -4.78 & 1.78$^{\dagger}$   & M2.5V & 9.47 & 1,22,23 \\
TOI-715\,b       & 0.083 & 42.40 & 19.288  & 1.980  & 231$^{\dagger}$ & 1.550 & 0.240 & 4.395$^{\dagger}$ & 3075 & 55.78  & -4.49$^{\dagger}$ & M4    & 11.81 & 1,2,24 \\
TRAPPIST-1\,d    & 0.022 & 12.43 & 4.049   & 0.815  & 263$^{\dagger}$ & 0.788 & 0.119 & 0.394$^{\dagger}$ & 2566 & -52.00 & 0.22$^{\dagger}$  & M8V  & 11.35 & 1,25,26,27,28 \\
TRAPPIST-1\,e    & 0.029 & 12.43 & 6.101   & 0.929  & 229$^{\dagger}$ & 0.920 & 0.119 & 0.671$^{\dagger}$ & 2566 & -52.00 & 0.22$^{\dagger}$  & M8V  & 11.35 & 1,25,26,27,28 \\
TRAPPIST-1\,f    & 0.038 & 12.43 & 9.208   & 1.048  & 200$^{\dagger}$ & 1.045 & 0.119 & 1.044$^{\dagger}$ & 2566 & -52.00 & 0.22$^{\dagger}$  & M8V  & 11.35 & 1,25,26,27,28 \\
TRAPPIST-1\,g    & 0.047 & 12.43 & 12.352  & 1.137  & 180$^{\dagger}$ & 1.129 & 0.119 & 1.373$^{\dagger}$ & 2566 & -52.00 & 0.22$^{\dagger}$  & M8V  & 11.35 & 1,25,26,27,28 \\
\enddata
\tablerefs{
(1) 2MASS Catalog; \citealt{Cutri2003};
(2) \citealt{Gaia2023};
(3) \citealt{Fukui2016};
(4) \citealt{Bonomo2023};
(5) \citealt{DiamondLowe2022};
(6) \citealt{Dressing2017};
(7) TESS Input Catalog: \citealt{Stassun2019};
(8) \citealt{Crossfield2016};
(9) \citealt{Vanderburg2020};
(10) \citealt{Hardegree-Ullman2019};
(11) \citealt{Torres2015};
(12) \citealt{Gajdos2019};
(13) \citealt{Morton2016};
(14) \citealt{Berger2018};
(15) \citealt{Quintana2014};
(16) \citealt{Jenkins2015};
(17) \citealt{Weiss2024};
(18) \citealt{Borucki2013};
(19) \citealt{Borucki2019};
(20) \citealt{Cadieux2024};
(21) \citealt{Delrez2022};
(22) \citealt{Gilbert2023};
(23) \citealt{SDSS2022};
(24) \citealt{Dransfield2024};
(25) \citealt{Agol2021};
(26) \citealt{Apogee2020};
(27) \citealt{Gizis2000};
(28) \citealt{Piaulet2025}.
}
\tablecomments{$\,a$ is the semi-major axis; $d_\ast$ is the distance to the star; $P$ is the orbital period; $T_\mathrm{14}$ is the transit duration; $T_\mathrm{eq}$ is the planetary equilibrium temperature; $R_p$ is the planet radius; $R_\ast$ is the host-star radius; $M_\mathrm{p}$ is the planet mass; $T_\ast$ is the host-star effective temperature; $v_{\rm sys}$ is the host-star systemic velocity; $v_{\rm bary}$ is the barycentric correction that maximizes $\left| v_{\rm sys} - v_{\rm bary} \right|$ over the course of a year; Sp. T is the host-star spectral type; and $m_J$ is the Vega magnitude of the host star in the J band. When $v_{\rm sys}$ is unavailable, a default value of $0\,\mathrm{km\,s^{-1}}$ is assumed as a simplifying approximation. Values marked with a dagger ($\dagger$) were calculated in this work. These data are obtained in part from the SIMBAD database and the NASA Exoplanet Archive.}
\end{deluxetable*}

\subsection{Atmospheric model: structure and composition\label{subsec:two_two}}

\begin{enumerate}
    \item \textbf{Temperature profile.} We adopt a two-layer temperature-pressure profile, a standard approximation in radiative-convective equilibrium models \citep[e.g.,][]{Kasting1993, Kopparapu2013}. The tropopause is set at 0.1\,bar \citep{Robinson2014}. Above this, the stratosphere is assumed to be isothermal at the atmospheric skin temperature, as described by, e.g., \citet{Pierrehumbert2010, Parmentier2014}:
\begin{equation}
T_{\mathrm{strat}} = T_{\mathrm{skin}} \approx 2^{-1/4} \, T_{\mathrm{eq}}
\end{equation}
Below the tropopause, we assume a moist-adiabatic temperature--pressure relation:
\begin{equation}
T(p) = T_{\mathrm{skin}}\left(\frac{p}{p_{\mathrm{trop}}}\right)^{\kappa_{\mathrm{moist}}}
\end{equation}
where $\kappa_{\mathrm{moist}} = 0.17$ approximates the lapse rate of a saturated Earth-like atmosphere and accounts for latent heat release during convection \citep{Robinson2012}. This prescription ensures a continuous transition between the radiative and convective layers. The surface temperature is the moist-adiabatic extension of $T_{\mathrm{skin}}$ to 1\,bar.

\begin{figure*}[ht!]
    \centering
    \includegraphics[width=0.8\textwidth]{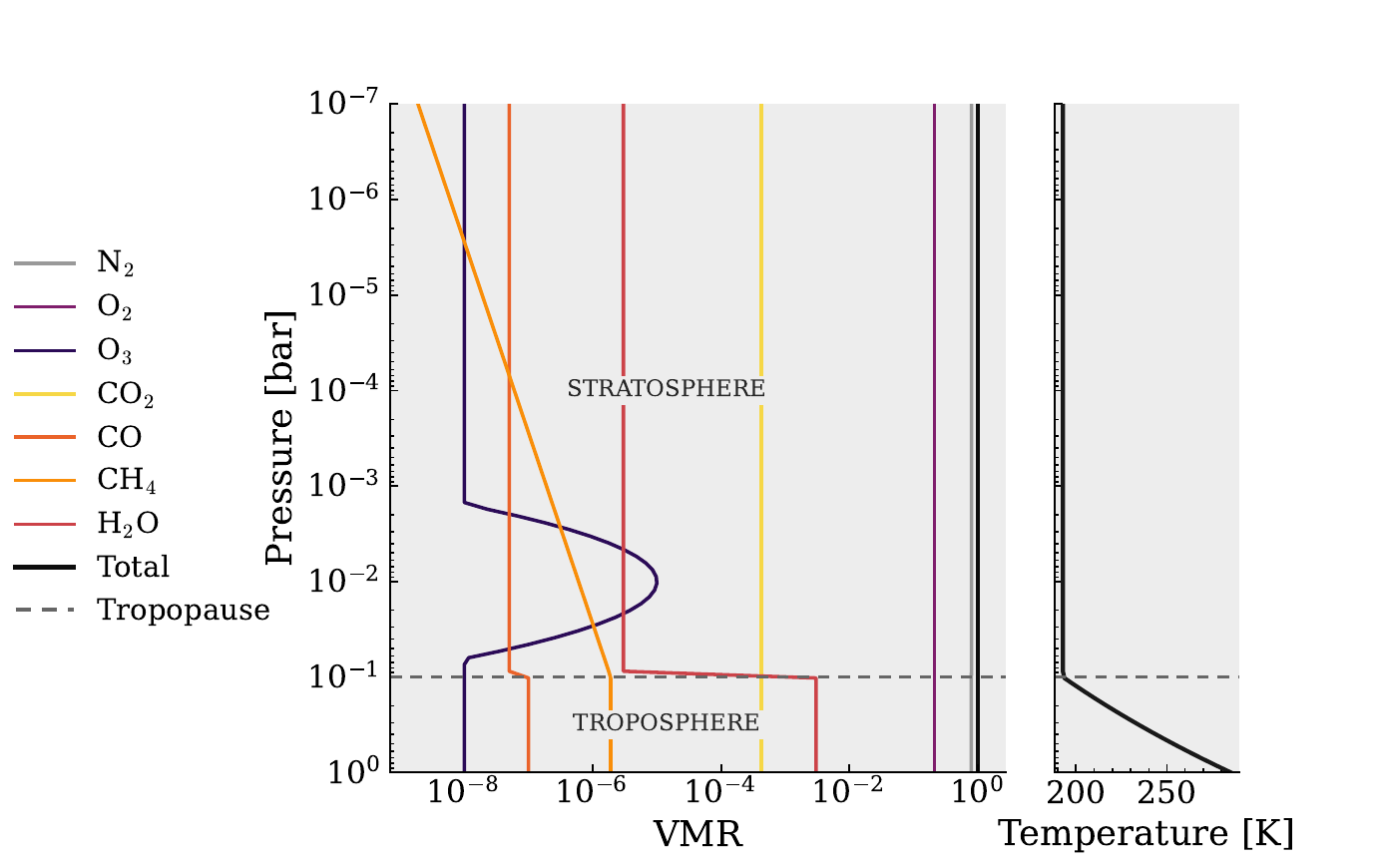}
    \caption{Vertical profiles of temperature and volume mixing ratios for the species included in our atmospheric model, simulated for TRAPPIST-1\,e. The profile behavior follows from the equations and assumptions described in Section~\ref{subsec:two_two}.}
    \label{fig:one}
\end{figure*}
\vspace{1cm}
    \item \textbf{Atmospheric VMRs.} The atmospheric composition is specified as follows:

\begin{itemize}
    \item H$_2$O: For simplicity, we adopt an idealized two-level water vapor profile. The stratosphere is fixed to a constant volume mixing ratio of 3\,ppmv, representative of Earth's dry stratosphere \citep[e.g.,][]{Prather1989,Oman2008}. In the troposphere, water vapor is assumed to be well mixed at a constant volume mixing ratio of 0.3\%, corresponding to a moderately humid Earth-like atmosphere. This value is not intended to reproduce the detailed vertical structure of terrestrial humidity, but is representative of the order of magnitude of Earth's tropospheric water vapor abundance \citep[e.g.,][]{Heise2006,Gentine2013,Vergados2018}.
    \item O$_3$: Described by a log-normal distribution in pressure as in \citet{vonParis2011, vonParis2013}, with peak mixing ratio 10\,ppmv centered at $p=10$\,mbar, width $\sigma_{\ln p}=0.5$, and a floor of 10\,ppbv elsewhere \citep[e.g.,][]{Bekki2009, Kuttippurath2024}.
    \item CH$_4$: Well-mixed in the troposphere at 1.9\,ppmv \citep{Lan2025}. Above the tropopause, the abundance is assumed to decrease exponentially with altitude, following \citet{Bainbridge1966}; in pressure coordinates, we approximate this as $(p/p_{\mathrm{trop}})^{0.5}$.
    \item CO: We simplify the CO treatment of \citet{Pan2004, ElAmraoui2014} to a two-level profile: 100\,ppbv in the troposphere and 50\,ppbv in the stratosphere.
    \item O$_2$: Taken as a constant value of 0.2095 at all pressures, corresponding to the standard dry Earth atmosphere.
    \item CO$_2$: Taken as a constant value of 420\,ppmv at all pressures \citep{Lan2025}.
    \item N$_2$: Fills the remaining fraction in each layer so that the total abundance sums to unity.
\end{itemize}
\end{enumerate}

O$_3$, CO, and N$_2$ are included for completeness in the atmospheric composition, but are not included as opacity sources in the single-species runs. At each layer, all mixing ratios are renormalized to ensure $\sum_i x_i = 1$, and the mean molecular mass $\mu(p)$ is computed consistently.

The resulting vertical profiles of temperature and volume mixing ratios (VMR) are displayed in Figure~\ref{fig:one}.

\subsection{Visibility constraints and planetary velocity\label{subsec:two_three}}

The 2011 ELT construction proposal\footnote{\url{https://www.eso.org/public/products/books/book_0046/}} specifies a minimum altitude cutoff of 20$^\circ$ (above the local horizon at the ELT site) to avoid rapid increases in atmospheric extinction, dome vignetting, and seeing degradation that arise at lower altitudes. In this work, we use this limit as a simplified visibility criterion to exclude targets that are strongly disfavored from the ELT site. 

To construct a coarse annual visibility metric for each host star, we generate a uniform grid of 366 epochs spanning January 1, 2025, to January 1, 2026, and transform its coordinates from the International Celestial Reference System (ICRS) to the ELT altitude--azimuth frame using \texttt{Astropy} \citep{Astropy2022}. At each sampled epoch, we evaluate the apparent altitude of the star and identify the epochs for which the altitude exceeds 20$^\circ$. Consequently, we retain only planets whose host stars satisfy the adopted 20$^\circ$ altitude criterion at least once within the sampled annual grid and exclude those for which no such sampled epoch exists. This cut therefore provides a consistent first-order visibility criterion for defining the target sample, while more detailed scheduling effects are left to future observation-specific planning.

\begin{figure}[ht!]
    \centering
    \includegraphics[width=\columnwidth]{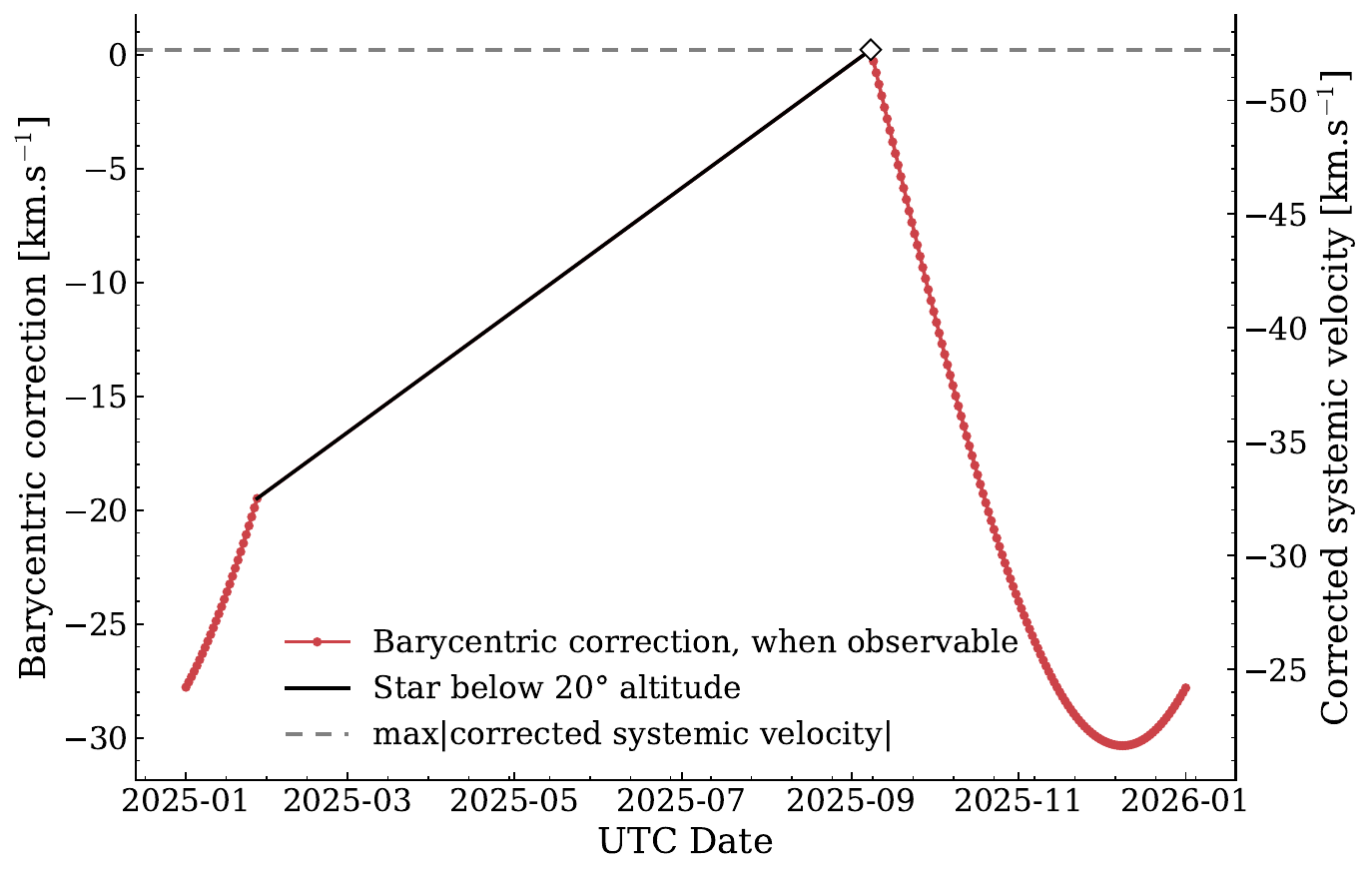}
    \caption{Barycentric velocity correction for TRAPPIST-1 from January 2025 to January 2026. In our screening framework, sampled epochs between February and September do not satisfy the adopted 20$^\circ$ criterion. Over the retained epochs, the barycentric correction varies between approximately $-30\,\mathrm{km\,s^{-1}}$ and $0\,\mathrm{km\,s^{-1}}$. The sinusoidal modulation reflects the projection of Earth's nearly circular orbital velocity onto the line of sight to the star, producing a radial component with an amplitude of about $\pm 30\,\mathrm{km\,s^{-1}}$ and a one-year period, phase-shifted according to the star's ecliptic position. The right-hand $y$-axis shows the relative Doppler separation $v_{\mathrm{sys}} - v_{\mathrm{bary}}(t)$, whose absolute value is maximized for TRAPPIST-1 at a barycentric velocity of $0.22\,\mathrm{km\,s^{-1}}$ within this framework.}
    \label{fig:two}
\end{figure}

In parallel, we compute the barycentric velocity correction at each sampled epoch using a DE430 ephemeris and the \texttt{barycorrpy} package \citep{Kanodia2018}, following the algorithm of \citet{Wright2014}. For each retained target, this yields an annual time series of barycentric corrections over the subset of sampled epochs that satisfy the altitude cut. An example is shown in Figure~\ref{fig:two}.

Among the sampled epochs that satisfy the screening criterion, we adopt the most optimistic Doppler-separation configuration by selecting the epoch at which the absolute difference between the systemic velocity and the barycentric correction is largest,
\begin{equation}
\begin{aligned}
v_{\mathrm{max}} &= v_{\mathrm{sys}} - v_{\mathrm{bary}}(t_{\mathrm{max}}),\\
t_{\mathrm{max}} &= \arg\max_t \left| v_{\mathrm{sys}} - v_{\mathrm{bary}}(t)\right|
\end{aligned}
\label{eq:barycentric_systemic}
\end{equation}

We use this quantity as a best-case proxy for the barycentric--systemic offset in the synthetic observations. The systemic velocities and the corresponding sampled barycentric corrections at maximum separation are summarized in Table~\ref{tab:inputs}.

\section{Building exoplanet transmission spectra}\label{sec:three}

We generate model transmission spectra with \texttt{petitRADTRANS} \citep{Molliere2019}, which performs radiative transfer for planetary atmospheres. For each target, we use published HWC system parameters together with our atmospheric model as inputs.

In all simulations, we include Rayleigh scattering by N$_2$; clouds are not modeled. We also neglect atmospheric refraction under the conditions relevant to our targets (temperate-warm, close-in rocky planets around K--M dwarfs). In this regime, the large stellar angular size pushes the refractive boundary to pressures/altitudes where gas and Rayleigh opacity dominate the slant optical depth, so refraction makes only a minor contribution and does not set the continuum for our targets. This behavior is captured by analytic refractive-boundary estimates and validated against ray-tracing comparisons \citep{Robinson2017}; for the Earth-Sun case, refraction excludes only the lowest $\sim$13\,km and changes the effective radius by at most a few tens of km across the optical/NIR \citep{Betremieux2013}.

Because our analysis relies on high-resolution spectroscopy, we adopt molecular opacity data at commensurate resolution. For H$_2$O, CO$_2$ and CO, we use the HITEMP line lists \citep{Rothman2010}; for CH$_4$ we use the HITEMP update by \citet{Hargreaves2020}; for O$_3$, the HITRAN database \citep{Rothman2013}; for O$_2$, the HITRAN compilation \citep{Gordon2022}; and for N$_2$, the WCCRMT database \citep{Western2018}. The H$_2$O, CO$_2$, CO, O$_3$, and CH$_4$ opacity files are obtained in HDF5 format from the Keeper knowledge base of the Max Planck Digital Library,\footnote{\url{https://keeper.mpdl.mpg.de/}} while those for O$_2$ and N$_2$ are binary files from the DACE Opacity Database,\footnote{\url{https://dace.unige.ch}} which employs the open-source \texttt{HELIOS-K} code \citep{Grimm2015,Grimm2021} to convert line lists into opacity coefficients suitable for atmospheric simulations. Each binary file corresponds to a specific temperature, pressure, and wavenumber range, which we merge into a single line-by-line file readable by \texttt{petitRADTRANS}.

We simulate a time series of spectra spanning twice the transit duration and centered on mid-transit. Prior to resampling, we apply instrumental broadening by convolving each spectrum with a Gaussian kernel whose width is set by the ANDES resolving power. The convolution is performed using the \texttt{PyAstronomy} library \citep{PyAstronomy}. We choose to extend the Gaussian kernel to five standard deviations to ensure computational efficiency while adequately representing the instrument response. After broadening, we convert vacuum wavelengths to air wavelengths using the \citet{Edlen1966} relation. The broadened spectra are then interpolated onto a constant-$R$ wavelength grid (described in Section~\ref{subsec:four_two}). We neglect other contributions to the broadening or Doppler shifting of the planet spectrum (e.g., planetary rotation and atmospheric dynamics), consistent with the deliberately simplified scope adopted here.

The stellar spectrum used in the synthetic observations is based on the stellar count spectrum returned by the ANDES Exposure Time Calculator (ETC), as described in Section~\ref{sec:four}. We Doppler-shift the stellar spectrum by $v_{\mathrm{max}}$ and the planet's transmission spectrum by
\begin{equation}
v_{\mathrm{tot}}(t) = v_{p}(t) + v_{\mathrm{max}}
\label{eq:v_tot}
\end{equation}
where $v_{\mathrm{tot}}(t)$ is the planet--ELT line-of-sight velocity during transit, $v_p(t)$ is the planet's radial velocity in the star--planet frame, and $v_{\mathrm{max}}$ is the constant barycentric-systemic offset at the epoch of maximum Doppler separation (Equation~\ref{eq:barycentric_systemic}). The planet's radial velocity is given by

\begin{equation}
v_p(t) = K_p \sin(2\pi \phi(t))
\end{equation}

where $\phi(t)$ is the orbital phase at time $t$ and $K_p$ is the radial velocity semiamplitude of the planet,

\begin{equation}
K_p = \frac{2\pi a \sin(i_p)}{P \sqrt{1 - e^2}}
\end{equation}

where $a$ is the orbital semi-major axis of the planet, $i_p$ is the orbital inclination, $P$ is the orbital period, and $e$ is the eccentricity. We assume circular orbits in our simulations.

We then inject the Doppler-shifted planetary transmission spectrum during the in-transit window. The observed spectrum model is

\begin{equation}
S_{\text{obs}}(\lambda) = S_{\text{star}}(\lambda) \times (1 - S_{\text{planet}})
\end{equation}
where $S_{\text{star}}$ is the ETC-derived stellar count spectrum Doppler-shifted by $v_{\mathrm{max}}$ (Equation~\ref{eq:barycentric_systemic}), and $S_{\text{planet}}$ is the transit depth shifted by $v_{\mathrm{tot}}$ (Equation~\ref{eq:v_tot}) and set to zero outside of transit. $S_{\text{obs}}$ is the composite planet--host-star spectrum.

\section{Synthetic observations with ELT/ANDES} \label{sec:four}
\subsection{Observing mode}\label{subsec:four_one}

The instrumental assumptions adopted here are based on the ANDES design available during the review of this paper in June 2026. Because the design is still evolving ahead of the Preliminary Design Review, these assumptions should be regarded as current best estimates. The instrument is a fiber-fed, high-resolution spectrograph for the ELT with several observing modes. The configuration relevant for this work is the seeing-limited point-source mode, which provides $R\sim100,000$ spectroscopy over the broad optical--near-infrared wavelength range used here, with separate object and sky apertures. ANDES also includes an adaptive-optics-assisted integral-field-unit (IFU) mode in the near infrared, designed for diffraction-limited observations and high-contrast/reflected-light applications. This IFU mode can use up to 61 spatial elements, or spaxels, with selectable spaxel scales in the 5--100\,mas range \citep{Marconi2024,Palle2025}. However, our simulations concern unresolved transmission spectroscopy of transiting planets, for which the star and planet are observed together as a point source. We therefore use the seeing-limited mode, not the AO-assisted IFU mode, and the relevant spatial factor in the detector-noise term is the effective number of detector pixels in the seeing-limited extraction aperture.

\subsection{S/N grid}\label{subsec:four_two}
\subsubsection{ETC S/N scale}\label{subsubsec:four_two_one}
Signal-to-noise ratios (S/N) under our observing conditions are obtained from the ANDES Exposure Time Calculator\footnote{\url{https://andes.inaf.it/instrument/exposure-time-calculator/}} \citep[ETC; ][]{Sanna2014,Sanna2021,Sanna2024}. We use the latest ETC release available at the time of revision, v5.2\footnote{Release: 2026-04-17, \url{https://andes-etc.brera.inaf.it/}}, to set the absolute S/N scale. The exported v5.2 wavelength points, however, are representative S/N samples and do not form a fully sampled high-resolution grid. To recover the dense wavelength sampling needed for the spectral simulations, we also use ETC v3.2\footnote{Release: 2024-09-09}, which returns a densely sampled S/N curve, including small-scale variations caused by telluric absorption and instrumental response. We therefore rescale the dense v3.2 S/N curve locally to match the v5.2 S/N anchors. Both ETC outputs are first evaluated for a 600\,s reference exposure (see Section~\ref{subsec:four_three}).

We denote by $\lambda_i^{\rm v3.2}$ and $({\rm S/N})_{\rm v3.2}(\lambda_i^{\rm v3.2})$ the dense wavelength grid and S/N values from v3.2. We denote by $\lambda_j^{\rm v5.2}$, $({\rm S/N})_{\rm v5.2}(\lambda_j^{\rm v5.2})$, and $N_{\star,{\rm v5.2}}(\lambda_j^{\rm v5.2})$ the v5.2 wavelength points, S/N values, and detected stellar signal for the same exposure time, respectively. $N_{\star,{\rm v5.2}}$ corresponds to the ETC ``object'' signal from the host star, expressed in $e^-$ per resolution element.

At each v5.2 wavelength point, we interpolate the v3.2 S/N onto $\lambda_j^{\rm v5.2}$ and compute the local scale factor
\begin{equation}
f_j =\frac{({\rm S/N})_{\rm v5.2}(\lambda_j^{\rm v5.2})}{({\rm S/N})_{\rm v3.2}(\lambda_j^{\rm v5.2})}
\label{eq:etc_scale_factor}
\end{equation}

\begin{figure}[ht!]
    \centering
    \includegraphics[width=\columnwidth]{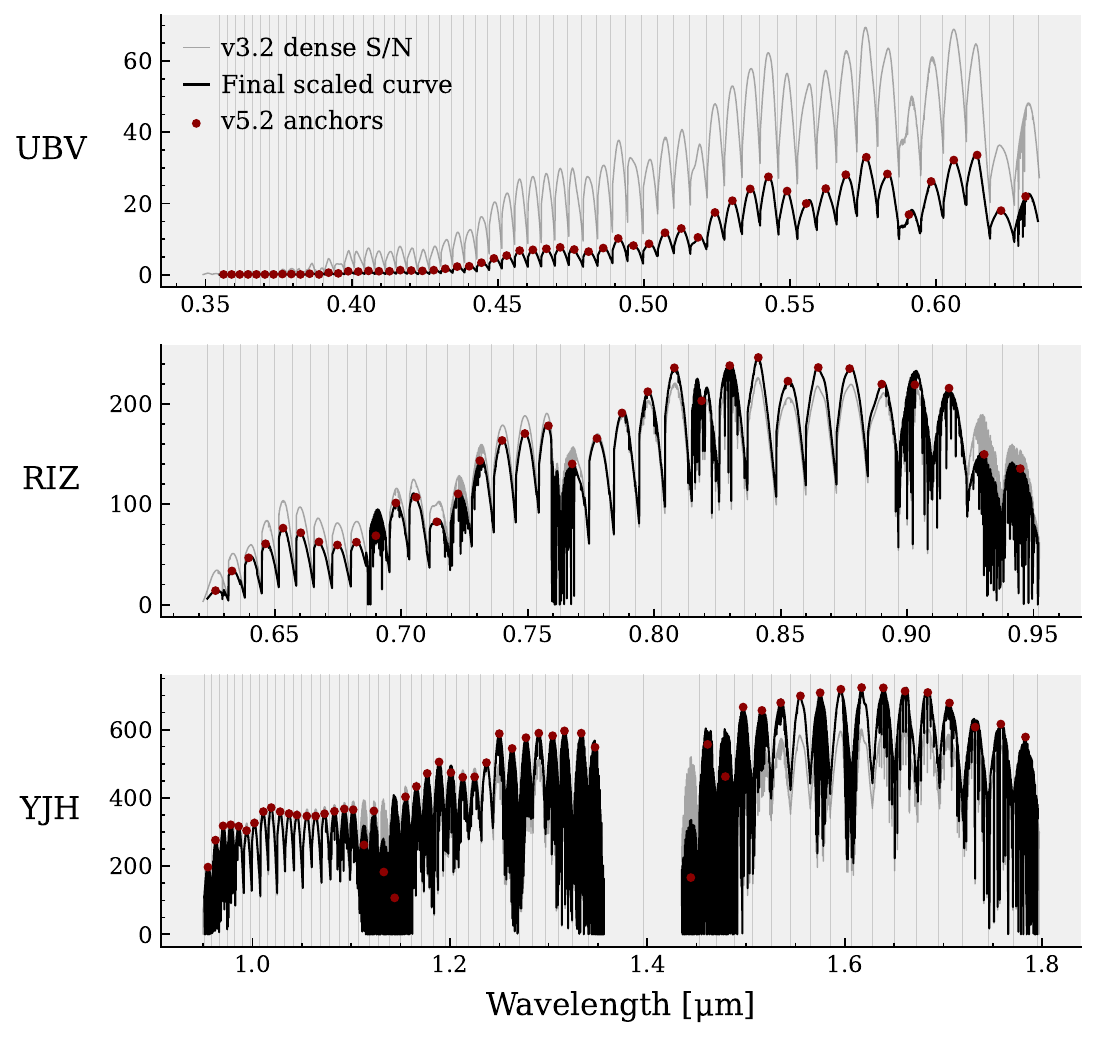}
    \caption{Construction of the ETC S/N curve, shown here for TRAPPIST-1 and a 600\,s reference exposure. The red points show the representative S/N anchors from v5.2, which set the absolute normalization. The gray curves show the densely sampled wavelength structure from v3.2. For each v5.2 anchor, we compute a local multiplicative scale factor (Equation~\ref{eq:etc_scale_factor}) and apply it to the corresponding v3.2 wavelength interval. The resulting scaled S/N curve is shown in black. Thin vertical lines indicate the wavelength intervals assigned to each v5.2 anchor point.}
    \label{fig:three}
\end{figure}

Each v5.2 wavelength point defines a bin centered on that point, with edges halfway to the adjacent v5.2 points. For a v5.2 wavelength point $\lambda_j^{\rm v5.2}$, the bin boundaries are
\begin{equation}
\lambda_{j\pm1/2}^{\rm v5.2}=\frac{1}{2}\left(\lambda_j^{\rm v5.2}+\lambda_{j\pm1}^{\rm v5.2}\right)
\label{eq:etc_segment_boundary}
\end{equation}
Within the interval associated with $\lambda_j^{\rm v5.2}$, we multiply the dense v3.2 S/N by the corresponding factor $f_j$:
\begin{equation}
({\rm S/N})_{\rm ETC}(\lambda)=f_j\,({\rm S/N})_{\rm v3.2}(\lambda),\qquad\lambda_{j-1/2}^{\rm v5.2}\leq\lambda<\lambda_{j+1/2}^{\rm v5.2}
\label{eq:etc_hybrid_snr}
\end{equation}
This produces an ETC S/N curve that follows the v5.2 normalization locally while retaining the dense wavelength structure available from v3.2. This construction is illustrated in Figure~\ref{fig:three}.

\subsubsection{Constant-resolution grid}\label{subsubsec:four_two_two}
After defining the S/N scale, we place the simulated spectra, S/N values, and stellar count spectrum on a common wavelength grid with one point per resolution element. This ensures that the model spectra are sampled at the same resolution-element scale at which the ETC reports the S/N and detected stellar signal, and at which the instrumental broadening is defined. The corresponding constant-$R$ sampling is defined by
\begin{equation}
\Delta\ln\lambda = \frac{1}{R}
\label{eq:constant_r_spacing}
\end{equation}
or, equivalently,
\begin{equation}
\lambda_{k+1}=\lambda_k\exp\left(\frac{1}{R}\right)
\label{eq:constant_r_grid}
\end{equation}
This grid is uniform in resolving power, $\Delta\lambda/\lambda \simeq 1/R$, and therefore provides one sample per resolution element over the wavelength range retained from the ETC output. The S/N curve from Equation~\ref{eq:etc_hybrid_snr} is interpolated onto this constant-$R$ grid. The v5.2 stellar count spectrum is interpolated onto the same grid and used as the stellar spectrum in the observation model.

\subsection{Assumptions and inputs to the ETC}\label{subsec:four_three}
Table~\ref{tab:etc} summarizes the main instrumental and observing parameters used in the ETC calculations. We assume an airmass of $1.0$ and adopt an integration time of 10\,min per exposure. This choice is short enough to avoid significant orbital smearing of the planetary absorption lines during individual exposures; using the orbital parameters in Table~\ref{tab:inputs}, we estimate the maximum line-of-sight velocity drift over a 10\,min exposure at mid-transit, where the velocity changes most rapidly: the largest drift in our sample occurs for TRAPPIST-1\,d, with $\Delta v_{\rm exp}\simeq0.64\,\mathrm{km\,s^{-1}}$. This is smaller than the ANDES pixel scale of approximately $1.0\,\mathrm{km\,s^{-1}}$, assuming $R \simeq 100,000$ and three pixels per resolution element. Therefore, the planetary lines remain within the same spectral pixel during an exposure, and orbital smearing is negligible for all targets in our sample. A target-by-target optimization of the exposure time, including readout overheads and detector systematics, is beyond the scope of this work. For each target, we provide the host star's $J$-band Vega magnitude and spectral type, as listed in Table~\ref{tab:inputs}. Because the ETC relies on a finite set of models from the \citet{Pickles1998} stellar library, we approximate each true spectral type by the closest available option.

\begin{deluxetable*}{ll}
\tabletypesize{\footnotesize}
\tablewidth{0pt}
\tablecaption{Main ANDES ETC parameters.}\label{tab:etc}
\tablehead{
\colhead{Quantity} & \colhead{Adopted value or treatment}
}
\startdata
Observing mode & Seeing-limited point-source mode \\
Resolving power & $100,000$ \\
Exposure time & 600\,s \\
Airmass & 1.0 \\
Magnitude & $J$-band Vega system (see Table~\ref{tab:inputs}) \\
Stellar templates & Closest available \citet{Pickles1998}-library spectral type \\
Seeing & 0.8\,arcsec \\
Sky transparency & PHOTOMETRIC \\
Moon phase & $0$ \\
Binning & $1\times1$ \\
Number of separate readouts & 1 \\
\enddata
\tablecomments{Parameters used for the ETC v5.2 calculations. ANDES includes an AO-assisted IFU, but the present transmission-spectroscopy simulations use the seeing-limited point-source mode.}
\end{deluxetable*}

Although readout noise is included in our calculations, we ignore the readout overhead; for the several-minute exposures considered here, the per-frame overhead is expected to be small (a few percent at most), so this approximation does not materially affect our results.

The sky-background term is taken directly from the wavelength- and airmass-dependent sky-background tables distributed with the ANDES ETC, separately for the UBV, RIZ, and YJH bands. The ETC tables used here do not expose separate OH-airglow, zodiacal-light, moonlight, or optical-sky components, so we cannot decompose the sky-background term into individual physical contributions. We therefore use the total ETC sky-background model directly.

To evaluate whether an additional OH-line mask would substantially reduce the usable wavelength range, we estimate the fractional coverage of OH-line cores using the ESO SOFI OH line list.\footnote{\url{https://www.eso.org/sci/facilities/lasilla/instruments/sofi/tools/OHskyLines.html}} We mask $\pm 2\lambda/R$ around each tabulated line at $R=100,000$, merge overlapping intervals, sum the resulting OH-covered wavelength intervals, and divide by the total wavelength span. This gives an OH-core coverage of only $\simeq 1.1\%$ over the near-infrared wavelength range covered by the ETC output, with approximate Y, J, and H fractions of 1.0\%, 1.7\%, and 2.0\%, respectively. This is not a full OH-emission model, but it shows that the loss of wavelength coverage would be small at ANDES resolution.

\subsection{Building the final spectra}\label{subsec:four_four}

Using the procedure described in Section~\ref{subsec:four_two}, we obtain three arrays for each band on the same reconstructed constant-$R$ wavelength grid: the wavelength array, the ETC S/N per resolution element, and the v5.2 stellar count spectrum. The band-specific arrays are concatenated to form a single spectrum and the effective exposure time is derived based on the assumed 10\,min integration time.

The ETC includes telluric absorption through a TAPAS-based atmospheric transmission spectrum \citep{Bertaux2014}. Thus, the wavelength dependence of the ETC-derived throughput and S/N already reflects atmospheric transmission losses. In our simulations, we do not apply an additional multiplicative telluric spectrum to the synthetic planet--star spectra. We instead assume that telluric lines can be corrected down to the photon-noise level, with the most affected regions removed by the S/N masks described below; the limitations of this optimistic assumption are discussed in Section~\ref{subsec:eight_four}.

We then apply a two-step S/N mask before adding synthetic noise. First, we discard the lowest-S/N 3\% of ETC wavelength elements over the full wavelength range. This global cut was chosen empirically to remove the strongest low-S/N outliers in the ETC wavelength grid while preserving nearly all of the usable spectral coverage. A comparison with TAPAS atmospheric-transmission models shows that these rejected pixels are strongly enriched in telluric absorption regions: across the planets tested, the median fraction of rejected pixels with TAPAS transmission below 0.8 is 37.0\%, compared with 1.7\% among retained pixels. The rejected pixels are therefore not selected from a predefined telluric or OH-line mask; they also include wavelengths at the edges of the instrumental range where the predicted S/N drops. We then divide the spectrum into $0.06\,\mu$m bins and, within each bin, discard the lowest 5\% of ETC wavelength elements by S/N, as well as any pixels with zero S/N. We then intersect these masks across all transit counts, retaining only pixels that satisfy the S/N threshold for every exposure. This ensures that all exposure times share an identical wavelength grid. We then add synthetic Gaussian noise on the retained wavelength grid using the corresponding ETC S/N values.

\begin{figure*}[ht!]
    \centering
    \includegraphics[width=0.85\textwidth]{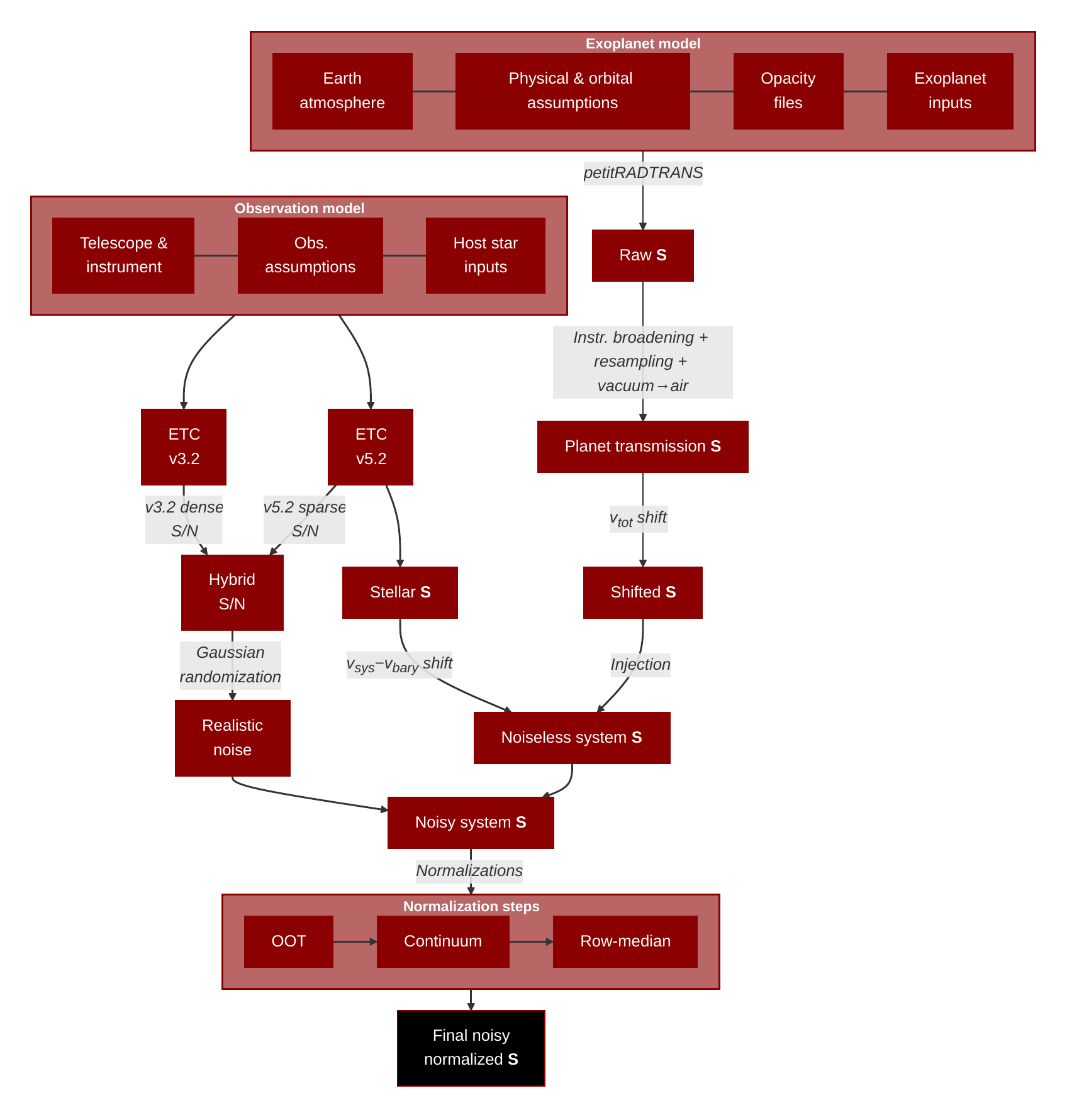}
    \caption{Workflow for constructing the high-resolution transmission spectrum used as input to the CCF analysis for a given molecule-planet pair. Here, $\mathbf{S}$ denotes the spectrum at each stage of the workflow.}
    \label{fig:four}
\end{figure*}

We normalize each simulated spectrum by sequentially dividing it by its out-of-transit time-averaged stellar baseline, and its median spectrum along wavelength to correct for throughput variations and stellar continuum fluctuations. The corresponding uncertainty from the finite out-of-transit baseline is propagated analytically, including covariance terms for out-of-transit exposures. The noisy spectrum is also continuum-normalized as described by \citet{Bello-Arufe2023}. To do so, we first fit the continuum using a Gaussian filter with a standard deviation of $\sim 500$\,pixels. We then divide the spectrum by the continuum such that we isolate high-frequency spectral features while suppressing the low-frequency continuum. We propagate the pixel uncertainties accordingly. This spectrum is then used as input for a cross-correlation analysis to evaluate the detection significance as a function of the number of observed transits. For clarity, the complete procedure for constructing this final spectrum is summarized in Figure~\ref{fig:four}.

\section{CCF analysis and limitations of peak-S/N estimates}\label{sec:five}

Having generated the noisy spectra, we next define the cross-correlation statistic used throughout this work. For each planet-molecule configuration, Monte Carlo realization, and assumed cumulative number of transits $n$, we simulate a phase-sampled sequence spanning twice the transit duration and centered on mid-transit (Section~\ref{sec:four}).

For the final detection step, we reduce each normalized sequence to a single representative central in-transit exposure and compute a one-dimensional CCF from that spectrum. The time series still enters the analysis through the construction of the out-of-transit reference spectrum and the propagated uncertainty budget. We adopt this choice as a simple and homogeneous approach across the large grid of planet-molecule configurations explored here. In particular, it avoids introducing additional assumptions about how to combine spectra obtained at different in-transit velocities and phases, and it places all targets on the same footing when comparing relative detectability. The CCF is therefore well suited to the comparative objectives of this study, although a phase-resolved combination of all in-transit spectra may affect the absolute detection strengths.

For each cumulative transit count $n$, we cross-correlate the baseline-corrected, continuum-normalized representative in-transit spectrum, $S_{\mathrm{obs}}(\lambda, n)$, with a noiseless molecular template, $S_T(\lambda,v)$, shifted from $-200$ to $200\,\mathrm{km\,s^{-1}}$ in steps of $1\,\mathrm{km\,s^{-1}}$. To ensure a common wavelength interval for all trial velocities, we truncate the spectrum to the subset of pixels that remains valid after Doppler shifting over the full velocity range. For each trial velocity, the shifted template is normalized to unit sum and mean-subtracted over the retained wavelength interval; the observed spectrum is mean-subtracted over that same interval. The CCF is inverse-variance weighted, using the per-pixel variances propagated through the out-of-transit, continuum, and median normalizations. Using the propagated variance $\sigma^2(\lambda, n)$, we compute
\begin{equation}
c(v,n)=\sum_{\lambda}\frac{\left[S_T(\lambda,v)-\overline{S_T}(v)\right]\left[S_{\mathrm{obs}}(\lambda, n)-\overline{S_{\mathrm{obs}}}(n)\right]}{\sigma^2(\lambda,n)}
\label{eq:ccf}
\end{equation}
where the overbars denote the mean over the retained wavelength interval.

In the CCF, the planetary contribution is expected near the known reference velocity $v_{\rm tot}$ (Equation~\ref{eq:v_tot}). A common approach in the literature is to estimate a detection significance by dividing the CCF peak amplitude by the standard deviation of CCF values far from the expected planetary signal \citep[e.g.,][]{Brogi2013,deKok2013,Bello-Arufe2022}. In our simulations, however, this approach is not sufficiently robust. Although we assume ideal telluric correction at the spectral level, wavelengths coincident with strong telluric absorption have lower S/N in the ETC output and therefore contribute noisier pixels to the CCF. Because these wavelengths are fixed in the observer frame, their contribution tends to produce enhanced variance, and occasionally localized positive or negative peaks, near $v=0\,\mathrm{km\,s^{-1}}$ in the CCF. These features must be interpreted as noise associated with telluric absorption bands, and are not an explicitly simulated residual telluric spectrum. The CCF baseline is also not flat, the variance changes with velocity, and neighboring velocity bins are correlated because nearby shifted templates overlap strongly in wavelength space. As a result, the far-from-peak standard deviation is not a reliable local noise estimate, and peak-based S/N values can lead to biased target rankings.

As a first mitigation, we apply the wavelength masks described in Section~\ref{subsec:four_four} before computing the CCF, removing the lowest-S/N pixels both globally and within local wavelength bins. Even after masking, however, the CCF retains velocity-dependent variance and covariance, especially near $v=0\,\mathrm{km\,s^{-1}}$, because some low-S/N pixels associated with telluric absorption bands remain even after our double-masking procedure. We therefore adopt a Bayesian treatment that models the full one-dimensional CCF as the data vector. Building on the HRCCS retrieval framework that recasts the CCF as a likelihood \citep{Brogi2019,Gibson2020}, we extend it with molecule-specific kernels to capture the expected CCF morphology and an autoregressive AR(1) term to model inter-bin covariance directly in velocity space. We then use \texttt{MultiNest} to compute model evidences, and the resulting Bayes factors provide the basis for assessing molecular detections.

\section{Bayesian evidence estimation of atmospheric signals}\label{sec:six}

To determine when a species is detected, we apply Bayesian model comparison to the 1-D CCF, contrasting a null model with a signal model incorporating the expected molecular signal from the planetary atmosphere, hereafter referred to as the ``signal'' model. Using \texttt{MultiNest}, we compute the evidences $Z_{\rm sig}$ and $Z_{\rm null}$, where

\begin{equation}
Z_{m} = \int \mathcal{L}_{m}(\mathbf{y}\mid \theta_{m})\pi_{m}(\theta_{m})\mathrm{d}\theta_{m}
\end{equation}
with likelihood $\mathcal{L}_{m}$ for the CCF data $\mathbf{y}$ and prior $\pi_{m}$ over parameters $\theta_{m}$ of model $m\in\{\mathrm{sig},\mathrm{null}\}$. We compare the two models via the Bayes factor $B$, calculated as
\begin{equation}
\log_{10} B = \log_{10} Z_{\rm sig} - \log_{10} Z_{\rm null}
\label{eq:log10B}
\end{equation}
and we map $\log_{10} B$ to Jeffreys categories for qualitative statements.

\subsection{CCF data vector and noise model}\label{subsec:six_one}

For each planet-molecule configuration, we start by generating $M=100$ independent trials: in trial $k$ we draw a new realization of the noise, process it through the simulation pipeline, and obtain a CCF array $\{y_i^{(k)}\}$ sampled at velocity bins $\{v_i\}$. This ensemble constitutes our Monte Carlo sample for that configuration. Each CCF spans $[-200, 200]\,\mathrm{km\,s^{-1}}$, sampled on $N=401$ velocity bins, typically with a candidate peak near the expected total planetary velocity $v_{\rm tot}$ (Equation~\ref{eq:v_tot}) and, in some cases, a residual telluric feature near $0\,\mathrm{km\,s^{-1}}$. We then estimate the empirical dispersion across trials,
\begin{equation}
\sigma_i = \sqrt{\frac{1}{M-1}\sum_{k=1}^{M}\left(y_i^{(k)} - \bar{y}_i\right)^2},\qquad
\bar{y}_i = \frac{1}{M}\sum_{k=1}^{M} y_i^{(k)}
\end{equation}
where $y_i^{(k)}$ is the CCF value at velocity bin $v_i$ in trial $k$. In practice, adjacent bins within a given CCF are correlated. The per-bin standard deviations $\sigma_i$, estimated across Monte Carlo realizations, quantify the empirical scatter at each velocity bin but do not encode the covariance between neighboring bins within a single CCF realization (Appendix~\ref{sub_app:B_one}). Rather than estimating the full CCF covariance matrix $\boldsymbol{\Sigma}$ directly from the spectra (Appendix~\ref{sub_app:B_two}), we approximate it by $\tilde{\boldsymbol{\Sigma}} = \mathbf{D}\,\mathbf{R}\,\mathbf{D}$, where $\mathbf{D}=\mathrm{diag}(\sigma_1,\dots,\sigma_N)$ contains the per-velocity standard deviations and $\mathbf{R}$ is a stationary AR(1) correlation matrix with elements $R_{ij}=\hat\rho_{\rm w}^{|i-j|}$ (Appendix~\ref{sub_app:B_three}).
The lag-1 correlation coefficient $\hat\rho_{\rm w}$ is estimated from off-peak velocity regions only, excluding both the planetary signal and the telluric residual near zero velocity, then computed using a Yule-Walker estimator for a stationary AR(1) process and clipped to $|\hat\rho_{\rm w}|<0.9$ (Appendix~\ref{sub_app:B_four}).

To prevent local variance anomalies from disproportionately weighting the signal-bearing bins, we replace the empirical $\sigma_i$ near $v_{\rm tot}$ and near $v=0$ by a single robust scale $\tilde{\sigma}$, defined as the median of $\sigma_i$ over off-peak velocities, while retaining the original $\sigma_i$ elsewhere. The resulting standard deviation array, together with $\hat\rho_{\rm w}$, is used in the AR(1) likelihood to account for correlated noise in velocity space.

\subsection{Null and signal models}\label{subsec:six_two}
\begin{figure}[ht!]
    \centering
    \includegraphics[width=\columnwidth]{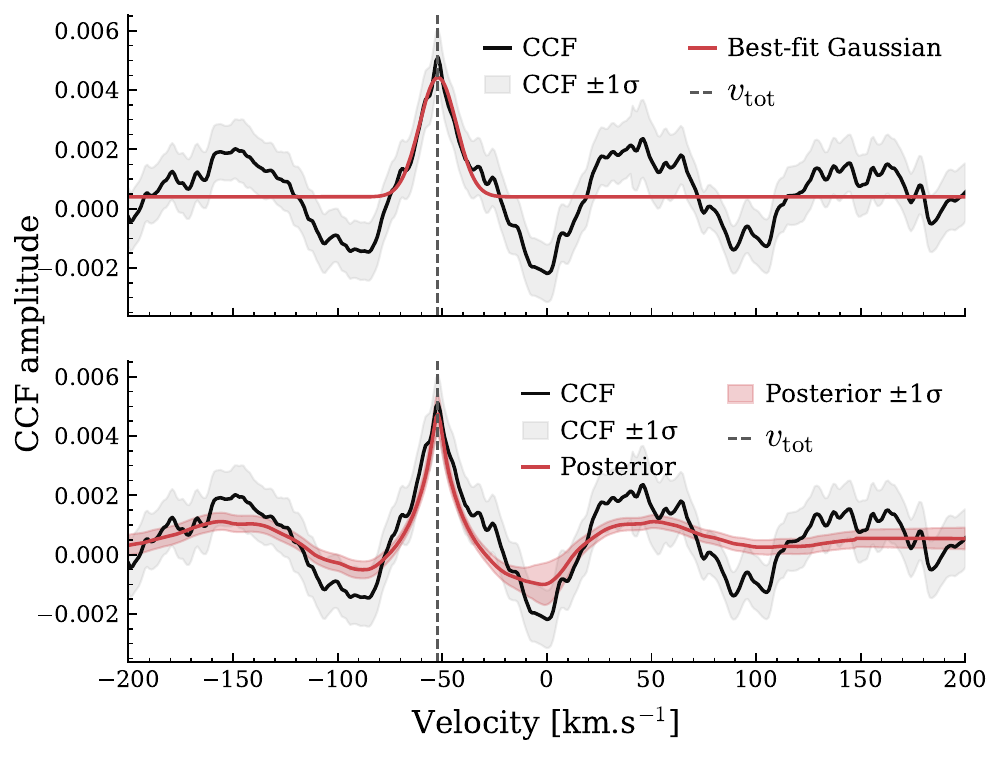}
    \caption{CCF for CO$_2$ on TRAPPIST-1\,e corresponding to 40 stacked transits, fitted with two different models. \textbf{Top:} Gaussian + constant baseline fitted by bounded non-linear least-squares (Trust-Region Reflective algorithm), minimizing the sum of squared residuals. It fails to capture the secondary lobes. \textbf{Bottom:} posterior predictive mean of the signal model, from our Bayesian framework. The molecule-specific kernel captures the full shape of the CCF, outperforming the simple Gaussian model.}
\label{fig:five}
\end{figure}

A straightforward approach would be to fit a Gaussian profile to the CCF maximum. However, such a fit would not address the telluric contamination discussed in Section~\ref{sec:five}. In addition, some of the species we study exhibit quasi-regular line spacings within a vibrational band, which generate secondary peaks in the CCF at characteristic lags; this is a generic consequence of structured-template autocorrelation in correlation analyses \citep[e.g., for a signal processing example,][]{Said2016}. O$_2$ and CO$_2$ exhibit structured band patterns in the spectral regions considered here \citep{Kiehl1985,Martin1932}, which can produce side lobes in the CCF through template autocorrelation, as illustrated in Figure~\ref{fig:five}. By contrast, CH$_4$ and H$_2$O possess dense and irregular spectra \citep[e.g.,][]{Yurchenko2014, Yurchenko2024, Barber2006, Tennyson2013}, leading to a single dominant CCF maximum with no discernible secondary structure. A simple Gaussian fit, with its flat baseline, cannot capture the side lobes. The baseline then absorbs mid-scale features, biasing the recovered amplitude $A$ toward lower values and reducing the inferred evidence. This behavior is illustrated in the top panel of Figure~\ref{fig:five}.

To address this, we introduce a molecule-specific model that more accurately represents the shape of the CCF. For the Bayesian comparison, both hypotheses (null and signal) share a quadratic baseline that captures the smooth, low-order curvature of the CCF across the $\pm 200\,\mathrm{km\,s^{-1}}$ range:
\begin{equation}
B(v) = y_0 + b(v - v_{\rm tot}) + c(v - v_{\rm tot})^2
\end{equation}
where $(y_0, b, c)$ are nuisance parameters and $v_{\rm tot}$ is the reference velocity defined in Equation~\ref{eq:v_tot}.

Both models also include a zero-velocity nuisance component,
\begin{equation}
T(v) = A_{\rm tel}\,\exp\left(-\frac{v^2}{2\sigma_{\rm tel}^2}\right)
\end{equation}
modeled as a Gaussian centered at zero velocity with free amplitude $A_{\rm tel}$ and width $\sigma_{\rm tel}$. 
This term captures localized CCF structure caused by noisy pixels in telluric absorption bands (Appendix~\ref{app:A}) and is treated as a nuisance component in both the signal and null hypotheses.

The null model is therefore
\begin{equation}
m_{\rm null}(v) = B(v) + T(v)
\end{equation}
while the signal model adds a molecular kernel centered at the expected planetary velocity,
\begin{equation}
m_{\rm sig}(v) = B(v) + A\,K(v - v_{\rm tot}) + T(v)
\end{equation}
where $A$ is the signal amplitude. The kernel $K$ depends on the molecule and is tabulated on the CCF velocity grid as a function of the lag $\Delta v = v - v_{\rm tot}$.

Operationally, the kernel is constructed as the autocorrelation of the molecular template $\mathcal{T}$,
\begin{equation}
K(\Delta v) \propto \int \mathcal{T}(u)\,\mathcal{T}(u+\Delta v)\,du
\end{equation}
$K$ is evaluated on the same velocity grid as the CCF. In practice, the kernel is mean-subtracted and normalized to unit $\ell_2$ norm before fitting. Mean subtraction ensures that the kernel captures only structured features in the CCF and does not absorb any constant offset, which is modeled separately by the baseline. Unit $\ell_2$ normalization fixes the overall scale of the kernel, so that the amplitude parameter $A$ directly measures the strength of the molecular signal on a common scale across all species. Using this molecule-specific kernel instead of a simple Gaussian preserves the characteristic secondary lobes present in the CCFs of species with quasi-regular line spacings. The resulting improvement over a Gaussian model is illustrated in the bottom panel of Figure~\ref{fig:five}.

\subsection{Priors}\label{subsec:six_three}

We choose weakly informative priors on the model parameters. In the implementation used for this work, the kernel centroid is fixed at the expected total planetary velocity $v_{\rm tot}$ (Equation~\ref{eq:v_tot}), and the kernel width is set by the template autocorrelation; we therefore do not fit additional centroid or stretch parameters.

\begin{itemize}
\item \textbf{Amplitude.}
From the Monte Carlo CCF ensemble we compute a per-velocity standard deviation $\sigma_i$ (Section~\ref{subsec:six_one}). We then define $\tilde{\sigma}$ as the median of $\sigma_i$ over ``off-peak'' velocities, excluding both the planetary core and the telluric region (i.e., $|v - v_{\rm tot}| < 40~\mathrm{km\,s^{-1}}$ and $|v| < 40~\mathrm{km\,s^{-1}}$). For each molecule we introduce an amplitude scale
\begin{equation}
A_\sigma = f_{\mathrm{mol}}\,\tilde{\sigma}
\end{equation}
and adopt a half-normal prior
\begin{equation}
A \sim \mathrm{HalfNormal}(A_\sigma)
\end{equation}
In this work we set $f_{\mathrm{mol}} = 7$ for all species.

\item \textbf{Baseline.} We place normal priors on the offset, slope and curvature of the baseline:
\begin{equation}
\begin{aligned}
y_0 &\sim \mathcal{N}\!\left(0,\tilde{\sigma}\right),\\
b &\sim \mathcal{N}\!\left(0, \frac{0.2A_\sigma}{\Delta v_{\rm span}}\right),\\
c &\sim \mathcal{N}\!\left(0, \frac{0.1A_\sigma}{\Delta v_{\rm span}^2}\right)
\end{aligned}
\end{equation}
where $\Delta v_{\rm span}$ denotes the full velocity span of the CCF.

\item \textbf{Telluric component.}
The telluric amplitude $A_{\rm tel}$ is assigned a zero-mean normal prior,
\begin{equation}
A_{\rm tel} \sim \mathcal{N}\left(0,\tilde{\sigma}\right)
\end{equation}
while the width is given a log-uniform prior,
\begin{equation}
\sigma_{\rm tel} \sim \mathcal{LU}\left(5,80\right)
\end{equation}
\end{itemize}

\subsection{Likelihood and nested sampling}\label{subsec:six_four}

All priors are built via inverse cumulative distribution function (CDF) transforms from unit-hypercube draws $u\sim\mathcal{U}(0,1)$ in \texttt{MultiNest} and numerically clipped to $\left[10^{-9},\,1-10^{-9}\right]$ to avoid infinities. Transforms are implemented with \texttt{SciPy} \citep{Virtanen2020}. For a parameter $\theta$:

\textit{(i) Normal prior.} For $\theta\sim\mathcal{N}(\mu,\sigma^2)$,
\begin{equation}
\theta = \mu + \sigma\sqrt{2}\,\operatorname{erf}^{-1}(2u - 1)
\end{equation}
where $\operatorname{erf}^{-1}$ is the inverse error function.

\textit{(ii) Half-normal prior.} For a non-negative parameter with scale $\sigma$,
\begin{equation}
\theta = \sigma\sqrt{2}\,\operatorname{erf}^{-1}(u), \qquad \theta \ge 0
\end{equation}

\textit{(iii) Log-uniform prior.} For bounds $[x_{\min},x_{\max}]$,
\begin{equation}
\theta = x_{\min}\left(\frac{x_{\max}}{x_{\min}}\right)^{u}
\end{equation}

These mappings ensure that \texttt{MultiNest} samples the intended prior measures directly from $u$ while maintaining numerical stability in the likelihood evaluation. 

The likelihood is derived from the multivariate normal distribution:
\begin{equation}
\begin{aligned}
\ln \mathcal{L}(\mathbf{y}\mid \boldsymbol{\theta})= -\tfrac{1}{2}\Big[&\ln\!\big(\det(\mathbf{\Sigma})\big) \\&+ (\mathbf{y}-\mathbf{m}(\boldsymbol{\theta}))^{\top}\mathbf{\Sigma}^{-1}(\mathbf{y}-\mathbf{m}(\boldsymbol{\theta})) \\&+ N\ln(2\pi)\Big]
\end{aligned}
\label{eq:multivariate-likelihood}
\end{equation}
where $\mathbf{y}=(y_1,\dots,y_N)^\top$ is the observed CCF sampled on the velocity grid $\{v_i\}_{i=1}^N$, $\mathbf{m}(\boldsymbol{\theta})$ is the model prediction evaluated on the same grid, $N$ is the number of velocity bins, $\mathbf{\Sigma}$ is the $N\times N$ covariance matrix of the residual vector, and $\det(\mathbf{\Sigma})$ is the determinant of $\mathbf{\Sigma}$.

Under the AR(1) approximation (Section~\ref{subsec:six_one} and Appendix~\ref{sub_app:B_three}) and using the clipped Yule-Walker estimator $\hat{\rho}_{\rm w}$ derived in Appendix~\ref{sub_app:B_four}, the likelihood admits the closed-form expression
\begin{align}
\ln \mathcal{L}(\mathbf{y}\mid\boldsymbol{\theta},\hat{\rho}_{\rm w})
= &- \sum_{i=1}^N \ln\sigma_i
-\frac{N-1}{2}\ln\!\left(1-\hat{\rho}_{\rm w}^2\right) \nonumber \\
&-\frac{1}{2}\Bigg[z_1^2
+ \frac{1}{1-\hat{\rho}_{\rm w}^2}
\sum_{i=2}^N (z_i-\hat{\rho}_{\rm w} z_{i-1})^2 \Bigg] \nonumber \\
&-\frac{N}{2}\ln(2\pi)
\label{eq:ar1-loglike-final}
\end{align}
where \(z_i = \left[y_i - m_i(\boldsymbol{\theta})\right]/\sigma_i\) denotes the standardized residual at velocity bin \(v_i\). It renders the likelihood computationally tractable within nested sampling while retaining the dominant velocity-bin correlation structure of the CCF. A full description and justification of the correlation model and estimator is provided in Appendix~\ref{app:B}.

We fit the null and atmospheric-signal models with \texttt{MultiNest} via \texttt{PyMultiNest} \citep{Buchner2016}, obtaining posterior samples and the log-evidence \(\ln Z\) for each model, and we compute the Bayes factor defined in Equation~\ref{eq:log10B}.

As a basic calibration test of the Bayesian framework, we also applied the same analysis to null templates containing no injected planetary signal. In these cases, the model comparison consistently favored the null hypothesis, with Bayes factors indicating no spurious preference for the signal model. This provides an initial validation that the framework is not predisposed to favor the signal model in the absence of injected signal.

\section{Application of the detection framework}\label{sec:seven}

We apply our analysis pipeline\footnote{The framework developed in this work, \texttt{ANDES-Forecast}, is publicly available at \url{https://github.com/EvannKurzawa/ANDES-Forecast}} to the 18 planets retained after the visibility and radius cuts described in Section~\ref{sec:two}. For each planet, we evaluate the detectability of H$_2$O, CH$_4$, CO$_2$, and O$_2$, using six cumulative transit counts corresponding to 3, 10, 25, 50, 75, and 100 transits, adopting 100 transits as a practical upper limit.

For each planet--molecule pair and each transit count, we generate $M \geq 100$ independent CCF realizations. We compute the Bayes factor for each one and use the median $\log_{10} B$ over these $M$ draws as the representative detection strength. The Monte Carlo stability of this median is assessed using the stopping criterion described in Appendix~\ref{app:C}. We then estimate the minimum number of transits required to reach ``decisive'' evidence (Jeffreys scale; $\log_{10} B \ge 2.0$), by linearly interpolating the median Bayes factors obtained across the transit-count grid.

\begin{figure}[t!]
    \centering
    \includegraphics[width=\columnwidth]{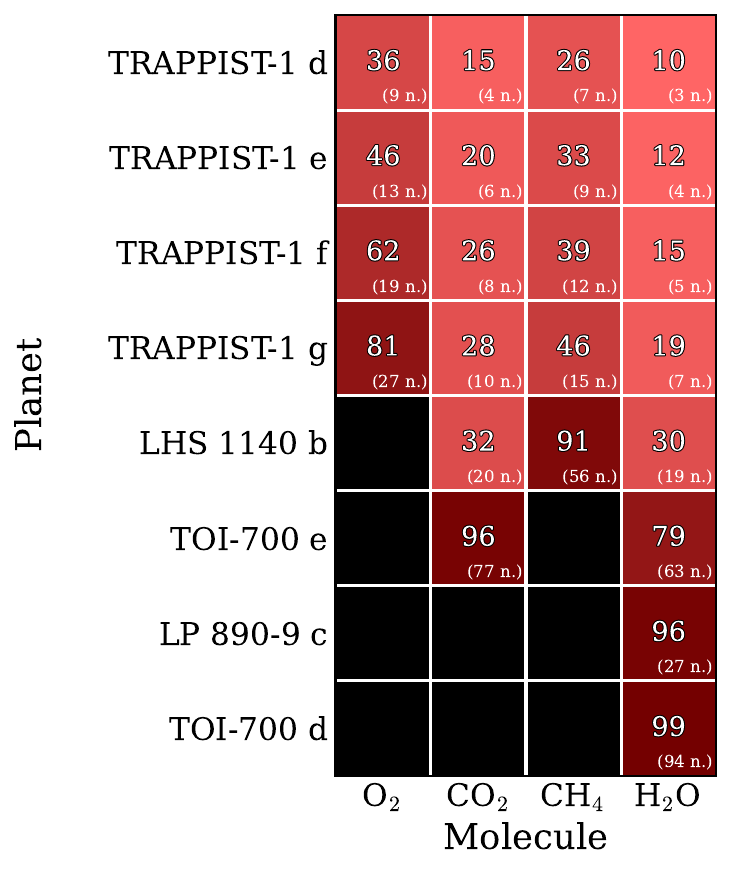}
    \caption{Estimated minimum number of transits required for decisive evidence ($\log_{10} B \geq 2.0$) of each gas with ANDES within our framework. Planets are ordered from top to bottom by (i) number of detectable molecules, and (ii) smaller total required transits (sum over detected molecules). Empty cells indicate that a decisive detection cannot be achieved in 100 transits under the assumptions adopted here. The number in parentheses gives the approximate equivalent observing time in 7\,hr nights, rounded up to the nearest integer, and represents idealized time only; it does not include weather or seasonal constraints.}
\label{fig:six}
\end{figure}

All planets that yield decisive evidence for at least one gas in fewer than 100 transits are listed in Figure~\ref{fig:six}. This criterion reduces the sample from eighteen to eight planets. Figure~\ref{fig:six} compiles these estimates and provides the most planning-relevant result: a concise view of the transit requirement for each gas and retained target. As a rough measure of observing cost, we also report in each cell the corresponding idealized number of 7\,hr observing nights. This quantity is computed as
\begin{equation}
N_{\rm nights} =\frac{2\,N_{\rm transits}\,T_{14}}{7\,{\rm hr}}
\end{equation}
where $T_{14}$ is the transit duration and the factor of two corresponds to the simulated observing sequence spanning one $T_{14}$ in transit plus an equal out-of-transit baseline. The resulting $N_{\rm nights}$ values are rounded up to the nearest integer. They should be interpreted as idealized observing-night equivalents, not as real scheduling estimates, since they do not include, e.g., weather losses and seasonal constraints. We discuss its implications in Section~\ref{subsec:eight_one}.

\section{Results and discussion}\label{sec:eight}
\subsection{Main detectability results}\label{subsec:eight_one}

From the initial pool of 70 targets in our sample, we exclude 29 nontransiting ones, 11 that are rejected by our screening proxy based on the adopted 20$^\circ$ altitude criterion, and 12 on the basis of planetary radius. Among the remaining 18 planets, 8 meet the Jeffreys decisive detection threshold for at least one molecule within 100 transits. These final detectability estimates are summarized in Figure~\ref{fig:six}, which provides the main synthesis of our results. Within the assumptions adopted here, H$_2$O emerges as the most accessible species, meeting the detectability criterion for all eight planets in less than 100 transits. CO$_2$ is detectable for six planets and CH$_4$ is detectable for five, while O$_2$, detectable for four targets, proves the most challenging.

The TRAPPIST-1 system yields particularly favorable results, with all four molecules detectable on each of the four planets considered. Among them, TRAPPIST-1\,d emerges as the most promising target overall in our ranking, requiring the fewest transits for every molecule and reaching full detectability within 36 transits. Three molecules are also detectable on LHS 1140\,b within the 100-transit limit, although the required transit counts remain operationally difficult.

First, the planet-to-planet differences in detectability are predominantly explained by variations in scale height. The amplitude of individual planetary lines varies approximately with the atmospheric scale height, as $\Delta D \propto H R_p / R_\star^2$, where $H = kT/(\mu g)$ increases with temperature and decreases with gravity and mean molecular weight \citep{Seager2000}. This explains why planets with higher temperature and lower gravity, such as TRAPPIST-1\,d, yield stronger features than cooler, higher-gravity planets like TRAPPIST-1\,e.

Second, the contrast between the molecular detectability patterns in, e.g., the TRAPPIST-1 and LHS 1140 systems is instead primarily driven by the spectral energy distributions of their host stars. CO$_2$ absorption is probed mainly near 1.6\,$\mu$m, whereas CH$_4$ and H$_2$O draw much of their information from red optical and near-infrared bands. Cooler stars like TRAPPIST-1 emit a larger fraction of their energy in the red optical and near-infrared than earlier M dwarfs such as LHS 1140, leading to a higher photon budget in the relevant CH$_4$/H$_2$O bands. Synthetic atmosphere models confirm that late-M dwarfs concentrate a larger share of their flux at longer wavelengths, with flux escaping primarily through the J and H bands in that order of decreasing strength \citep{Allard1995}. As a result, CH$_4$ and H$_2$O are easier to detect in TRAPPIST-1 than in the earlier-type system LHS 1140, even though CO$_2$ detectability remains comparable across both.

Finally, although TOI-700\,d, TOI-700\,e, and LP 890-9\,c (also known as SPECULOOS-2\,c) reach a decisive detection within the 100-transit limit adopted here, the required number of transits remains large enough to make such observations operationally impractical.

\subsection{Comparison with previous studies}\label{subsec:eight_two}

Establishing direct comparisons with previous studies is challenging. While a limited number of works have explored the potential of the ELT (or other extremely large telescopes) for characterizing the atmospheres of potentially habitable rocky planets around M dwarfs \citep[e.g.,][]{Currie2023, Wunderlich2020}, these typically consider hypothetical planets at distances that differ from those of our sample (often 5--15\,pc). Most of these studies targeted individual absorption bands, and none specifically investigated the performance of ANDES. Our comparisons should therefore be interpreted as approximate contextual checks rather than as strict one-to-one validations.

For comparison purposes only, we provide in Appendix~\ref{app:D} a band-limited version of our analysis for TRAPPIST-1\,e, together with the corresponding transit requirements for selected molecular bands. This appendix uses the heuristic threshold $\log_{10} B = 5.43$ solely to facilitate comparison with studies reporting 5$\sigma$ estimates; these values are not used for the main target rankings or for the transit requirements shown in Figure~\ref{fig:six}.

\citet{Wunderlich2020} explored a range of atmospheric scenarios for TRAPPIST-1\,e and f, including CO$_2$-rich cases, and estimated the number of transits required for a 5$\sigma$ detection in several individual spectral bands. For the O$_2$ band at 1.24--1.30\,$\mu$m, they reported that up to 910 transits could be required. Applying our band-limited mode to TRAPPIST-1\,e over the same O$_2$ band (Appendix~\ref{app:D}; Table~\ref{tab:t1e}), we do not reach the heuristic ``5$\sigma$'' threshold within 300 transits. This remains qualitatively consistent with the conclusion that O$_2$ is extremely challenging to detect in transmission, although the numerical values cannot be compared directly because of differences in atmospheric assumptions, noise treatment, and detection metrics. \citet{Wunderlich2020} also reported 26 transits for CH$_4$ in the 2.12--2.50\,$\mu$m band; however, this wavelength interval lies outside the range retained in our ETC-based simulations.

Later, \citet{Currie2023} simulated transmission spectroscopy of an Earth-like planet transiting an M dwarf, with an orbital configuration analogous to TRAPPIST-1\,e. They reported that the ELT could detect CO$_2$ at 1.56\,$\mu$m in 34 transits, CH$_4$ at 1.6\,$\mu$m in 33 transits, O$_2$ at 1.27\,$\mu$m in 200 transits, and H$_2$O at 0.9\,$\mu$m in $>300$ transits (all at a 5$\sigma$ significance level). Using our band-limited mode applied to TRAPPIST-1\,e up to 300 transits, we find that CO$_2$ at 1.56--1.62\,$\mu$m reaches our heuristic ``5$\sigma$'' threshold after 151 transits, while CH$_4$ at 1.62--1.70\,$\mu$m, O$_2$ at 1.24--1.30\,$\mu$m, and H$_2$O at 0.90--0.98\,$\mu$m do not reach this threshold within 300 transits. Overall, the comparison remains qualitatively consistent in identifying CO$_2$ as one of the most accessible of these band-limited cases and H$_2$O as very challenging in the 0.9\,$\mu$m band, but our transit requirements are generally less favorable.

The numerical discrepancies with previous estimates likely arise in part because our simulations adopt the most recent ANDES ETC outputs, which reflect an updated instrumental design with lower throughput, and therefore lower S/N, in some wavelength regions. We propagate this updated S/N scale into the simulated spectra by locally rescaling the dense v3.2 wavelength structure to the v5.2 normalization and interpolating the result onto a constant-$R$ grid. This treatment of the ETC outputs differs from both the simpler feature-based S/N estimates of \citet{Wunderlich2020} and the SPECTR-based simulated observations of \citet{Currie2023}, which use a separate implementation of the ELT/instrument noise budget.

\citet{Palle2025} presented an ANDES forward-modeling estimate for TRAPPIST-1\,d and reported that a wet N$_2$-dominated atmosphere containing 400\,ppm CO$_2$ could be recovered at $>12\sigma$ in 5 transits under photon-limited assumptions. This result is broadly consistent with the idea that molecular absorption features in TRAPPIST-1\,d could be accessible to ANDES under optimistic assumptions. A direct numerical comparison, however, is difficult. The \citet{Palle2025} calculation uses a multi-species atmospheric template and a peak-based cross-correlation significance, whereas our analysis treats each absorber separately and uses Bayesian model comparison with a correlated-noise CCF likelihood. In our framework, H$_2$O and CO$_2$ reach decisive evidence for TRAPPIST-1\,d within 15 transits, while CH$_4$ requires 26 transits. The difference likely reflects several conservative aspects of our setup: the use of the updated ETC v5.2 normalization, the absence of seeing-limited K-band information, and the use of a Bayesian evidence metric with correlated CCF noise.

Finally, comparison with previous target-prioritization studies reveals strong convergence on the most amenable targets for atmospheric characterization. Both \citet{Hill2023} and \citet{Bohl2025} identify TRAPPIST-1\,d as one of the most favorable habitable-zone planets, consistent with our findings, where TRAPPIST-1\,d also ranks highest (see Figure~\ref{fig:six}). In our framework, the remaining potentially habitable TRAPPIST-1 planets (e, f, g) follow closely, similar to the sequence reported by \citet{Bohl2025}. Other planets that appear prominently in at least one of these works, as well as in ours, include LHS 1140\,b, TOI-700\,d, and LP 890-9\,c, reinforcing their status as strong secondary candidates. However, the low density of LHS 1140\,b may indicate a water world composition instead of a purely rocky one \citep{Damiano2024, Cadieux2024}, although the planet might still retain habitable conditions on its surface.

\subsection{Preliminary JWST results for these targets}\label{subsec:eight_three}

Nearly all of the planets identified in our final sample have been or will be observed by the James Webb Space Telescope (JWST), and these early results provide valuable context to test the assumptions of our atmospheric modeling. Initial JWST observations appear inconsistent with significant atmospheres on most warm-to-hot rocky exoplanets around M dwarfs (e.g., \citealt{gillon2025,meiervaldes2025,allen2025,fortune2025,luque2025}; but see also \citealt{gressier2024,august2025,belloarufe2025}), but it is still unclear whether this trend extends to the temperate population \citep{Damiano2024,cadieux2024b,Espinoza2025,Piaulet2025,Glidden2025}.

For TRAPPIST-1\,d, our top-ranked target, JWST/NIRSpec transmission spectra disfavor a modern Earth-like, cloud-free atmosphere like the one we model in this work \citep{Piaulet2025}. Instead, the data are consistent with either a tenuous Mars-like atmosphere, a high-altitude Venus-like cloud deck, or no atmosphere at all. In contrast, existing JWST observations of TRAPPIST-1\,e rule out H$_2$-dominated atmospheres but remain consistent with high-mean-molecular-weight atmospheres, including N$_2$-dominated compositions with trace CO$_2$ or CH$_4$ \citep{Espinoza2025,Glidden2025}. Meanwhile, LHS 1140\,b has a featureless or weakly sloped transmission spectrum that disfavors a clear H$_2$/He-dominated envelope and is consistent with a high-mean-molecular-weight atmosphere or water-rich scenario \citep{Damiano2024,cadieux2024b}.

Taken together, early JWST results suggest that modern, cloud-free Earth-like atmospheres remain plausible for nearly all targets in our sample, including LHS 1140\,b and TRAPPIST-1\,e. While our simulations present lower limits on the number of transits required for atmospheric detection with ANDES, the ongoing synergy between JWST and the ELT will be crucial in refining our understanding of atmospheric diversity and biosignature detectability on rocky exoplanets.
 
\subsection{Important caveats}\label{subsec:eight_four}

Below, we summarize the most relevant caveats.

\begin{enumerate}
    \item \textbf{Telluric correction.} We assume that telluric absorption lines can be removed down to the photon-noise level. This remains a significant challenge, particularly in the near-infrared where telluric contamination is most severe and time-variable. Current methods such as \texttt{molecfit} \citep{Smette2015, Kausch2015} and data-driven principal component analysis (PCA; e.g., \citealt{Meech2022}) can achieve excellent results under favorable conditions, but full correction at the noise limit is rarely attained. The achievable precision depends on the instrument stability, the accuracy of the telluric models, and the temporal variability of the atmosphere during the observing sequence. Residual telluric structures may therefore introduce additional systematic noise that increases the effective number of required transits.

    \item \textbf{Planet-star spectral separation.} 
    We assume that the planetary spectrum can be isolated from the stellar and telluric signals without degrading the exoplanetary features. This is a strong assumption, especially for the terrestrial planets considered here, whose radial-velocity drift during transit is typically only a few tenths to a few kilometers per second. Traditional detrending techniques, including SYSREM \citep{Tamuz2005, Mazeh2007} or PCA-based filtering, rely on the assumption that the planetary lines shift sufficiently in velocity space to avoid being removed together with the quasi-stationary stellar and telluric lines. In practice, this assumption can fail for terrestrial systems, causing significant attenuation of the planetary signal. To address this limitation, \citet{Piskunov2025} recently introduced \emph{TSD} (Transmission Spectroscopy Decomposition), an inverse-problem framework that simultaneously solves for the stellar, telluric, and planetary spectra directly from the time series, thereby avoiding template cross-correlation and minimizing signal loss. Although promising, the method requires resolving the planet's velocity change during transit and may therefore remain challenging for slowly orbiting rocky planets. Complementary approaches based on machine learning are also emerging; for instance, \citet{Kjaersgaard2023} proposed \emph{TAU}, a neural-network-based telluric and stellar correction framework that can disentangle overlapping spectral components without relying on linear decomposition. These techniques represent an important step beyond classical detrending but have yet to be extensively validated in the low-velocity, low-signal-to-noise regime relevant to temperate terrestrial planets.
    
    \item \textbf{Clouds and hazes.}
    We assume cloud-free atmospheres, although clouds and hazes may be present on these planets \citep[e.g.,][]{yang2013}, potentially muting spectral features and complicating atmospheric characterization \citep{Lustig-Yaeger2019}. High-resolution transmission spectroscopy, however, probes the cores of individual lines, which form at low pressures, and is therefore less sensitive to clouds and hazes than its low-resolution counterpart (e.g., JWST) \citep{Gandhi2020}.

    \item \textbf{Stacking and noise scaling.} We assume that stacking multiple transits improves the signal-to-noise ratio as $\sqrt{N_\mathrm{transits}}$, but it is unclear whether this will be true for ELT/ANDES. The effective scaling will need to be established empirically during commissioning and early surveys; until then, our estimates should be considered optimistic lower limits.

    \item \textbf{Stellar contamination.} We do not explicitly account for stellar variability, active regions, or flares, which are common among M dwarfs. Such activity can alter the effective photospheric spectrum during transit and bias molecular detections (the so-called transit light source effect, \citealt{rackham2018,rathcke2025}). While high-resolution spectroscopy provides some mitigation by leveraging the planet's orbital motion to disentangle the planetary and stellar components \citep{cauley2018}, the impact of stellar contamination on high-resolution transmission spectroscopy of M dwarf systems remains poorly constrained.

    \item \textbf{Single-species treatment.} We compute transmission spectra for one target absorber at a time rather than performing full multi-species radiative transfer. Molecular overlap and masking are therefore neglected, so the reported transit requirements should be interpreted as optimistic estimates for isolated species.

    \item \textbf{Screening proxy versus true observability.}
    Our target selection and estimated transit counts are based on a simplified altitude-based visibility proxy rather than a full transit-by-transit observability analysis. The sampled epochs used to evaluate the altitude criterion and barycentric correction are not tied to actual transit times.

    \item \textbf{Additional simplifying assumptions,} including circular orbits, Earth-like atmospheric composition and surface pressure, and a negligible readout overhead compared to the 10\,min integration time. In addition, because we apply the optimal barycentric correction to maximize Doppler separation between exoplanetary and telluric spectral lines, our results represent an optimistic scenario. In practice, observers are encouraged to schedule observations such that the exoplanet spectral lines are Doppler-shifted away from the noisy telluric lines in order to increase the S/N of molecular features.

\end{enumerate}

Overall, these assumptions imply that our results likely underestimate the true number of transits required for detection. The goal of this work is to provide optimistic estimates of molecular detectability with ELT/ANDES under idealized observing conditions, and to estimate the minimum transit counts required for the most favorable transiting targets. Future efforts could aim to relax these assumptions and refine the physical modeling and pipeline implementation, and we encourage the community to focus on these challenges before ANDES starts operations, to maximize scientific return.

\section{Summary and Conclusions}\label{sec:nine}

We developed a framework to estimate ELT/ANDES transit requirements for detecting four key atmospheric molecules, including two biosignature gases, on potentially habitable exoplanets. For each planet, we modeled a cloud-free, Earth-like, two-layer atmosphere and simulated the transmission spectra with \texttt{petitRADTRANS}, adding realistic noise with the ANDES ETC. We cross-correlated the resulting spectra with model templates, and we quantified the detection significance using a Bayesian framework featuring a custom kernel scheme to emulate the CCFs and an autoregressive AR(1) model to mitigate correlation across velocity bins. We then applied the full pipeline to the most favorable planets in our sample, which yielded evidence scores and the required transit numbers to reach detectability. 

Our simulations indicate that water vapor and carbon dioxide are the most accessible species for ground-based detection with ANDES, requiring fewer than $\sim$20 transits for the most favorable targets (e.g., TRAPPIST-1\,d and e). Methane is moderately accessible, while oxygen remains the most challenging species to detect. Eight planets show at least one molecule detectable within $\lesssim$100 transits: TRAPPIST-1\,d, e, f, g; LHS 1140\,b; LP 890-9\,c; and TOI-700\,d and e. TRAPPIST-1\,d emerges as the most favorable target overall, with all four gases detectable in fewer than 36 transits. These results should be regarded as optimistic lower limits, contingent on photon-noise performance, perfect telluric and stellar removal, and cloud-free atmospheres. 

In summary, this study highlights the potential of ELT/ANDES for atmospheric characterization of nearby transiting terrestrial exoplanets in the photon-limited regime. Under these assumptions, the detection of key molecules central to the search for life, such as CO$_2$, CH$_4$, and H$_2$O on the TRAPPIST-1 planets may be feasible within campaigns of $\lesssim 30$ transits. Achieving this performance will require substantial progress in telluric and stellar contamination removal, which remain the dominant barriers to reaching the photon-limited regime, particularly for slowly moving planets, where spectral disentanglement is most challenging.

Finally, although this work focuses on transiting planets, our results also highlight the importance of complementary searches around nearby nontransiting systems. Transmission spectroscopy benefits from a well-defined observing geometry, but it is restricted to the small subset of habitable-zone planets that transit and, as shown here, may require large observing campaigns even under optimistic assumptions. Reflected-light observations combining high-contrast imaging and high-resolution spectroscopy offer an alternative route for ANDES, because they can target the nearest habitable-zone planets regardless of whether they transit \citep{Snellen2015,Wang2017,Mawet2017,Vaughan2024,Palle2025}. Such systems may therefore be at least as important as transiting planets for future ELT biosignature searches, and deserve a dedicated analysis beyond the transmission-spectroscopy study presented here.

\clearpage
\onecolumngrid

\begin{acknowledgments}
E.K.F. gratefully acknowledges the Région Auvergne-Rhône-Alpes and the Fondation Ailes de France for their financial support. 
This research was carried out at the Jet Propulsion Laboratory, California Institute of Technology, under a contract with the National Aeronautics and Space Administration (80NM0018D0004). 
Part of the High Performance Computing resources used in this investigation were provided by funding from the JPL Information and Technology Solutions Directorate. We also thank the ANDES ETC team, and in particular Nicoletta Sanna and Alessandro Marconi, for their helpful clarifications regarding the ETC outputs and the evolving ANDES instrumental assumptions.

This research has made use of the NASA Exoplanet Archive, which is operated by the California Institute of Technology, under contract with the National Aeronautics and Space Administration under the Exoplanet Exploration Program. It has also made use of the SIMBAD database, operated at CDS, Strasbourg Astronomical Observatory, France; the Habitable Worlds Catalog \hwc{PHL}; the DACE platform (Data \& Analysis Center for Exoplanets) operated at the University of Geneva within the Swiss National Centre of Competence in Research (NCCR); the Keeper knowledge base of the Max Planck Digital Library; the TESS Input Catalog; the USNO-B Catalog; data products from the Two Micron All Sky Survey (2MASS); and public resources provided by the European Southern Observatory (ESO), including the ELT construction proposal, the ANDES documentation, and the ANDES Exposure Time Calculator.
\end{acknowledgments}

\software{
\texttt{Astropy} \citep{Astropy2022},
\texttt{SciPy} \citep{Virtanen2020},
\texttt{PyAstronomy} \citep{PyAstronomy},
\texttt{barycorrpy} \citep{Kanodia2018},
\texttt{petitRADTRANS} \citep{Molliere2019},
\texttt{SPRIGHT} \citep{Parviainen2024},
\texttt{PyMultiNest} \citep{Buchner2016},
ANDES Exposure Time Calculator \citep{Sanna2024},
\texttt{ANDES-Forecast} (\url{https://github.com/EvannKurzawa/ANDES-Forecast})}

\appendix
\restartappendixnumbering

\makeatletter
\renewcommand{\theHequation}{\thesection.\arabic{equation}}
\renewcommand{\theHfigure}{\thesection.\arabic{figure}}
\renewcommand{\theHtable}{\thesection.\arabic{table}}
\makeatother

\section{Zero-velocity artifact}\label{app:A}
\begin{figure}[hbp!]
    \centering
    \includegraphics[width=0.6\columnwidth]{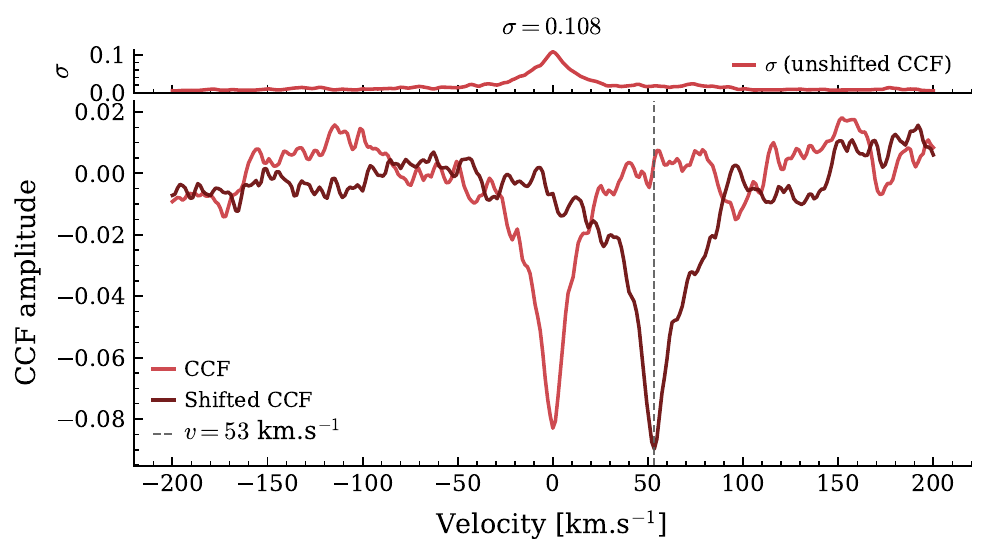}
    \caption{CCFs for Kepler-1649\,c with low-S/N pixel masking. Two CCFs are shown: the standard one, and one where the stellar, planetary, and telluric contributions are shifted independently. In the shifted case, the feature follows the telluric-noise contribution at $53\,\mathrm{km\,s^{-1}}$, as discussed in Appendix~\ref{app:A}. For illustration, we display the standard deviation curve from the unshifted case, which shows a pronounced uncertainty peak at $0\,\mathrm{km\,s^{-1}}$. This supports the interpretation that the feature is caused by noisy pixels associated with telluric absorption bands.}
\label{fig:a1}
\end{figure}

We employ a one-dimensional CCF analysis, which in several cases reveals a narrow peak at $v \approx 0\,\mathrm{km\,s^{-1}}$ in the observer frame. This manifests as either a positive or negative peak, as shown in Figure~\ref{fig:a1}, and accounts for most of the variance near zero lag. Consequently, ``far-from-signal'' standard-deviation estimates cease to be representative of the baseline, and classical S/N metrics become unreliable.

This zero-velocity behavior is a consequence of how telluric absorption enters our simulations. We assume ideal telluric correction at the spectral level and do not inject a deterministic residual telluric spectrum; however, the ANDES ETC assigns lower S/N to wavelengths affected by telluric absorption. The effect can be understood as follows:

\begin{enumerate}
    \item \textbf{Stationarity in the observer frame.} Wavelengths affected by telluric absorption are fixed in the observer's rest frame. Even after ideal telluric correction, these wavelengths retain larger uncertainties in the ETC noise model. Their noise contribution can therefore produce enhanced CCF variance near $v \approx 0\,\mathrm{km\,s^{-1}}$, including localized positive or negative peaks in individual realizations.
    \item \textbf{Molecular overlap with the target gases.} The dense line lists of O$_2$, H$_2$O, CO$_2$, and CH$_4$ overlap strongly with Earth's atmosphere, so even small residual mismatches can mimic the spectral patterns used to search for exoplanet signals.
\end{enumerate}

We perform the following diagnostics, which directly link the zero-velocity feature to telluric-noise effects:

\begin{enumerate}
    \item \textbf{Velocity shift experiment.}  
    The observed spectrum can be decomposed into contributions from the star, the planet, and Earth's atmosphere. To test their individual contributions, we artificially shift each component by arbitrary velocities: $53\,\mathrm{km\,s^{-1}}$ for the telluric spectrum, $29\,\mathrm{km\,s^{-1}}$ for the stellar spectrum, and $97\,\mathrm{km\,s^{-1}}$ for the planetary spectrum. In the resulting CCFs, the spurious feature shifted exactly with the telluric component ($53\,\mathrm{km\,s^{-1}}$), whereas structures associated with the stellar and planetary spectra did not follow. This isolates the terrestrial component as the origin of the artifact.

    \item \textbf{Monte Carlo uncertainty analysis.}  
    We construct the CCF standard deviation array by repeating the calculation hundreds of times with Monte Carlo sampling. This reveals a localized peak in uncertainty centered at $0\,\mathrm{km\,s^{-1}}$, consistent with excess variance introduced by telluric features, supporting the interpretation of a telluric origin.
\end{enumerate}

A representative example of these two diagnostics is shown in Figure~\ref{fig:a1}.

In the Bayesian analysis, this zero-velocity CCF structure is marginalized over by including a Gaussian nuisance component centered at zero velocity in both the signal and null models.

\clearpage
\section{Correlation between CCF velocity bins: origin, ideal treatment, and our practical approach}\label{app:B}

Here we derive the CCF correlation structure, using classical time-series results on AR(1) processes \citep[e.g.,][]{Hamilton1994, Wilks2011}.

\subsection{Correlation origin and behavior with $\Delta v$}\label{sub_app:B_one}
Let the continuum-normalized spectrum at pixel index $p$ be
\begin{equation}
d_p = s_p + n_p
\end{equation}
with signal $s_p$ and noise $n_p$ (zero mean). For a velocity $v$, we correlate the data with a Doppler-shifted template $t_p(v)$, applying inverse-variance weights $w_p$ to each pixel:
\begin{equation}
y(v) = \sum_{p=1}^{P} \underbrace{\big[w_p\,t_p(v)\big]}_{\displaystyle \phi_p(v)} d_p = \boldsymbol{\phi}(v)^{\!\top}\mathbf{d}
\end{equation}
Here $\boldsymbol{\phi}(v)\in\mathbb{R}^P$ is the velocity-dependent weight vector, $y(v)$ is the CCF between the data and the template, and $\mathbf{d}$ is the data vector.
Sampling $y(v)$ on a velocity grid $\{v_i\}_{i=1}^N$ yields the CCF vector $\mathbf{y}\in\mathbb{R}^N$ with components $y_i = y(v_i)$. Because nearby weight vectors $\boldsymbol{\phi}(v_i)$ and $\boldsymbol{\phi}(v_j)$ are shifted versions of the same template, they overlap strongly. Using the covariance of linear combinations of a random vector, this overlap couples $y_i$ and $y_j$:
\begin{equation}
\mathrm{Cov}(y_i,y_j) = \mathrm{Cov}\big(\boldsymbol{\phi}_i^{\top}\mathbf{d}, \boldsymbol{\phi}_j^{\top}\mathbf{d}\big) = \boldsymbol{\phi}_i^{\top}\mathbf{C}_d\,\boldsymbol{\phi}_j
\label{eq:covyiyj}
\end{equation}
where $\mathbf{C}_d=\mathrm{Cov}(\mathbf{d})$ is the data covariance matrix, and $\boldsymbol{\phi}_i = \boldsymbol{\phi}(v_i)$.

If we consider white noise (same variance $\sigma^2$ for each pixel, no correlations) for the data, then the data covariance matrix is diagonal:
$\mathbf{C}_d=\sigma^2 \mathbf{I}$, hence
\begin{equation}
\mathrm{Cov}(y_i,y_j) = \sigma^2\, \boldsymbol{\phi}_i^{\top}\boldsymbol{\phi}_j,
\qquad
\mathrm{Var}(y_i)=\sigma^2\|\boldsymbol{\phi}_i\|^2
\end{equation}
Using the Pearson correlation coefficient, the correlation between CCF bins is
\begin{equation}
\rho_{ij} = \frac{\mathrm{Cov}(y_i,y_j)}{\sqrt{\mathrm{Var}(y_i)\,\mathrm{Var}(y_j)}} = \frac{\boldsymbol{\phi}_i^{\top}\boldsymbol{\phi}_j}{\|\boldsymbol{\phi}_i\|\,\|\boldsymbol{\phi}_j\|}
\label{eq:rho_discrete}
\end{equation}
Thus, $\rho_{ij}$ is the normalized dot product between two weight vectors, such that $\rho_{ii}=1$. If $\boldsymbol{\phi}_j$ is a velocity-shifted version of $\boldsymbol{\phi}_i$, the correlation is the normalized overlap of the template with itself at that shift. Writing a continuous weight function $\phi(x)$ and a velocity lag $\Delta v = v_j-v_i$, the continuum analogue of Equation~\ref{eq:rho_discrete} is the normalized inner product:
\begin{equation}
\rho(\Delta v) = \frac{\int \phi(x)\phi(x-\Delta v)dx}{\int \phi(x)^2dx}
\end{equation}

For a Gaussian-shaped weight, let $\phi(x)=\exp[-x^2/(2\sigma^2)]$. Then
\begin{align}
\int \phi(x)\phi(x-\Delta v)\,dx
&=\int \exp\left[-\frac{x^2+(x-\Delta v)^2}{2\sigma^2}\right]dx \\
&=\exp\left[-\frac{(\Delta v)^2}{4\sigma^2}\right]\int \exp\left[-\frac{(x-\Delta v/2)^2}{\sigma^2}\right]dx \nonumber\\
\end{align}
and since shifting the center of a Gaussian does not change its total area, the integral is independent of $\Delta v$. Therefore,
\begin{equation}
\boxed{\rho(\Delta v) \propto \exp\left[-\frac{(\Delta v)^2}{4\sigma^2}\right]}
\label{eq:gauss_rho}
\end{equation}
In practice, our kernel is the autocorrelation of the molecular template, not an ideal Gaussian. For CH$_4$ and H$_2$O, the shape is close to Gaussian, while for CO$_2$ and O$_2$ it resembles a Gaussian with secondary lobes (see Section~\ref{subsec:six_two}). Nevertheless, \(\rho(\Delta v)\) still decays with increasing \(\Delta v\), as expected from the diminishing overlap between shifted kernels, with the Gaussian case of Equation~\ref{eq:gauss_rho} providing a useful approximation to the general behavior.

\subsection{Ideal treatment: full covariance across velocity bins}\label{sub_app:B_two}
For a given realization, let $\mathbf{y}\in\mathbb{R}^N$ denote the CCF vector sampled on the $N$-point velocity grid, and let $\mathbf{m}(\boldsymbol{\theta})$ be the corresponding model prediction, with $i$th component $m_i(\boldsymbol{\theta})$. Since the CCF bins are linear combinations of Gaussian noise contributions at the pixel level, they are themselves Gaussian. The correct distribution for a vector of correlated Gaussian variables is the multivariate normal distribution, with covariance matrix $\mathbf{\Sigma}$. The likelihood is then the standard multivariate normal form \citep[e.g.,][]{Bishop2006}:
\begin{equation}
\ln \mathcal{L}(\mathbf{y}\mid \boldsymbol{\theta})
= -\tfrac{1}{2}\Big[
\ln(\det(\mathbf{\Sigma}))
+ (\mathbf{y}-\mathbf{m}(\boldsymbol{\theta}))^{\top}\mathbf{\Sigma}^{-1}(\mathbf{y}-\mathbf{m}(\boldsymbol{\theta}))
+ N\ln(2\pi)
\Big]
\label{eq:full-likelihood}
\end{equation}
where $\det(\mathbf{\Sigma})$ corresponds to the determinant of $\mathbf{\Sigma}$.
From Equation~\ref{eq:covyiyj}, let us consider multiple exposures indexed by $e$.
Each exposure has a matrix $\mathbf{F}^{(e)}$, whose $i$-th row is $\boldsymbol{\phi}^{(e)}(v_i)^{\top}$, and a noise covariance matrix $\mathbf{C}^{(e)}$ for its spectrum.
The combined CCF covariance is then
\begin{equation}
\mathbf{\Sigma} = \sum_{e}\mathbf{F}^{(e)}\mathbf{C}^{(e)}\mathbf{F}^{(e)\top}
\end{equation}

This expression follows directly from the linearity of the CCF, and therefore holds analytically. It naturally incorporates both variations in noise across pixels and correlations between neighboring pixels.

Evaluating Equation~\ref{eq:full-likelihood} requires applying $\mathbf{\Sigma}^{-1}$ and computing $\ln(\det(\mathbf{\Sigma}))$. A straightforward Cholesky factorization $\mathbf{\Sigma}=\mathbf{L}\mathbf{L}^{\top}$ is $\mathcal{O}(N^3)$ per likelihood call \citep[e.g.,][]{Krishnamoorthy2011}, which is too expensive to use inside nested sampling for every CCF realization. Instead of constructing $\mathbf{\Sigma}$ explicitly from the spectra, we approximate it at the CCF level with a simple AR(1) correlation model, which still captures the dominant neighboring-bin correlations while admitting an analytic, $\mathcal{O}(N)$ evaluation of the likelihood.

\subsection{Practical approach: AR(1) likelihood derivation}
\label{sub_app:B_three}

Starting from the full multivariate Gaussian likelihood (Equation~\ref{eq:full-likelihood}), we derive an explicit closed-form expression for the likelihood under an AR(1) correlation model.

\subsubsection{Covariance structure}\label{sub_sub_app:B_three_one}
Let
\begin{equation}
e_i = y_i - m_i(\boldsymbol{\theta})
\end{equation}
denote the residuals on the velocity grid $\{v_i\}_{i=1}^N$. We assume that the residuals follow a stationary AR(1) process,
\begin{equation}
e_i = \rho\,e_{i-1} + \varepsilon_i, \qquad |\rho|<1
\end{equation}
where $\varepsilon_i$ is a zero-mean white-noise innovation with variance
$\sigma_\varepsilon^2$. 
From the variance recursion, we obtain
\begin{equation}
\mathrm{Var}(e_i) = \sigma^2 = \frac{\sigma_\varepsilon^2}{1-\rho^2}
\end{equation}
and by induction on $h$, the lag-$h$ correlation is
\begin{equation}
\mathrm{Corr}(e_i, e_{i+h}) = \rho^{h}
\end{equation}

By iterating the AR(1) relation, we further obtain
\begin{equation}
\mathrm{Cov}(e_i,e_j) = \sigma^2 \rho^{|i-j|}
\end{equation}

Allowing for heteroscedastic noise across the velocity grid, we define an
approximate covariance matrix for the residual vector $\mathbf{e}$ as
\begin{equation}
\tilde{\mathbf{\Sigma}} = \mathbf{D}\,\mathbf{R}\,\mathbf{D}
\approx \mathbf{\Sigma}
\end{equation}

where $\mathbf{D}=\mathrm{diag}(\sigma_1,\dots,\sigma_N)$ with
$\sigma_i=\sqrt{\mathrm{Var}(e_i)}$ collects the per-bin standard deviations. The matrix $\mathbf{R}$ is the correlation matrix of the
standardized residuals $z_i=e_i/\sigma_i$, whose elements are
$R_{ij}=\rho^{|i-j|}$ and satisfy $\mathrm{Var}(z_i)=1$.

\subsubsection{Determinant of the covariance matrix}\label{sub_sub_app:B_three_two}
Using the multiplicative property of determinants, we have
\begin{equation}
\det(\tilde{\mathbf{\Sigma}})
= \det(\mathbf{D}\mathbf{R}\mathbf{D})
= \det(\mathbf{D})^2\,\det(\mathbf{R})
\end{equation}
Since $\det(\mathbf{D})=\prod_{i=1}^N\sigma_i$, it follows that
\begin{equation}
\ln(\det(\tilde{\mathbf{\Sigma}})) = 2\sum_{i=1}^N\ln(\sigma_i) + \ln(\det(\mathbf{R}))
\end{equation}

\subsubsection{Standardized residuals}\label{sub_sub_app:B_three_three}
Defining the residual vector
$\mathbf{z}=(z_1,\dots,z_N)^\top$, the quadratic form becomes
\begin{equation}
(\mathbf{y}-\mathbf{m}(\boldsymbol{\theta}))^{\top}
\tilde{\mathbf{\Sigma}}^{-1}
(\mathbf{y}-\mathbf{m}(\boldsymbol{\theta}))
= \mathbf{z}^{\top}\mathbf{R}^{-1}\mathbf{z}
\end{equation}

\subsubsection{Innovation representation}\label{sub_sub_app:B_three_four}
For an AR(1) process, correlations between neighboring bins can be removed by an explicit lower-triangular linear transformation acting on the standardized residual vector. We define the lower-triangular matrix
\begin{equation}
\mathbf{J} =
\begin{pmatrix}
\sqrt{1-\rho^2} & 0 & 0 & \cdots & 0 \\
-\rho & 1 & 0 & \cdots & 0 \\
0 & -\rho & 1 & \cdots & 0 \\
\vdots & \vdots & \vdots & \ddots & \vdots \\
0 & 0 & 0 & -\rho & 1
\end{pmatrix}
\end{equation}

Applying this transformation to the standardized residual vector $\mathbf{z}$ defines the innovation vector $\boldsymbol{\eta} = \mathbf{J}\mathbf{z}$, whose components are independent Gaussian variables with variance $(1-\rho^2)$. Explicitly,
\begin{equation}
\eta_1 = \sqrt{1-\rho^2}\,z_1, \qquad \eta_i = z_i - \rho z_{i-1}, \quad i \ge 2
\end{equation}

It follows that
\begin{equation}
\mathbf{R}^{-1} = (1-\rho^2)^{-1}\mathbf{J}^\top\mathbf{J}
\label{eq:inverted_r}
\end{equation}

Using this identity, the quadratic form in the likelihood can be written as
\begin{equation}
\mathbf{z}^\top \mathbf{R}^{-1} \mathbf{z}
= (1-\rho^2)^{-1}\, \mathbf{z}^\top \mathbf{J}^\top\mathbf{J}\,\mathbf{z}
= (1-\rho^2)^{-1}\, \boldsymbol{\eta}^\top \boldsymbol{\eta}
\end{equation}

Since $\boldsymbol{\eta}^\top\boldsymbol{\eta} = \|\boldsymbol{\eta}\|_2^2$
is the squared Euclidean norm of the innovation vector, we obtain
\begin{equation}
\boldsymbol{\eta}^\top \boldsymbol{\eta}
= \eta_1^2 + \sum_{i=2}^N \eta_i^2
= (1-\rho^2)z_1^2 + \sum_{i=2}^N (z_i - \rho z_{i-1})^2
\end{equation}

and therefore
\begin{equation}
\mathbf{z}^\top\mathbf{R}^{-1}\mathbf{z} = z_1^2
+ \frac{1}{1-\rho^2}\sum_{i=2}^N \bigl(z_i - \rho z_{i-1}\bigr)^2
\label{eq:ar1-qf}
\end{equation}

\subsubsection{Determinant of the correlation matrix}\label{sub_sub_app:B_three_five}
From the triangular structure of $\mathbf{J}$, its determinant is given by
the product of its diagonal elements,
\begin{equation}
\det(\mathbf{J}) = \sqrt{1-\rho^2}
\end{equation}

Using Equation~\ref{eq:inverted_r} and the multiplicativity of the determinant, we obtain
\begin{equation}
\det(\mathbf{R}^{-1})
= (1-\rho^2)^{-N}\,\det(\mathbf{J}^\top)\,\det(\mathbf{J})
= (1-\rho^2)^{-N}\,\det(\mathbf{J})^2
= (1-\rho^2)^{-(N-1)}
\end{equation}

Taking the inverse gives
\begin{equation}
\det(\mathbf{R}) = (1-\rho^2)^{N-1}, \qquad \ln(\det(\mathbf{R})) = (N-1)\ln(1-\rho^2)
\end{equation}

\subsubsection{Final AR(1) log-likelihood}\label{sub_sub_app:B_three_six}
Substituting the above expressions into the full likelihood yields
\begin{equation}
\boxed{\ln \mathcal{L}(\mathbf{y}\mid\boldsymbol{\theta},\rho)
= - \sum_{i=1}^N \ln\sigma_i-\frac{N-1}{2}\ln(1-\rho^2)-\frac{1}{2}\Bigg[z_1^2 + \frac{1}{1-\rho^2}\sum_{i=2}^N (z_i-\rho z_{i-1})^2 \Bigg]-\frac{N}{2}\ln(2\pi)}
\label{eq:ar1-loglike-final-appendix}
\end{equation}

This expression is exact for the AR(1) covariance model and can be evaluated
in $\mathcal{O}(N)$ time without explicitly constructing or inverting
the covariance matrix.

\subsection{Implementation details and estimation of the AR(1) coefficient}
\label{sub_app:B_four}

In the implementation used for the Bayesian model comparison, the constant term $-\tfrac{N}{2}\ln(2\pi)$ in the log-likelihood is dropped, as it cancels in Bayes factors and does not affect posterior inferences.

\subsubsection{Estimating the per-bin scatter}\label{sub_sub_app:B_four_one}
The standard deviations $\sigma_i$ at each velocity bin are estimated empirically from the Monte Carlo ensemble of CCF realizations for a fixed planet, exposure time, and molecule. To avoid letting localized structures in the planetary and telluric regions dominate the noise estimate, we replace $\sigma_i$ within $|v-v_{\rm tot}|\le 40~\mathrm{km\,s^{-1}}$ and $|v|\le 40~\mathrm{km\,s^{-1}}$ by the median standard deviation measured outside these intervals. Elsewhere, the full velocity dependence of $\sigma_i$ is retained.

\subsubsection{Estimating the correlation parameter}\label{sub_sub_app:B_four_two}
The correlation parameter $\rho$ is estimated using the same off-peak velocity regions defined above, i.e., excluding $\pm40~\mathrm{km\,s^{-1}}$ around $v_{\rm tot}$ and zero velocity, so that neither planetary signal nor telluric residuals contribute to the estimator. From the $N_{\rm off}$ off-peak velocity bins, we form
\begin{equation}
u_k = y_{i_k} - \mathrm{median}\!\left(\{y_{i_k}\}_{k=1}^{N_{\rm off}}\right)
\end{equation}
where $\{i_k\}_{k=1}^{N_{\rm off}}$ are the indices of the retained off-peak velocity bins in the full CCF grid. To limit the influence of outliers,
we compute the median absolute deviation (MAD) of $\{u_k\}_{k=1}^{N_{\rm off}}$ and winsorize the sequence at $\pm 3\,\mathrm{MAD}$,
\begin{equation}
u_k^{(w)} = \mathrm{clip}\bigl(u_k,\,-3\,\mathrm{MAD},\,3\,\mathrm{MAD}\bigr)
\end{equation}

The lag-1 correlation coefficient is estimated using the Yule-Walker estimator for an AR(1) process \citep{Yule1927, Walker1931, Brockwell1991};
\begin{equation}
\gamma(1) = \mathrm{Cov}(u_k^{(w)},u_{k-1}^{(w)}) = \rho\,\gamma(0),
\qquad
\gamma(0) = \mathrm{Var}(u_k^{(w)})
\end{equation}
implying
\begin{equation}
\rho = \frac{\gamma(1)}{\gamma(0)}
     = \frac{\mathrm{Cov}(u_k^{(w)},u_{k-1}^{(w)})}{\mathrm{Var}(u_k^{(w)})}
\end{equation}

Replacing the population moments by their empirical counterparts,
\begin{equation}
\hat{\gamma}(1)
= \frac{1}{N_{\rm off}-1}\sum_{k=2}^{N_{\rm off}} u_k^{(w)}\,u_{k-1}^{(w)},\qquad
\hat{\gamma}(0)
= \frac{1}{N_{\rm off}}\sum_{k=1}^{N_{\rm off}} \bigl(u_k^{(w)}\bigr)^2
\end{equation}
yields the practical Yule-Walker estimator
\begin{equation}
\boxed{\hat{\rho}_{\rm w}= \frac{\hat{\gamma}(1)}{\hat{\gamma}(0)}= \frac{N_{\rm off}}{N_{\rm off}-1}\frac{\sum_{k=2}^{N_{\rm off}} u_k^{(w)}\,u_{k-1}^{(w)}}{\sum_{k=1}^{N_{\rm off}} \bigl(u_k^{(w)}\bigr)^2}}
\end{equation}

\subsubsection{Final likelihood evaluation}\label{sub_sub_app:B_four_three}
For a given planet, exposure time, and molecule, the estimated $\hat{\rho}_{\rm w}$ is treated as fixed and is not sampled as a free parameter. The covariance matrix is constructed as
$\tilde{\mathbf{\Sigma}}=\mathbf{D}\mathbf{R}(\hat{\rho}_{\rm w})\mathbf{D}$, and the AR(1) log-likelihood derived in Section~\ref{sub_sub_app:B_three_six} is evaluated for each CCF realization.
This approach accounts explicitly for velocity-bin correlations, reduces the effective number of independent samples when $\hat{\rho}_{\rm w}>0$, and yields deliberately conservative Bayes factors, while avoiding the computational cost of full pixel-space covariance treatments.

\section{Monte Carlo stability and stopping criterion}\label{app:C}

To quantify the Monte Carlo uncertainty in the final Bayes factors, we run the pipeline $M$ times per configuration. Here, we assess whether $M=100$ is sufficiently precise for our purposes.

We estimate the standard error (SE) of the sample median and its associated margin of error (MoE) at 95\% confidence. Let $s$ denote the Monte Carlo standard deviation, i.e., the between-realization scatter of the detection statistic. As a working approximation, when the underlying distribution is close to normal, the standard error of the sample median is
\begin{equation}
\mathrm{SE} \approx 1.253\,\frac{s}{\sqrt{M}}
\end{equation}
which is the standard asymptotic normal-approximation result for the variance of the sample median \citep[e.g.,][]{Walck1996}. For a two-sided 95\% confidence interval, we then define
\begin{equation}
\mathrm{MoE}_{95} = 1.96\,\mathrm{SE}
\label{eq:moe95}
\end{equation}

For each configuration, we compute $\mathrm{MoE}_{95}$ and define a relative, scale-aware metric,
\begin{equation}
\mathrm{rMoE}_{95} = \frac{\mathrm{MoE}_{95}}{\max\bigl(|\tilde{x}|,0.5\bigr)}
\label{eq:rmoe}
\end{equation}
where $\tilde{x}$ is the sample median of $\log_{10} B$. We adopt the following stopping criterion: if $\mathrm{rMoE}_{95}<0.5$, the median $\log_{10} B$ is considered sufficiently precise for assigning a Jeffreys category; otherwise, we increase $M$ and recompute until the condition is satisfied. The threshold $\mathrm{rMoE}_{95}<0.5$ is an operational choice adopted in this work.

This definition scales the 95\% margin of error for $\tilde{x}$ either to the signal amplitude itself (for large $|\tilde{x}|$) or to one Jeffreys step (0.5) when $|\tilde{x}|$ is small. As a result, strong signals are not over-penalized, while weak signals do not render Equation~\ref{eq:rmoe} unstable near zero. This helps ensure that Jeffreys-category assignments are not driven by Monte Carlo noise.

\begin{figure}[htbp!]
    \centering
    \includegraphics[width=0.65\columnwidth]{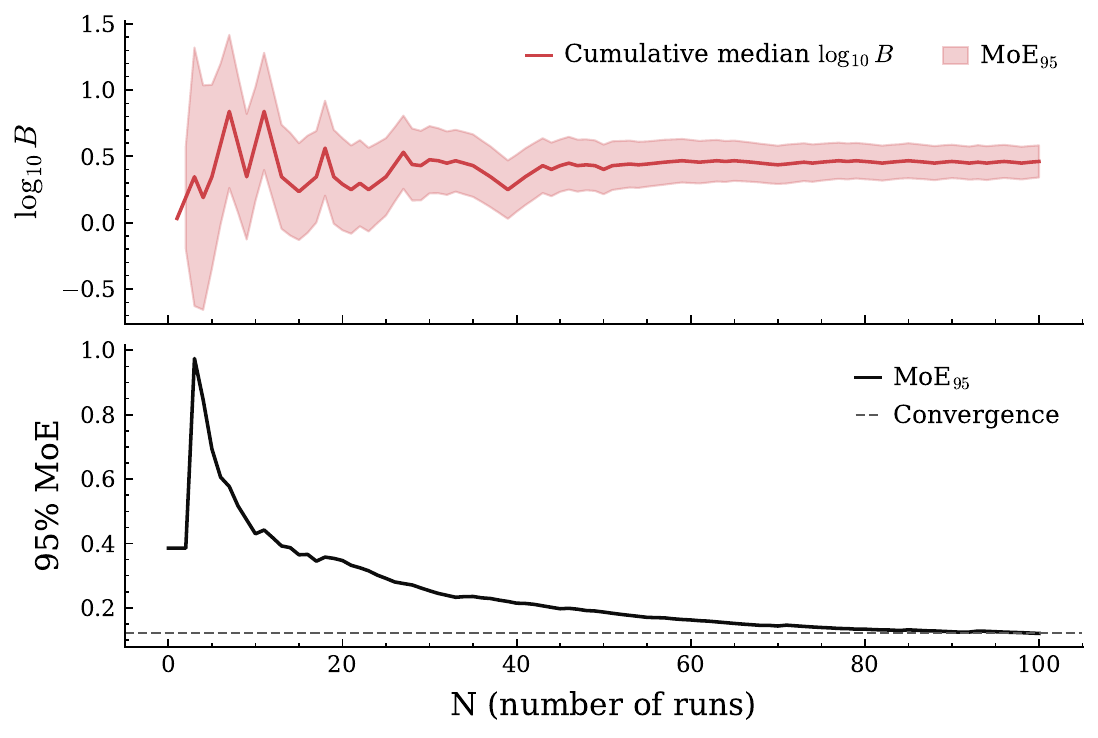}
    \caption{Results from $M=100$ independent simulation realizations followed by Bayesian analysis, shown for H$_2$O, on LP 890-9\,c, after 20 transits. \textbf{Top:} The cumulative median of $\log_{10} B$ stabilizes by $M\!\approx\!80$ realizations while the 95\% confidence interval narrows. \textbf{Bottom:} The 95\% margin of error on $\log_{10} B$ decreases approximately as $M^{-1/2}$ and stabilizes by $M\!\approx\!90$. At that point the convergence criterion is met, yielding a statistically stable estimate of $\log_{10} B$.}
\label{fig:e2}
\end{figure}

Since $\mathrm{MoE}_{95}\propto M^{-1/2}$ for fixed variance, $\mathrm{rMoE}_{95}$ decreases on average with the number of realizations $M$. In practice, the procedure reaches the target precision after a finite number of realizations for the cases considered here, as illustrated in Figure~\ref{fig:e2}.

\clearpage
\section{Band-limited mode and comparisons}\label{app:D}
\FloatBarrier

While no published study appears to provide an equally detailed ANDES-specific analysis for potentially habitable rocky exoplanets, several works have investigated this topic using general parameters representative of a high-resolution spectrograph on an extremely large telescope \citep[e.g.,][]{Fujii2018, Currie2023}. In our analysis, we obtain one CCF per realization and exposure setting over the full wavelength range of ANDES. This approach is appropriate because ANDES will use its entire spectral range simultaneously\footnote{\url{https://elt.eso.org/instrument/ANDES/}}. However, a full-range detection metric is only directly comparable to the initial forward-model detectability forecasts of \citet{Palle2025} (see Section~\ref{subsec:eight_two}), whereas most previous studies assess detectability within individual molecular bands \citep[e.g.,][]{Wunderlich2020, Currie2023}. We therefore consider, in addition to the full-wavelength-range analysis, a band-limited mode that applies the same masking and processing steps but restricts the cross-correlation to spectral intervals corresponding to relevant molecular bands. This allows us to evaluate molecule-specific spectral windows independently, even though ANDES can in principle be used over the full wavelength range simultaneously.

In light of the extensive literature on the TRAPPIST-1 system, we use this band-limited mode and compile Table~\ref{tab:t1e}, which reports the number of transits required to reach a detection threshold of $\log_{10} B=5.43$ for various bands on TRAPPIST-1\,e. Values are reported at 20, 50, 100, 200, 250, and 300 transits using the interpolation in Section~\ref{sec:seven}, and the table contains comparisons to previous studies. The specific threshold of $\log_{10} B=5.43$ corresponds, via the heuristic relation $n_\sigma \approx \sqrt{2\ln B}$ as written by \citet{Kipping2025} and motivated by the asymptotic connection between Bayes factors and likelihood-ratio statistics discussed by \citet{Kass1995}, to roughly 5$\sigma$. We adopt this only as a heuristic reference to facilitate comparison with earlier work that reported results in $\sigma$-units. We stress that such conversions between Bayes factors and frequentist significance are not formally defined and should not be interpreted literally, as discussed by \citet{Kipping2025} and \citet{Thorngren2026}. Our analysis itself is expressed entirely in terms of Bayes factors. For interpretation, readers should rely on the Bayes factors themselves; the $\sigma$ proxy is provided solely for approximate comparability.

This comparison also illustrates the impact of wavelength coverage on transit requirements. A limited spectral range such as the H band can contain much of the information for some species, especially CO$_2$ near 1.6\,$\mu$m, but it does not provide an equally effective basis for all molecules. CH$_4$ and H$_2$O draw important information from multiple NIR regions, including longer wavelengths, while O$_2$ relies on bands at shorter wavelengths and remains strongly affected by telluric overlap. Therefore, the advantage of broad simultaneous wavelength coverage is not that every wavelength contributes equally, but that different molecules draw their detectability from different parts of the optical--NIR range. In the ETC setup adopted in this work, however, the retained seeing-limited wavelength range does not include the K band, which is particularly relevant for molecules with strong longer-wavelength NIR bands, such as CH$_4$ and H$_2$O.

\begin{deluxetable*}{lccc}
\tabletypesize{\footnotesize}
\tablewidth{0pt}
\tablecaption{Transit requirements for detecting different molecular bands on TRAPPIST-1\,e.\label{tab:t1e}}
\tablehead{
  \colhead{Band} & \colhead{Transits required (this work)} & \colhead{\citet{Currie2023}} & \colhead{\citet{Wunderlich2020}} \\ \colhead{\scriptsize ($\mu$m)} & \colhead{\scriptsize $(\log_{10} B=5.43)$} & \colhead{\scriptsize $(5\sigma)$} & \colhead{\scriptsize $(5\sigma)$}
}
\startdata
CO$_2$; 1.56--1.62 & 151 & 34 & - \\
O$_2$; 1.24--1.30 & $>300$ & 200 & 910 \\
CH$_4$; 1.62--1.70 & $>300$ & 33 & - \\
H$_2$O; 0.90--0.98 & $>300$ & $>300$ & -
\enddata
\tablecomments{Number of transits required to reach a band detection threshold of $\log_{10} B=5.43$ (used here only as a rough comparability proxy to ``5$\sigma$'' studies; see Appendix~\ref{app:D}). Values are reported for each relevant spectral band from our band-limited mode, with additional columns comparing to estimates from \citet{Currie2023} and \citet{Wunderlich2020}. As discussed in Appendix~\ref{app:D}, such $\sigma$-based comparisons should be treated with caution: we interpret results via Bayes factors, whereas the cited works adopt frequentist significance.}
\end{deluxetable*}

\bibliography{andes_hwc}{}

@INPROCEEDINGS{Sanna2024,
       author = {{Sanna}, N. and {Canto Martins}, B.~L. and {Martins}, A.~M. and {Oliva}, E. and {Le{\~a}o}, I.~C. and {Turchi}, A. and {De Medeiros}, J.~R. and {Rossi}, F. and {Brucalassi}, A. and {Chazelas}, B. and {Di Marcantonio}, P. and {Gaessler}, W. and {Landoni}, M. and {Lanotte}, A.~A. and {Lee}, D. and {Marconi}, A. and {Mason}, E. and {Monteiro}, M.~A. and {Origlia}, L. and {Scaudo}, A. and {Weber}, M. and {Zanutta}, A.},
        title = "{ANDES, the high-resolution spectrograph for the ELT: the exposure time calculator}",
    booktitle = {Ground-based and Airborne Instrumentation for Astronomy X},
         year = 2024,
       editor = {{Bryant}, Julia J. and {Motohara}, Kentaro and {Vernet}, Jo{\"e}l. R.~D.},
       series = {Society of Photo-Optical Instrumentation Engineers (SPIE) Conference Series},
       volume = {13096},
        month = jul,
          eid = {130964G},
        pages = {130964G},
          doi = {10.1117/12.3018720},
       adsurl = {https://ui.adsabs.harvard.edu/abs/2024SPIE13096E..4GS}
}

@ARTICLE{Currie2023,
       author = {{Currie}, Miles H. and {Meadows}, Victoria S. and {Rasmussen}, Kaitlin C.},
        title = "{There's More to Life than O$_{2}$: Simulating the Detectability of a Range of Molecules for Ground-based, High-resolution Spectroscopy of Transiting Terrestrial Exoplanets}",
      journal = {\psj},
         year = 2023,
        month = may,
       volume = {4},
       number = {5},
          eid = {83},
        pages = {83},
          doi = {10.3847/PSJ/accf86},
archivePrefix = {arXiv},
       eprint = {2304.10683},
 primaryClass = {astro-ph.EP},
       adsurl = {https://ui.adsabs.harvard.edu/abs/2023PSJ.....4...83C}
}

@ARTICLE{Delrez2022,
       author = {{Delrez}, L. and {Murray}, C.~A. and {Pozuelos}, F.~J. and {Narita}, N. and {Ducrot}, E. and {Timmermans}, M. and {Watanabe}, N. and {Burgasser}, A.~J. and {Hirano}, T. and {Rackham}, B.~V. and {Stassun}, K.~G. and {Van Grootel}, V. and {Aganze}, C. and {Cointepas}, M. and {Howell}, S. and {Kaltenegger}, L. and {Niraula}, P. and {Sebastian}, D. and {Almenara}, J.~M. and {Barkaoui}, K. and {Baycroft}, T.~A. and {Bonfils}, X. and {Bouchy}, F. and {Burdanov}, A. and {Caldwell}, D.~A. and {Charbonneau}, D. and {Ciardi}, D.~R. and {Collins}, K.~A. and {Daylan}, T. and {Demory}, B. -O. and {de Wit}, J. and {Dransfield}, G. and {Fajardo-Acosta}, S.~B. and {Fausnaugh}, M. and {Fukui}, A. and {Furlan}, E. and {Garcia}, L.~J. and {Gnilka}, C.~L. and {G{\'o}mez Maqueo Chew}, Y. and {G{\'o}mez-Mu{\~n}oz}, M.~A. and {G{\"u}nther}, M.~N. and {Harakawa}, H. and {Heng}, K. and {Hooton}, M.~J. and {Hori}, Y. and {Ikoma}, M. and {Jehin}, E. and {Jenkins}, J.~M. and {Kagetani}, T. and {Kawauchi}, K. and {Kimura}, T. and {Kodama}, T. and {Kotani}, T. and {Krishnamurthy}, V. and {Kudo}, T. and {Kunovac}, V. and {Kusakabe}, N. and {Latham}, D.~W. and {Littlefield}, C. and {McCormac}, J. and {Melis}, C. and {Mori}, M. and {Murgas}, F. and {Palle}, E. and {Pedersen}, P.~P. and {Queloz}, D. and {Ricker}, G. and {Sabin}, L. and {Schanche}, N. and {Schroffenegger}, U. and {Seager}, S. and {Shiao}, B. and {Sohy}, S. and {Standing}, M.~R. and {Tamura}, M. and {Theissen}, C.~A. and {Thompson}, S.~J. and {Triaud}, A.~H.~M.~J. and {Vanderspek}, R. and {Vievard}, S. and {Wells}, R.~D. and {Winn}, J.~N. and {Zou}, Y. and {Z{\'u}{\~n}iga-Fern{\'a}ndez}, S. and {Gillon}, M.},
        title = "{Two temperate super-Earths transiting a nearby late-type M dwarf}",
      journal = {\aap},
         year = 2022,
        month = nov,
       volume = {667},
          eid = {A59},
        pages = {A59},
          doi = {10.1051/0004-6361/202244041},
archivePrefix = {arXiv},
       eprint = {2209.02831},
 primaryClass = {astro-ph.EP},
       adsurl = {https://ui.adsabs.harvard.edu/abs/2022A&A...667A..59D}
}

@ARTICLE{Molliere2019,
       author = {{Molli{\`e}re}, P. and {Wardenier}, J.~P. and {van Boekel}, R. and {Henning}, Th. and {Molaverdikhani}, K. and {Snellen}, I.~A.~G.},
        title = "{petitRADTRANS. A Python radiative transfer package for exoplanet characterization and retrieval}",
      journal = {\aap},
         year = 2019,
        month = jul,
       volume = {627},
          eid = {A67},
        pages = {A67},
          doi = {10.1051/0004-6361/201935470},
archivePrefix = {arXiv},
       eprint = {1904.11504},
 primaryClass = {astro-ph.EP},
       adsurl = {https://ui.adsabs.harvard.edu/abs/2019A&A...627A..67M}
}

@ARTICLE{Maiolino2013,
       author = {{Maiolino}, R. and {Haehnelt}, M. and {Murphy}, M.~T. and {Queloz}, D. and {Origlia}, L. and {Alcala}, J. and {Alibert}, Y. and {Amado}, P.~J. and {Allende Prieto}, C. and {Ammler-von Eiff}, M. and {Asplund}, M. and {Barstow}, M. and {Becker}, G. and {Bonfils}, X. and {Bouchy}, F. and {Bragaglia}, A. and {Burleigh}, M.~R. and {Chiavassa}, A. and {Cimatti}, D.~A. and {Cirasuolo}, M. and {Cristiani}, S. and {D'Odorico}, V. and {Dravins}, D. and {Emsellem}, E. and {Farihi}, J. and {Figueira}, P. and {Fynbo}, J. and {Gansicke}, B.~T. and {Gillon}, M. and {Gustafsson}, B. and {Hill}, V. and {Israelyan}, G. and {Korn}, A. and {Larsen}, S. and {De Laverny}, P. and {Liske}, J. and {Lovis}, C. and {Marconi}, A. and {Martins}, C. and {Molaro}, P. and {Nisini}, B. and {Oliva}, E. and {Petitjean}, P. and {Pettini}, M. and {Recio Blanco}, A. and {Rebolo}, R. and {Reiners}, A. and {Rodriguez-Lopez}, C. and {Ryde}, N. and {Santos}, N.~C. and {Savaglio}, S. and {Snellen}, I. and {Strassmeier}, K. and {Tanvir}, N. and {Testi}, L. and {Tolstoy}, E. and {Triaud}, A. and {Vanzi}, L. and {Viel}, M. and {Volonteri}, M.},
        title = "{A Community Science Case for E-ELT HIRES}",
      journal = {arXiv e-prints},
         year = 2013,
        month = oct,
          eid = {arXiv:1310.3163},
        pages = {arXiv:1310.3163},
          doi = {10.48550/arXiv.1310.3163},
archivePrefix = {arXiv},
       eprint = {1310.3163},
 primaryClass = {astro-ph.IM},
       adsurl = {https://ui.adsabs.harvard.edu/abs/2013arXiv1310.3163M}
}

@misc{PHL,
  author       = {{PHL @ UPR Arecibo}},
  title        = {Habitable Worlds Catalog},
  year         = {2024},
  url          = {http://phl.upr.edu/hwc},
  note         = {Accessed 2026 March 18}
}

@ARTICLE{Parviainen2024,
       author = {{Parviainen}, Hannu and {Luque}, Rafael and {Palle}, Enric},
        title = "{SPRIGHT: a probabilistic mass-density-radius relation for small planets}",
      journal = {\mnras},
         year = 2024,
        month = jan,
       volume = {527},
       number = {3},
        pages = {5693-5716},
          doi = {10.1093/mnras/stad3504},
archivePrefix = {arXiv},
       eprint = {2311.07255},
 primaryClass = {astro-ph.EP},
       adsurl = {https://ui.adsabs.harvard.edu/abs/2024MNRAS.527.5693P}
}

@ARTICLE{Gilbert2023,
       author = {{Gilbert}, Emily A. and {Vanderburg}, Andrew and {Rodriguez}, Joseph E. and {Hord}, Benjamin J. and {Clement}, Matthew S. and {Barclay}, Thomas and {Quintana}, Elisa V. and {Schlieder}, Joshua E. and {Kane}, Stephen R. and {Jenkins}, Jon M. and {Twicken}, Joseph D. and {Kunimoto}, Michelle and {Vanderspek}, Roland and {Arney}, Giada N. and {Charbonneau}, David and {G{\"u}nther}, Maximilian N. and {Huang}, Chelsea X. and {Isopi}, Giovanni and {Kostov}, Veselin B. and {Kristiansen}, Martti H. and {Latham}, David W. and {Mallia}, Franco and {Mamajek}, Eric E. and {Mireles}, Ismael and {Quinn}, Samuel N. and {Ricker}, George R. and {Schulte}, Jack and {Seager}, S. and {Suissa}, Gabrielle and {Winn}, Joshua N. and {Youngblood}, Allison and {Zapparata}, Aldo},
        title = "{A Second Earth-sized Planet in the Habitable Zone of the M Dwarf, TOI-700}",
      journal = {\apjl},
         year = 2023,
        month = feb,
       volume = {944},
       number = {2},
          eid = {L35},
        pages = {L35},
          doi = {10.3847/2041-8213/acb599},
archivePrefix = {arXiv},
       eprint = {2301.03617},
 primaryClass = {astro-ph.EP},
       adsurl = {https://ui.adsabs.harvard.edu/abs/2023ApJ...944L..35G}
}

@ARTICLE{Vanderburg2020,
       author = {{Vanderburg}, Andrew and {Rowden}, Pamela and {Bryson}, Steve and {Coughlin}, Jeffrey and {Batalha}, Natalie and {Collins}, Karen A. and {Latham}, David W. and {Mullally}, Susan E. and {Col{\'o}n}, Knicole D. and {Henze}, Chris and {Huang}, Chelsea X. and {Quinn}, Samuel N.},
        title = "{A Habitable-zone Earth-sized Planet Rescued from False Positive Status}",
      journal = {\apjl},
         year = 2020,
        month = apr,
       volume = {893},
       number = {1},
          eid = {L27},
        pages = {L27},
          doi = {10.3847/2041-8213/ab84e5},
archivePrefix = {arXiv},
       eprint = {2004.06725},
 primaryClass = {astro-ph.EP},
       adsurl = {https://ui.adsabs.harvard.edu/abs/2020ApJ...893L..27V}
}

@ARTICLE{Agol2021,
       author = {{Agol}, Eric and {Dorn}, Caroline and {Grimm}, Simon L. and {Turbet}, Martin and {Ducrot}, Elsa and {Delrez}, Laetitia and {Gillon}, Micha{\"e}l and {Demory}, Brice-Olivier and {Burdanov}, Artem and {Barkaoui}, Khalid and {Benkhaldoun}, Zouhair and {Bolmont}, Emeline and {Burgasser}, Adam and {Carey}, Sean and {de Wit}, Julien and {Fabrycky}, Daniel and {Foreman-Mackey}, Daniel and {Haldemann}, Jonas and {Hernandez}, David M. and {Ingalls}, James and {Jehin}, Emmanuel and {Langford}, Zachary and {Leconte}, J{\'e}r{\'e}my and {Lederer}, Susan M. and {Luger}, Rodrigo and {Malhotra}, Renu and {Meadows}, Victoria S. and {Morris}, Brett M. and {Pozuelos}, Francisco J. and {Queloz}, Didier and {Raymond}, Sean N. and {Selsis}, Franck and {Sestovic}, Marko and {Triaud}, Amaury H.~M.~J. and {Van Grootel}, Valerie},
        title = "{Refining the Transit-timing and Photometric Analysis of TRAPPIST-1: Masses, Radii, Densities, Dynamics, and Ephemerides}",
      journal = {\psj},
         year = 2021,
        month = feb,
       volume = {2},
       number = {1},
          eid = {1},
        pages = {1},
          doi = {10.3847/PSJ/abd022},
archivePrefix = {arXiv},
       eprint = {2010.01074},
 primaryClass = {astro-ph.EP},
       adsurl = {https://ui.adsabs.harvard.edu/abs/2021PSJ.....2....1A}
}

@ARTICLE{Dressing2017,
       author = {{Dressing}, Courtney D. and {Vanderburg}, Andrew and {Schlieder}, Joshua E. and {Crossfield}, Ian J.~M. and {Knutson}, Heather A. and {Newton}, Elisabeth R. and {Ciardi}, David R. and {Fulton}, Benjamin J. and {Gonzales}, Erica J. and {Howard}, Andrew W. and {Isaacson}, Howard and {Livingston}, John and {Petigura}, Erik A. and {Sinukoff}, Evan and {Everett}, Mark and {Horch}, Elliott and {Howell}, Steve B.},
        title = "{Characterizing K2 Candidate Planetary Systems Orbiting Low-mass Stars. II. Planetary Systems Observed During Campaigns 1-7}",
      journal = {\aj},
         year = 2017,
        month = nov,
       volume = {154},
       number = {5},
          eid = {207},
        pages = {207},
          doi = {10.3847/1538-3881/aa89f2},
archivePrefix = {arXiv},
       eprint = {1703.07416},
 primaryClass = {astro-ph.EP},
       adsurl = {https://ui.adsabs.harvard.edu/abs/2017AJ....154..207D}
}

@ARTICLE{Crossfield2016,
       author = {{Crossfield}, Ian J.~M. and {Ciardi}, David R. and {Petigura}, Erik A. and {Sinukoff}, Evan and {Schlieder}, Joshua E. and {Howard}, Andrew W. and {Beichman}, Charles A. and {Isaacson}, Howard and {Dressing}, Courtney D. and {Christiansen}, Jessie L. and {Fulton}, Benjamin J. and {L{\'e}pine}, S{\'e}bastien and {Weiss}, Lauren and {Hirsch}, Lea and {Livingston}, John and {Baranec}, Christoph and {Law}, Nicholas M. and {Riddle}, Reed and {Ziegler}, Carl and {Howell}, Steve B. and {Horch}, Elliott and {Everett}, Mark and {Teske}, Johanna and {Martinez}, Arturo O. and {Obermeier}, Christian and {Benneke}, Bj{\"o}rn and {Scott}, Nic and {Deacon}, Niall and {Aller}, Kimberly M. and {Hansen}, Brad M.~S. and {Mancini}, Luigi and {Ciceri}, Simona and {Brahm}, Rafael and {Jord{\'a}n}, Andr{\'e}s and {Knutson}, Heather A. and {Henning}, Thomas and {Bonnefoy}, Micha{\"e}l and {Liu}, Michael C. and {Crepp}, Justin R. and {Lothringer}, Joshua and {Hinz}, Phil and {Bailey}, Vanessa and {Skemer}, Andrew and {Defrere}, Denis},
        title = "{197 Candidates and 104 Validated Planets in K2{\textquoteright}s First Five Fields}",
      journal = {\apjs},
         year = 2016,
        month = sep,
       volume = {226},
       number = {1},
          eid = {7},
        pages = {7},
          doi = {10.3847/0067-0049/226/1/7},
archivePrefix = {arXiv},
       eprint = {1607.05263},
 primaryClass = {astro-ph.EP},
       adsurl = {https://ui.adsabs.harvard.edu/abs/2016ApJS..226....7C}
}

@ARTICLE{Torres2015,
       author = {{Torres}, Guillermo and {Kipping}, David M. and {Fressin}, Francois and {Caldwell}, Douglas A. and {Twicken}, Joseph D. and {Ballard}, Sarah and {Batalha}, Natalie M. and {Bryson}, Stephen T. and {Ciardi}, David R. and {Henze}, Christopher E. and {Howell}, Steve B. and {Isaacson}, Howard T. and {Jenkins}, Jon M. and {Muirhead}, Philip S. and {Newton}, Elisabeth R. and {Petigura}, Erik A. and {Barclay}, Thomas and {Borucki}, William J. and {Crepp}, Justin R. and {Everett}, Mark E. and {Horch}, Elliott P. and {Howard}, Andrew W. and {Kolbl}, Rea and {Marcy}, Geoffrey W. and {McCauliff}, Sean and {Quintana}, Elisa V.},
        title = "{Validation of 12 Small Kepler Transiting Planets in the Habitable Zone}",
      journal = {\apj},
         year = 2015,
        month = feb,
       volume = {800},
       number = {2},
          eid = {99},
        pages = {99},
          doi = {10.1088/0004-637X/800/2/99},
archivePrefix = {arXiv},
       eprint = {1501.01101},
 primaryClass = {astro-ph.EP},
       adsurl = {https://ui.adsabs.harvard.edu/abs/2015ApJ...800...99T}
}

@ARTICLE{DiamondLowe2022,
       author = {{Diamond-Lowe}, Hannah and {Kreidberg}, Laura and {Harman}, C.~E. and {Kempton}, Eliza M. -R. and {Rogers}, Leslie A. and {Joyce}, Simon R.~G. and {Eastman}, Jason D. and {King}, George W. and {Kopparapu}, Ravi and {Youngblood}, Allison and {Kosiarek}, Molly R. and {Livingston}, John H. and {Hardegree-Ullman}, Kevin K. and {Crossfield}, Ian J.~M.},
        title = "{The K2-3 System Revisited: Testing Photoevaporation and Core-powered Mass Loss with Three Small Planets Spanning the Radius Valley}",
      journal = {\aj},
         year = 2022,
        month = nov,
       volume = {164},
       number = {5},
          eid = {172},
        pages = {172},
          doi = {10.3847/1538-3881/ac7807},
archivePrefix = {arXiv},
       eprint = {2207.12755},
 primaryClass = {astro-ph.EP},
       adsurl = {https://ui.adsabs.harvard.edu/abs/2022AJ....164..172D}
}

@ARTICLE{Fukui2016,
       author = {{Fukui}, Akihiko and {Livingston}, John and {Narita}, Norio and {Hirano}, Teruyuki and {Onitsuka}, Masahiro and {Ryu}, Tsuguru and {Kusakabe}, Nobuhiko},
        title = "{Ground-based Transit Observation of the Habitable-zone Super-Earth K2-3d}",
      journal = {\aj},
         year = 2016,
        month = dec,
       volume = {152},
       number = {6},
          eid = {171},
        pages = {171},
          doi = {10.3847/0004-6256/152/6/171},
archivePrefix = {arXiv},
       eprint = {1610.01333},
 primaryClass = {astro-ph.EP},
       adsurl = {https://ui.adsabs.harvard.edu/abs/2016AJ....152..171F}
}

@ARTICLE{Bonomo2023,
       author = {{Bonomo}, A.~S. and {Dumusque}, X. and {Massa}, A. and {Mortier}, A. and {Bongiolatti}, R. and {Malavolta}, L. and {Sozzetti}, A. and {Buchhave}, L.~A. and {Damasso}, M. and {Haywood}, R.~D. and {Morbidelli}, A. and {Latham}, D.~W. and {Molinari}, E. and {Pepe}, F. and {Poretti}, E. and {Udry}, S. and {Affer}, L. and {Boschin}, W. and {Charbonneau}, D. and {Cosentino}, R. and {Cretignier}, M. and {Ghedina}, A. and {Lega}, E. and {L{\'o}pez-Morales}, M. and {Margini}, M. and {Mart{\'\i}nez Fiorenzano}, A.~F. and {Mayor}, M. and {Micela}, G. and {Pedani}, M. and {Pinamonti}, M. and {Rice}, K. and {Sasselov}, D. and {Tronsgaard}, R. and {Vanderburg}, A.},
        title = "{Cold Jupiters and improved masses in 38 Kepler and K2 small planet systems from 3661 HARPS-N radial velocities. No excess of cold Jupiters in small planet systems}",
      journal = {\aap},
         year = 2023,
        month = sep,
       volume = {677},
          eid = {A33},
        pages = {A33},
          doi = {10.1051/0004-6361/202346211},
archivePrefix = {arXiv},
       eprint = {2304.05773},
 primaryClass = {astro-ph.EP},
       adsurl = {https://ui.adsabs.harvard.edu/abs/2023A&A...677A..33B}
}

@ARTICLE{Dransfield2024,
       author = {{Dransfield}, Georgina and {Timmermans}, Mathilde and {Triaud}, Amaury H.~M.~J. and {D{\'e}vora-Pajares}, Mart{\'\i}n and {Aganze}, Christian and {Barkaoui}, Khalid and {Burgasser}, Adam J. and {Collins}, Karen A. and {Cointepas}, Marion and {Ducrot}, Elsa and {G{\"u}nther}, Maximilian N. and {Howell}, Steve B. and {Murray}, Catriona A. and {Niraula}, Prajwal and {Rackham}, Benjamin V. and {Sebastian}, Daniel and {Stassun}, Keivan G. and {Z{\'u}{\~n}iga-Fern{\'a}ndez}, Sebasti{\'a}n and {Almenara}, Jos{\'e} Manuel and {Bonfils}, Xavier and {Bouchy}, Fran{\c{c}}ois and {Burke}, Christopher J. and {Charbonneau}, David and {Christiansen}, Jessie L. and {Delrez}, Laetitia and {Gan}, Tianjun and {Garc{\'\i}a}, Lionel J. and {Gillon}, Micha{\"e}l and {G{\'o}mez Maqueo Chew}, Yilen and {Hesse}, Katharine M. and {Hooton}, Matthew J. and {Isopi}, Giovanni and {Jehin}, Emmanu{\"e}l and {Jenkins}, Jon M. and {Latham}, David W. and {Mallia}, Franco and {Murgas}, Felipe and {Pedersen}, Peter P. and {Pozuelos}, Francisco J. and {Queloz}, Didier and {Rodriguez}, David R. and {Schanche}, Nicole and {Seager}, Sara and {Srdoc}, Gregor and {Stockdale}, Chris and {Twicken}, Joseph D. and {Vanderspek}, Roland and {Wells}, Robert and {Winn}, Joshua N. and {de Wit}, Julien and {Zapparata}, Aldo},
        title = "{A 1.55 R$_{{\ensuremath{\oplus}}}$ habitable-zone planet hosted by TOI-715, an M4 star near the ecliptic South Pole}",
      journal = {\mnras},
         year = 2024,
        month = jan,
       volume = {527},
       number = {1},
        pages = {35-52},
          doi = {10.1093/mnras/stad1439},
archivePrefix = {arXiv},
       eprint = {2305.06206},
 primaryClass = {astro-ph.EP},
       adsurl = {https://ui.adsabs.harvard.edu/abs/2024MNRAS.527...35D}
}

@ARTICLE{Borucki2013,
       author = {{Borucki}, William J. and {Agol}, Eric and {Fressin}, Francois and {Kaltenegger}, Lisa and {Rowe}, Jason and {Isaacson}, Howard and {Fischer}, Debra and {Batalha}, Natalie and {Lissauer}, Jack J. and {Marcy}, Geoffrey W. and {Fabrycky}, Daniel and {D{\'e}sert}, Jean-Michel and {Bryson}, Stephen T. and {Barclay}, Thomas and {Bastien}, Fabienne and {Boss}, Alan and {Brugamyer}, Erik and {Buchhave}, Lars A. and {Burke}, Chris and {Caldwell}, Douglas A. and {Carter}, Josh and {Charbonneau}, David and {Crepp}, Justin R. and {Christensen-Dalsgaard}, J{\o}rgen and {Christiansen}, Jessie L. and {Ciardi}, David and {Cochran}, William D. and {DeVore}, Edna and {Doyle}, Laurance and {Dupree}, Andrea K. and {Endl}, Michael and {Everett}, Mark E. and {Ford}, Eric B. and {Fortney}, Jonathan and {Gautier}, Thomas N. and {Geary}, John C. and {Gould}, Alan and {Haas}, Michael and {Henze}, Christopher and {Howard}, Andrew W. and {Howell}, Steve B. and {Huber}, Daniel and {Jenkins}, Jon M. and {Kjeldsen}, Hans and {Kolbl}, Rea and {Kolodziejczak}, Jeffery and {Latham}, David W. and {Lee}, Brian L. and {Lopez}, Eric and {Mullally}, Fergal and {Orosz}, Jerome A. and {Prsa}, Andrej and {Quintana}, Elisa V. and {Sanchis-Ojeda}, Roberto and {Sasselov}, Dimitar and {Seader}, Shawn and {Shporer}, Avi and {Steffen}, Jason H. and {Still}, Martin and {Tenenbaum}, Peter and {Thompson}, Susan E. and {Torres}, Guillermo and {Twicken}, Joseph D. and {Welsh}, William F. and {Winn}, Joshua N.},
        title = "{Kepler-62: A Five-Planet System with Planets of 1.4 and 1.6 Earth Radii in the Habitable Zone}",
      journal = {Science},
         year = 2013,
        month = may,
       volume = {340},
       number = {6132},
        pages = {587-590},
          doi = {10.1126/science.1234702},
archivePrefix = {arXiv},
       eprint = {1304.7387},
 primaryClass = {astro-ph.EP},
       adsurl = {https://ui.adsabs.harvard.edu/abs/2013Sci...340..587B}
}

@ARTICLE{Morton2016,
       author = {{Morton}, Timothy D. and {Bryson}, Stephen T. and {Coughlin}, Jeffrey L. and {Rowe}, Jason F. and {Ravichandran}, Ganesh and {Petigura}, Erik A. and {Haas}, Michael R. and {Batalha}, Natalie M.},
        title = "{False Positive Probabilities for all Kepler Objects of Interest: 1284 Newly Validated Planets and 428 Likely False Positives}",
      journal = {\apj},
         year = 2016,
        month = may,
       volume = {822},
       number = {2},
          eid = {86},
        pages = {86},
          doi = {10.3847/0004-637X/822/2/86},
archivePrefix = {arXiv},
       eprint = {1605.02825},
 primaryClass = {astro-ph.EP},
       adsurl = {https://ui.adsabs.harvard.edu/abs/2016ApJ...822...86M}
}

@ARTICLE{Jenkins2015,
       author = {{Jenkins}, Jon M. and {Twicken}, Joseph D. and {Batalha}, Natalie M. and {Caldwell}, Douglas A. and {Cochran}, William D. and {Endl}, Michael and {Latham}, David W. and {Esquerdo}, Gilbert A. and {Seader}, Shawn and {Bieryla}, Allyson and {Petigura}, Erik and {Ciardi}, David R. and {Marcy}, Geoffrey W. and {Isaacson}, Howard and {Huber}, Daniel and {Rowe}, Jason F. and {Torres}, Guillermo and {Bryson}, Stephen T. and {Buchhave}, Lars and {Ramirez}, Ivan and {Wolfgang}, Angie and {Li}, Jie and {Campbell}, Jennifer R. and {Tenenbaum}, Peter and {Sanderfer}, Dwight and {Henze}, Christopher E. and {Catanzarite}, Joseph H. and {Gilliland}, Ronald L. and {Borucki}, William J.},
        title = "{Discovery and Validation of Kepler-452b: A 1.6 R $_{⨁}$ Super Earth Exoplanet in the Habitable Zone of a G2 Star}",
      journal = {\aj},
         year = 2015,
        month = aug,
       volume = {150},
       number = {2},
          eid = {56},
        pages = {56},
          doi = {10.1088/0004-6256/150/2/56},
archivePrefix = {arXiv},
       eprint = {1507.06723},
 primaryClass = {astro-ph.EP},
       adsurl = {https://ui.adsabs.harvard.edu/abs/2015AJ....150...56J}
}

@ARTICLE{Berger2018,
       author = {{Berger}, Travis A. and {Huber}, Daniel and {Gaidos}, Eric and {van Saders}, Jennifer L.},
        title = "{Revised Radii of Kepler Stars and Planets Using Gaia Data Release 2}",
      journal = {\apj},
         year = 2018,
        month = oct,
       volume = {866},
       number = {2},
          eid = {99},
        pages = {99},
          doi = {10.3847/1538-4357/aada83},
archivePrefix = {arXiv},
       eprint = {1805.00231},
 primaryClass = {astro-ph.EP},
       adsurl = {https://ui.adsabs.harvard.edu/abs/2018ApJ...866...99B}
}

@ARTICLE{Cadieux2024,
       author = {{Cadieux}, Charles and {Plotnykov}, Mykhaylo and {Doyon}, Ren{\'e} and {Valencia}, Diana and {Jahandar}, Farbod and {Dang}, Lisa and {Turbet}, Martin and {Fauchez}, Thomas J. and {Cloutier}, Ryan and {Cherubim}, Collin and {Artigau}, {\'E}tienne and {Cook}, Neil J. and {Edwards}, Billy and {Hallatt}, Tim and {Charnay}, Benjamin and {Bouchy}, Fran{\c{c}}ois and {Allart}, Romain and {Mignon}, Lucile and {Baron}, Fr{\'e}d{\'e}rique and {Barros}, Susana C.~C. and {Benneke}, Bj{\"o}rn and {Canto Martins}, B.~L. and {Cowan}, Nicolas B. and {De Medeiros}, J.~R. and {Delfosse}, Xavier and {Delgado-Mena}, Elisa and {Dumusque}, Xavier and {Ehrenreich}, David and {Frensch}, Yolanda G.~C. and {Gonz{\'a}lez Hern{\'a}ndez}, J.~I. and {Hara}, Nathan C. and {Lafreni{\`e}re}, David and {Lo Curto}, Gaspare and {Malo}, Lison and {Melo}, Claudio and {Mounzer}, Dany and {Passeger}, Vera Maria and {Pepe}, Francesco and {Poulin-Girard}, Anne-Sophie and {Santos}, Nuno C. and {Sosnowska}, Danuta and {Su{\'a}rez Mascare{\~n}o}, Alejandro and {Thibault}, Simon and {Vaulato}, Valentina and {Wade}, Gregg A. and {Wildi}, Fran{\c{c}}ois},
        title = "{New Mass and Radius Constraints on the LHS 1140 Planets: LHS 1140 b Is either a Temperate Mini-Neptune or a Water World}",
      journal = {\apjl},
         year = 2024,
        month = jan,
       volume = {960},
       number = {1},
          eid = {L3},
        pages = {L3},
          doi = {10.3847/2041-8213/ad1691},
archivePrefix = {arXiv},
       eprint = {2310.15490},
 primaryClass = {astro-ph.EP},
       adsurl = {https://ui.adsabs.harvard.edu/abs/2024ApJ...960L...3C}
}

@ARTICLE{Kawahara2012,
       author = {{Kawahara}, Hajime and {Matsuo}, Taro and {Takami}, Michihiro and {Fujii}, Yuka and {Kotani}, Takayuki and {Murakami}, Naoshi and {Tamura}, Motohide and {Guyon}, Olivier},
        title = "{Can Ground-based Telescopes Detect the Oxygen 1.27 {\ensuremath{\mu}}m Absorption Feature as a Biomarker in Exoplanets?}",
      journal = {\apj},
         year = 2012,
        month = oct,
       volume = {758},
       number = {1},
          eid = {13},
        pages = {13},
          doi = {10.1088/0004-637X/758/1/13},
archivePrefix = {arXiv},
       eprint = {1206.0558},
 primaryClass = {astro-ph.EP},
       adsurl = {https://ui.adsabs.harvard.edu/abs/2012ApJ...758...13K}
}

@ARTICLE{Snellen2013,
       author = {{Snellen}, I.~A.~G. and {de Kok}, R.~J. and {le Poole}, R. and {Brogi}, M. and {Birkby}, J.},
        title = "{Finding Extraterrestrial Life Using Ground-based High-dispersion Spectroscopy}",
      journal = {\apj},
         year = 2013,
        month = feb,
       volume = {764},
       number = {2},
          eid = {182},
        pages = {182},
          doi = {10.1088/0004-637X/764/2/182},
archivePrefix = {arXiv},
       eprint = {1302.3251},
 primaryClass = {astro-ph.EP},
       adsurl = {https://ui.adsabs.harvard.edu/abs/2013ApJ...764..182S}
}

@ARTICLE{Pidhorodetska2020,
       author = {{Pidhorodetska}, Daria and {Fauchez}, Thomas J. and {Villanueva}, Geronimo L. and {Domagal-Goldman}, Shawn D. and {Kopparapu}, Ravi K.},
        title = "{Detectability of Molecular Signatures on TRAPPIST-1e through Transmission Spectroscopy Simulated for Future Space-based Observatories}",
      journal = {\apjl},
         year = 2020,
        month = aug,
       volume = {898},
       number = {2},
          eid = {L33},
        pages = {L33},
          doi = {10.3847/2041-8213/aba4a1},
archivePrefix = {arXiv},
       eprint = {2001.01338},
 primaryClass = {astro-ph.EP},
       adsurl = {https://ui.adsabs.harvard.edu/abs/2020ApJ...898L..33P}
}

@ARTICLE{Lopez-Morales2019,
       author = {{L{\'o}pez-Morales}, Mercedes and {Ben-Ami}, Sagi and {Gonzalez-Abad}, Gonzalo and {Garc{\'\i}a-Mej{\'\i}a}, Juliana and {Dietrich}, Jamie and {Szentgyorgyi}, Andrew},
        title = "{Optimizing Ground-based Observations of O$_{2}$ in Earth Analogs}",
      journal = {\aj},
         year = 2019,
        month = jul,
       volume = {158},
       number = {1},
          eid = {24},
        pages = {24},
          doi = {10.3847/1538-3881/ab21d7},
archivePrefix = {arXiv},
       eprint = {1905.05862},
 primaryClass = {astro-ph.EP},
       adsurl = {https://ui.adsabs.harvard.edu/abs/2019AJ....158...24L}
}

@ARTICLE{Lustig-Yaeger2019,
       author = {{Lustig-Yaeger}, Jacob and {Meadows}, Victoria S. and {Lincowski}, Andrew P.},
        title = "{The Detectability and Characterization of the TRAPPIST-1 Exoplanet Atmospheres with JWST}",
      journal = {\aj},
         year = 2019,
        month = jul,
       volume = {158},
       number = {1},
          eid = {27},
        pages = {27},
          doi = {10.3847/1538-3881/ab21e0},
archivePrefix = {arXiv},
       eprint = {1905.07070},
 primaryClass = {astro-ph.EP},
       adsurl = {https://ui.adsabs.harvard.edu/abs/2019AJ....158...27L}
}

@ARTICLE{Hardegree-Ullman2023,
       author = {{Hardegree-Ullman}, Kevin K. and {Apai}, D{\'a}niel and {Bergsten}, Galen J. and {Pascucci}, Ilaria and {L{\'o}pez-Morales}, Mercedes},
        title = "{Bioverse: A Comprehensive Assessment of the Capabilities of Extremely Large Telescopes to Probe Earth-like O$_{2}$ Levels in Nearby Transiting Habitable-zone Exoplanets}",
      journal = {\aj},
         year = 2023,
        month = jun,
       volume = {165},
       number = {6},
          eid = {267},
        pages = {267},
          doi = {10.3847/1538-3881/acd1ec},
archivePrefix = {arXiv},
       eprint = {2304.12490},
 primaryClass = {astro-ph.EP},
       adsurl = {https://ui.adsabs.harvard.edu/abs/2023AJ....165..267H}
}

@ARTICLE{Zhang2024,
       author = {{Zhang}, Huihao and {Wang}, Ji and {Plummer}, Michael K.},
        title = "{Detecting Biosignatures in Nearby Rocky Exoplanets Using High-contrast Imaging and Medium-resolution Spectroscopy with the Extremely Large Telescope}",
      journal = {\aj},
         year = 2024,
        month = jan,
       volume = {167},
       number = {1},
          eid = {37},
        pages = {37},
          doi = {10.3847/1538-3881/ad109e},
archivePrefix = {arXiv},
       eprint = {2311.18117},
 primaryClass = {astro-ph.EP},
       adsurl = {https://ui.adsabs.harvard.edu/abs/2024AJ....167...37Z}
}

@ARTICLE{Rodler2014,
       author = {{Rodler}, Florian and {L{\'o}pez-Morales}, Mercedes},
        title = "{Feasibility Studies for the Detection of O$_{2}$ in an Earth-like Exoplanet}",
      journal = {\apj},
         year = 2014,
        month = jan,
       volume = {781},
       number = {1},
          eid = {54},
        pages = {54},
          doi = {10.1088/0004-637X/781/1/54},
archivePrefix = {arXiv},
       eprint = {1312.1585},
 primaryClass = {astro-ph.EP},
       adsurl = {https://ui.adsabs.harvard.edu/abs/2014ApJ...781...54R}
}

@ARTICLE{Serindag2019,
       author = {{Serindag}, Dilovan B. and {Snellen}, Ignas A.~G.},
        title = "{Testing the Detectability of Extraterrestrial O$_{2}$ with the Extremely Large Telescopes Using Real Data with Real Noise}",
      journal = {\apjl},
         year = 2019,
        month = jan,
       volume = {871},
       number = {1},
          eid = {L7},
        pages = {L7},
          doi = {10.3847/2041-8213/aafa1f},
archivePrefix = {arXiv},
       eprint = {1901.02469},
 primaryClass = {astro-ph.EP},
       adsurl = {https://ui.adsabs.harvard.edu/abs/2019ApJ...871L...7S}
}

@article{Rukdee2024,
author = {Rukdee, Surangkhana},
year = {2024},
month = {11},
pages = {},
title = {Instrumentation prospects for rocky exoplanet atmospheres studies with high resolution spectroscopy},
volume = {14},
journal = {Scientific Reports},
doi = {10.1038/s41598-024-78071-5}
}

@ARTICLE{Palle2025,
       author = {{Palle}, Enric and {Biazzo}, Katia and {Bolmont}, Emeline and {Molli{\`e}re}, Paul and {Poppenhaeger}, Katja and {Birkby}, Jayne and {Brogi}, Matteo and {Chauvin}, Gael and {Chiavassa}, Andrea and {Hoeijmakers}, Jens and {Lellouch}, Emmanuel and {Lovis}, Christophe and {Maiolino}, Roberto and {Nortmann}, Lisa and {Parviainen}, Hannu and {Pino}, Lorenzo and {Turbet}, Martin and {Weder}, Jesse and {Albrecht}, Simon and {Antoniucci}, Simone and {Barros}, Susana C. and {Beaudoin}, Andre and {Benneke}, Bjorn and {Boisse}, Isabelle and {Bonomo}, Aldo S. and {Borsa}, Francesco and {Brandeker}, Alexis and {Brandner}, Wolfgang and {Buchhave}, Lars A. and {Cheffot}, Anne-Laure and {Deborde}, Robin and {Debras}, Florian and {Doyon}, Rene and {Di Marcantonio}, Paolo and {Giacobbe}, Paolo and {Gonz{\'a}lez Hern{\'a}ndez}, Jonay I. and {Helled}, Ravit and {Kreidberg}, Laura and {Machado}, Pedro and {Maldonado}, Jesus and {Marconi}, Alessandro and {Martins}, B.~L. Canto and {Miceli}, Adriano and {Mordasini}, Christoph and {N'Diaye}, Mamadou and {Niedzielski}, Andrzej and {Nisini}, Brunella and {Origlia}, Livia and {Peroux}, Celine and {Pietrow}, Alexander G.~M. and {Pinna}, Enrico and {Rauscher}, Emily and {Reffert}, Sabine and {Rodr{\'\i}guez-L{\'o}pez}, Cristina and {Rousselot}, Philippe and {Sanna}, Nicoletta and {Santos}, Nuno C. and {Simonnin}, Adrien and {Su{\'a}rez Mascare{\~n}o}, Alejandro and {Zanutta}, Alessio and {Zapatero-Osorio}, Maria Rosa and {Zechmeister}, Mathias},
        title = "{Ground-breaking exoplanet science with the ANDES spectrograph at the ELT}",
      journal = {Experimental Astronomy},
         year = 2025,
        month = jun,
       volume = {59},
       number = {3},
          eid = {29},
        pages = {29},
          doi = {10.1007/s10686-025-10000-4},
archivePrefix = {arXiv},
       eprint = {2311.17075},
 primaryClass = {astro-ph.IM},
       adsurl = {https://ui.adsabs.harvard.edu/abs/2025ExA....59...29P}
}

@ARTICLE{Wunderlich2020,
       author = {{Wunderlich}, Fabian and {Scheucher}, Markus and {Godolt}, M. and {Grenfell}, J.~L. and {Schreier}, F. and {Schneider}, P.~C. and {Wilson}, D.~J. and {S{\'a}nchez-L{\'o}pez}, A. and {L{\'o}pez-Puertas}, M. and {Rauer}, H.},
        title = "{Distinguishing between Wet and Dry Atmospheres of TRAPPIST-1 e and f}",
      journal = {\apj},
         year = 2020,
        month = oct,
       volume = {901},
       number = {2},
          eid = {126},
        pages = {126},
          doi = {10.3847/1538-4357/aba59c},
archivePrefix = {arXiv},
       eprint = {2006.11349},
 primaryClass = {astro-ph.EP},
       adsurl = {https://ui.adsabs.harvard.edu/abs/2020ApJ...901..126W}
}

@ARTICLE{Gordon2022,
       author = {{Gordon}, I.~E. and {Rothman}, L.~S. and {Hargreaves}, R.~J. and {Hashemi}, R. and {Karlovets}, E.~V. and {Skinner}, F.~M. and {Conway}, E.~K. and {Hill}, C. and {Kochanov}, R.~V. and {Tan}, Y. and {Wcis{\l}o}, P. and {Finenko}, A.~A. and {Nelson}, K. and {Bernath}, P.~F. and {Birk}, M. and {Boudon}, V. and {Campargue}, A. and {Chance}, K.~V. and {Coustenis}, A. and {Drouin}, B.~J. and {Flaud}, J. -M. and {Gamache}, R.~R. and {Hodges}, J.~T. and {Jacquemart}, D. and {Mlawer}, E.~J. and {Nikitin}, A.~V. and {Perevalov}, V.~I. and {Rotger}, M. and {Tennyson}, J. and {Toon}, G.~C. and {Tran}, H. and {Tyuterev}, V.~G. and {Adkins}, E.~M. and {Baker}, A. and {Barbe}, A. and {Can{\`e}}, E. and {Cs{\'a}sz{\'a}r}, A.~G. and {Dudaryonok}, A. and {Egorov}, O. and {Fleisher}, A.~J. and {Fleurbaey}, H. and {Foltynowicz}, A. and {Furtenbacher}, T. and {Harrison}, J.~J. and {Hartmann}, J. -M. and {Horneman}, V. -M. and {Huang}, X. and {Karman}, T. and {Karns}, J. and {Kassi}, S. and {Kleiner}, I. and {Kofman}, V. and {Kwabia-Tchana}, F. and {Lavrentieva}, N.~N. and {Lee}, T.~J. and {Long}, D.~A. and {Lukashevskaya}, A.~A. and {Lyulin}, O.~M. and {Makhnev}, V. Yu. and {Matt}, W. and {Massie}, S.~T. and {Melosso}, M. and {Mikhailenko}, S.~N. and {Mondelain}, D. and {M{\"u}ller}, H.~S.~P. and {Naumenko}, O.~V. and {Perrin}, A. and {Polyansky}, O.~L. and {Raddaoui}, E. and {Raston}, P.~L. and {Reed}, Z.~D. and {Rey}, M. and {Richard}, C. and {T{\'o}bi{\'a}s}, R. and {Sadiek}, I. and {Schwenke}, D.~W. and {Starikova}, E. and {Sung}, K. and {Tamassia}, F. and {Tashkun}, S.~A. and {Vander Auwera}, J. and {Vasilenko}, I.~A. and {Vigasin}, A.~A. and {Villanueva}, G.~L. and {Vispoel}, B. and {Wagner}, G. and {Yachmenev}, A. and {Yurchenko}, S.~N.},
        title = "{The HITRAN2020 molecular spectroscopic database}",
      journal = {\jqsrt},
         year = 2022,
        month = jan,
       volume = {277},
          eid = {107949},
        pages = {107949},
          doi = {10.1016/j.jqsrt.2021.107949},
       adsurl = {https://ui.adsabs.harvard.edu/abs/2022JQSRT.27707949G}
}

@ARTICLE{Grimm2015,
       author = {{Grimm}, Simon L. and {Heng}, Kevin},
        title = "{HELIOS-K: An Ultrafast, Open-source Opacity Calculator for Radiative Transfer}",
      journal = {\apj},
         year = 2015,
        month = aug,
       volume = {808},
       number = {2},
          eid = {182},
        pages = {182},
          doi = {10.1088/0004-637X/808/2/182},
archivePrefix = {arXiv},
       eprint = {1503.03806},
 primaryClass = {astro-ph.EP},
       adsurl = {https://ui.adsabs.harvard.edu/abs/2015ApJ...808..182G}
}

@ARTICLE{Grimm2021,
       author = {{Grimm}, Simon L. and {Malik}, Matej and {Kitzmann}, Daniel and {Guzm{\'a}n-Mesa}, Andrea and {Hoeijmakers}, H. Jens and {Fisher}, Chloe and {Mendon{\c{c}}a}, Jo{\~a}o M. and {Yurchenko}, Sergey N. and {Tennyson}, Jonathan and {Alesina}, Fabien and {Buchschacher}, Nicolas and {Burnier}, Julien and {Segransan}, Damien and {Kurucz}, Robert L. and {Heng}, Kevin},
        title = "{HELIOS-K 2.0 Opacity Calculator and Open-source Opacity Database for Exoplanetary Atmospheres}",
      journal = {\apjs},
         year = 2021,
        month = mar,
       volume = {253},
       number = {1},
          eid = {30},
        pages = {30},
          doi = {10.3847/1538-4365/abd773},
archivePrefix = {arXiv},
       eprint = {2101.02005},
 primaryClass = {astro-ph.EP},
       adsurl = {https://ui.adsabs.harvard.edu/abs/2021ApJS..253...30G}
}

@ARTICLE{Hargreaves2020,
       author = {{Hargreaves}, Robert J. and {Gordon}, Iouli E. and {Rey}, Michael and {Nikitin}, Andrei V. and {Tyuterev}, Vladimir G. and {Kochanov}, Roman V. and {Rothman}, Laurence S.},
        title = "{An Accurate, Extensive, and Practical Line List of Methane for the HITEMP Database}",
      journal = {\apjs},
         year = 2020,
        month = apr,
       volume = {247},
       number = {2},
          eid = {55},
        pages = {55},
          doi = {10.3847/1538-4365/ab7a1a},
archivePrefix = {arXiv},
       eprint = {2001.05037},
 primaryClass = {astro-ph.EP},
       adsurl = {https://ui.adsabs.harvard.edu/abs/2020ApJS..247...55H}
}

@ARTICLE{Rothman2010,
       author = {{Rothman}, L.~S. and {Gordon}, I.~E. and {Barber}, R.~J. and {Dothe}, H. and {Gamache}, R.~R. and {Goldman}, A. and {Perevalov}, V.~I. and {Tashkun}, S.~A. and {Tennyson}, J.},
        title = "{HITEMP, the high-temperature molecular spectroscopic database}",
      journal = {\jqsrt},
         year = 2010,
        month = oct,
       volume = {111},
        pages = {2139-2150},
          doi = {10.1016/j.jqsrt.2010.05.001},
       adsurl = {https://ui.adsabs.harvard.edu/abs/2010JQSRT.111.2139R}
}

@ARTICLE{Western2018,
       author = {{Western}, Colin M. and {Carter-Blatchford}, Luke and {Crozet}, Patrick and {Ross}, Amanda J. and {Morville}, J{\'e}r{\^o}me and {Tokaryk}, Dennis W.},
        title = "{The spectrum of N$_{2}$ from 4,500 to 15,700 cm$^{-1}$ revisited with PGOPHER}",
      journal = {\jqsrt},
         year = 2018,
        month = nov,
       volume = {219},
        pages = {127-141},
          doi = {10.1016/j.jqsrt.2018.07.017},
archivePrefix = {arXiv},
       eprint = {1905.02528},
 primaryClass = {physics.chem-ph},
       adsurl = {https://ui.adsabs.harvard.edu/abs/2018JQSRT.219..127W}
}

@ARTICLE{Rothman2013,
       author = {{Rothman}, L.~S. and {Gordon}, I.~E. and {Babikov}, Y. and {Barbe}, A. and {Chris Benner}, D. and {Bernath}, P.~F. and {Birk}, M. and {Bizzocchi}, L. and {Boudon}, V. and {Brown}, L.~R. and {Campargue}, A. and {Chance}, K. and {Cohen}, E.~A. and {Coudert}, L.~H. and {Devi}, V.~M. and {Drouin}, B.~J. and {Fayt}, A. and {Flaud}, J. -M. and {Gamache}, R.~R. and {Harrison}, J.~J. and {Hartmann}, J. -M. and {Hill}, C. and {Hodges}, J.~T. and {Jacquemart}, D. and {Jolly}, A. and {Lamouroux}, J. and {Le Roy}, R.~J. and {Li}, G. and {Long}, D.~A. and {Lyulin}, O.~M. and {Mackie}, C.~J. and {Massie}, S.~T. and {Mikhailenko}, S. and {M{\"u}ller}, H.~S.~P. and {Naumenko}, O.~V. and {Nikitin}, A.~V. and {Orphal}, J. and {Perevalov}, V. and {Perrin}, A. and {Polovtseva}, E.~R. and {Richard}, C. and {Smith}, M.~A.~H. and {Starikova}, E. and {Sung}, K. and {Tashkun}, S. and {Tennyson}, J. and {Toon}, G.~C. and {Tyuterev}, Vl. G. and {Wagner}, G.},
        title = "{The HITRAN2012 molecular spectroscopic database}",
      journal = {\jqsrt},
         year = 2013,
        month = nov,
       volume = {130},
        pages = {4-50},
          doi = {10.1016/j.jqsrt.2013.07.002},
       adsurl = {https://ui.adsabs.harvard.edu/abs/2013JQSRT.130....4R}
}

@misc{PyAstronomy,
       author = {{Czesla}, Stefan and {Schr{\"o}ter}, Sebastian and {Schneider}, Christian P. and {Huber}, Klaus F. and {Pfeifer}, Fabian and {Andreasen}, Daniel T. and {Zechmeister}, Mathias},
        title = "{PyA: Python astronomy-related packages}",
 howpublished = {Astrophysics Source Code Library, record ascl:1906.010},
         year = 2019,
        month = jun,
          eid = {ascl:1906.010},
       adsurl = {https://ui.adsabs.harvard.edu/abs/2019ascl.soft06010C}
}

@ARTICLE{Stassun2019,
       author = {{Stassun}, Keivan G. and {Oelkers}, Ryan J. and {Paegert}, Martin and {Torres}, Guillermo and {Pepper}, Joshua and {De Lee}, Nathan and {Collins}, Kevin and {Latham}, David W. and {Muirhead}, Philip S. and {Chittidi}, Jay and {Rojas-Ayala}, B{\'a}rbara and {Fleming}, Scott W. and {Rose}, Mark E. and {Tenenbaum}, Peter and {Ting}, Eric B. and {Kane}, Stephen R. and {Barclay}, Thomas and {Bean}, Jacob L. and {Brassuer}, C.~E. and {Charbonneau}, David and {Ge}, Jian and {Lissauer}, Jack J. and {Mann}, Andrew W. and {McLean}, Brian and {Mullally}, Susan and {Narita}, Norio and {Plavchan}, Peter and {Ricker}, George R. and {Sasselov}, Dimitar and {Seager}, S. and {Sharma}, Sanjib and {Shiao}, Bernie and {Sozzetti}, Alessandro and {Stello}, Dennis and {Vanderspek}, Roland and {Wallace}, Geoff and {Winn}, Joshua N.},
        title = "{The Revised TESS Input Catalog and Candidate Target List}",
      journal = {\aj},
         year = 2019,
        month = oct,
       volume = {158},
       number = {4},
          eid = {138},
        pages = {138},
          doi = {10.3847/1538-3881/ab3467},
archivePrefix = {arXiv},
       eprint = {1905.10694},
 primaryClass = {astro-ph.SR},
       adsurl = {https://ui.adsabs.harvard.edu/abs/2019AJ....158..138S}
}

@ARTICLE{Gajdos2019,
       author = {{Gajdo{\v{s}}}, Pavol and {Va{\v{n}}ko}, Martin and {Parimucha}, {\v{S}}tefan},
        title = "{Transit timing variations and linear ephemerides of confirmed Kepler transiting exoplanets}",
      journal = {Research in Astronomy and Astrophysics},
         year = 2019,
        month = mar,
       volume = {19},
       number = {3},
          eid = {041},
        pages = {041},
          doi = {10.1088/1674-4527/19/3/41},
archivePrefix = {arXiv},
       eprint = {1809.11104},
 primaryClass = {astro-ph.EP},
       adsurl = {https://ui.adsabs.harvard.edu/abs/2019RAA....19...41G}
}

@ARTICLE{Weiss2024,
       author = {{Weiss}, Lauren M. and {Isaacson}, Howard and {Howard}, Andrew W. and {Fulton}, Benjamin J. and {Petigura}, Erik A. and {Fabrycky}, Daniel and {Jontof-Hutter}, Daniel and {Steffen}, Jason H. and {Schlichting}, Hilke E. and {Wright}, Jason T. and {Beard}, Corey and {Brinkman}, Casey L. and {Chontos}, Ashley and {Giacalone}, Steven and {Hill}, Michelle L. and {Kosiarek}, Molly R. and {MacDougall}, Mason G. and {Mo{\v{c}}nik}, Teo and {Polanski}, Alex S. and {Turtelboom}, Emma V. and {Tyler}, Dakotah and {Van Zandt}, Judah},
        title = "{The Kepler Giant Planet Search. I. A Decade of Kepler Planet-host Radial Velocities from W. M. Keck Observatory}",
      journal = {\apjs},
         year = 2024,
        month = jan,
       volume = {270},
       number = {1},
          eid = {8},
        pages = {8},
          doi = {10.3847/1538-4365/ad0cab},
archivePrefix = {arXiv},
       eprint = {2304.00071},
 primaryClass = {astro-ph.EP},
       adsurl = {https://ui.adsabs.harvard.edu/abs/2024ApJS..270....8W}
}

@ARTICLE{Borucki2019,
       author = {{Borucki}, William and {Thompson}, Susan E. and {Agol}, Eric and {Hedges}, Christina},
        title = "{Kepler-62f: Kepler's First Small Planet in the Habitable Zone, but Is It Real?}",
      journal = {arXiv e-prints},
         year = 2019,
        month = may,
          eid = {arXiv:1905.05719},
        pages = {arXiv:1905.05719},
          doi = {10.48550/arXiv.1905.05719},
archivePrefix = {arXiv},
       eprint = {1905.05719},
 primaryClass = {astro-ph.EP},
       adsurl = {https://ui.adsabs.harvard.edu/abs/2019arXiv190505719B}
}

@ARTICLE{Bello-Arufe2022,
       author = {{Bello-Arufe}, Aaron and {Cabot}, Samuel H.~C. and {Mendon{\c{c}}a}, Jo{\~a}o M. and {Buchhave}, Lars A. and {Rathcke}, Alexander D.},
        title = "{Mining the Ultrahot Skies of HAT-P-70b: Detection of a Profusion of Neutral and Ionized Species}",
      journal = {\aj},
         year = 2022,
        month = feb,
       volume = {163},
       number = {2},
          eid = {96},
        pages = {96},
          doi = {10.3847/1538-3881/ac402e},
archivePrefix = {arXiv},
       eprint = {2112.03292},
 primaryClass = {astro-ph.EP},
       adsurl = {https://ui.adsabs.harvard.edu/abs/2022AJ....163...96B}
}

@ARTICLE{Bello-Arufe2023,
       author = {{Bello-Arufe}, Aaron and {Knutson}, Heather A. and {Mendon{\c{c}}a}, Jo{\~a}o M. and {Zhang}, Michael M. and {Cabot}, Samuel H.~C. and {Rathcke}, Alexander D. and {Ulla}, Ana and {Vissapragada}, Shreyas and {Buchhave}, Lars A.},
        title = "{Transmission Spectroscopy of the Lowest-density Gas Giant: Metals and a Potential Extended Outflow in HAT-P-67b}",
      journal = {\aj},
         year = 2023,
        month = aug,
       volume = {166},
       number = {2},
          eid = {69},
        pages = {69},
          doi = {10.3847/1538-3881/acd935},
archivePrefix = {arXiv},
       eprint = {2307.06356},
 primaryClass = {astro-ph.EP},
       adsurl = {https://ui.adsabs.harvard.edu/abs/2023AJ....166...69B}
}

@ARTICLE{Gaia2023,
       author = {{Gaia Collaboration}},
        title = "{Gaia Data Release 3. Summary of the content and survey properties}",
      journal = {\aap},
         year = 2023,
        month = jun,
       volume = {674},
          eid = {A1},
        pages = {A1},
          doi = {10.1051/0004-6361/202243940},
archivePrefix = {arXiv},
       eprint = {2208.00211},
 primaryClass = {astro-ph.GA},
       adsurl = {https://ui.adsabs.harvard.edu/abs/2023A&A...674A...1G},
      note = {Full author list available in the original publication}
}

@article{Cutri2003,
author = {Cutri, R. and Skrutskie, M. and Dyk, S. and Beichman, C. and Carpenter, J. and Chester, Tom and Cambresy, L. and Evans, T. and Fowler, John and Gizis, J. and Howard, E. and Huchra, J.},
year = {2003},
month = {03},
pages = {},
title = {2MASS All-Sky Catalog of Point Sources (Cutri+ 2003)},
journal = {VizieR Online Data Catalog}
}

@ARTICLE{SDSS2022,
       author = {{Abdurro'uf} and {Accetta}, Katherine and {Aerts}, Conny and {Silva Aguirre}, V{\'\i}ctor and {Ahumada}, Romina and {Ajgaonkar}, Nikhil and {Filiz Ak}, N. and {Alam}, Shadab and {Allende Prieto}, Carlos and {Almeida}, Andr{\'e}s and {Anders}, Friedrich and {Anderson}, Scott F. and {Andrews}, Brett H. and {Anguiano}, Borja and {Aquino-Ort{\'\i}z}, Erik and {Arag{\'o}n-Salamanca}, Alfonso and {Argudo-Fern{\'a}ndez}, Maria and {Ata}, Metin and {Aubert}, Marie and {Avila-Reese}, Vladimir and {Badenes}, Carles and {Barb{\'a}}, Rodolfo H. and {Barger}, Kat and {Barrera-Ballesteros}, Jorge K. and {Beaton}, Rachael L. and {Beers}, Timothy C. and {Belfiore}, Francesco and {Bender}, Chad F. and {Bernardi}, Mariangela and {Bershady}, Matthew A. and {Beutler}, Florian and {Bidin}, Christian Moni and {Bird}, Jonathan C. and {Bizyaev}, Dmitry and {Blanc}, Guillermo A. and {Blanton}, Michael R. and {Boardman}, Nicholas Fraser and {Bolton}, Adam S. and {Boquien}, M{\'e}d{\'e}ric and {Borissova}, Jura and {Bovy}, Jo and {Brandt}, W.~N. and {Brown}, Jordan and {Brownstein}, Joel R. and {Brusa}, Marcella and {Buchner}, Johannes and {Bundy}, Kevin and {Burchett}, Joseph N. and {Bureau}, Martin and {Burgasser}, Adam and {Cabang}, Tuesday K. and {Campbell}, Stephanie and {Cappellari}, Michele and {Carlberg}, Joleen K. and {Wanderley}, F{\'a}bio Carneiro and {Carrera}, Ricardo and {Cash}, Jennifer and {Chen}, Yan-Ping and {Chen}, Wei-Huai and {Cherinka}, Brian and {Chiappini}, Cristina and {Choi}, Peter Doohyun and {Chojnowski}, S. Drew and {Chung}, Haeun and {Clerc}, Nicolas and {Cohen}, Roger E. and {Comerford}, Julia M. and {Comparat}, Johan and {da Costa}, Luiz and {Covey}, Kevin and {Crane}, Jeffrey D. and {Cruz-Gonzalez}, Irene and {Culhane}, Connor and {Cunha}, Katia and {Dai}, Y. Sophia and {Damke}, Guillermo and {Darling}, Jeremy and {Davidson}, Jr., James W. and {Davies}, Roger and {Dawson}, Kyle and {De Lee}, Nathan and {Diamond-Stanic}, Aleksandar M. and {Cano-D{\'\i}az}, Mariana and {S{\'a}nchez}, Helena Dom{\'\i}nguez and {Donor}, John and {Duckworth}, Chris and {Dwelly}, Tom and {Eisenstein}, Daniel J. and {Elsworth}, Yvonne P. and {Emsellem}, Eric and {Eracleous}, Mike and {Escoffier}, Stephanie and {Fan}, Xiaohui and {Farr}, Emily and {Feng}, Shuai and {Fern{\'a}ndez-Trincado}, Jos{\'e} G. and {Feuillet}, Diane and {Filipp}, Andreas and {Fillingham}, Sean P. and {Frinchaboy}, Peter M. and {Fromenteau}, Sebastien and {Galbany}, Llu{\'\i}s and {Garc{\'\i}a}, Rafael A. and {Garc{\'\i}a-Hern{\'a}ndez}, D.~A. and {Ge}, Junqiang and {Geisler}, Doug and {Gelfand}, Joseph and {G{\'e}ron}, Tobias and {Gibson}, Benjamin J. and {Goddy}, Julian and {Godoy-Rivera}, Diego and {Grabowski}, Kathleen and {Green}, Paul J. and {Greener}, Michael and {Grier}, Catherine J. and {Griffith}, Emily and {Guo}, Hong and {Guy}, Julien and {Hadjara}, Massinissa and {Harding}, Paul and {Hasselquist}, Sten and {Hayes}, Christian R. and {Hearty}, Fred and {Hern{\'a}ndez}, Jes{\'u}s and {Hill}, Lewis and {Hogg}, David W. and {Holtzman}, Jon A. and {Horta}, Danny and {Hsieh}, Bau-Ching and {Hsu}, Chin-Hao and {Hsu}, Yun-Hsin and {Huber}, Daniel and {Huertas-Company}, Marc and {Hutchinson}, Brian and {Hwang}, Ho Seong and {Ibarra-Medel}, H{\'e}ctor J. and {Chitham}, Jacob Ider and {Ilha}, Gabriele S. and {Imig}, Julie and {Jaekle}, Will and {Jayasinghe}, Tharindu and {Ji}, Xihan and {Johnson}, Jennifer A. and {Jones}, Amy and {J{\"o}nsson}, Henrik and {Katkov}, Ivan and {Khalatyan}, Dr., Arman and {Kinemuchi}, Karen and {Kisku}, Shobhit and {Knapen}, Johan H. and {Kneib}, Jean-Paul and {Kollmeier}, Juna A. and {Kong}, Miranda and {Kounkel}, Marina and {Kreckel}, Kathryn and {Krishnarao}, Dhanesh and {Lacerna}, Ivan and {Lane}, Richard R. and {Langgin}, Rachel and {Lavender}, Ramon and {Law}, David R. and {Lazarz}, Daniel and {Leung}, Henry W. and {Leung}, Ho-Hin and {Lewis}, Hannah M. and {Li}, Cheng and {Li}, Ran and {Lian}, Jianhui and {Liang}, Fu-Heng and {Lin}, Lihwai and {Lin}, Yen-Ting and {Lin}, Sicheng and {Lintott}, Chris and {Long}, Dan and {Longa-Pe{\~n}a}, Pen{\'e}lope and {L{\'o}pez-Cob{\'a}}, Carlos and {Lu}, Shengdong and {Lundgren}, Britt F. and {Luo}, Yuanze and {Mackereth}, J. Ted and {de la Macorra}, Axel and {Mahadevan}, Suvrath and {Majewski}, Steven R. and {Manchado}, Arturo and {Mandeville}, Travis and {Maraston}, Claudia and {Margalef-Bentabol}, Berta and {Masseron}, Thomas and {Masters}, Karen L. and {Mathur}, Savita and {McDermid}, Richard M. and {Mckay}, Myles and {Merloni}, Andrea and {Merrifield}, Michael and {Meszaros}, Szabolcs and {Miglio}, Andrea and {Di Mille}, Francesco and {Minniti}, Dante and {Minsley}, Rebecca and {Monachesi}, Antonela},
        title = "{The Seventeenth Data Release of the Sloan Digital Sky Surveys: Complete Release of MaNGA, MaStar, and APOGEE-2 Data}",
      journal = {\apjs},
         year = 2022,
        month = apr,
       volume = {259},
       number = {2},
          eid = {35},
        pages = {35},
          doi = {10.3847/1538-4365/ac4414},
archivePrefix = {arXiv},
       eprint = {2112.02026},
 primaryClass = {astro-ph.GA},
       adsurl = {https://ui.adsabs.harvard.edu/abs/2022ApJS..259...35A}
}

@ARTICLE{Apogee2020,
       author = {{J{\"o}nsson}, Henrik and {Holtzman}, Jon A. and {Allende Prieto}, Carlos and {Cunha}, Katia and {Garc{\'\i}a-Hern{\'a}ndez}, D.~A. and {Hasselquist}, Sten and {Masseron}, Thomas and {Osorio}, Yeisson and {Shetrone}, Matthew and {Smith}, Verne and {Stringfellow}, Guy S. and {Bizyaev}, Dmitry and {Edvardsson}, Bengt and {Majewski}, Steven R. and {M{\'e}sz{\'a}ros}, Szabolcs and {Souto}, Diogo and {Zamora}, Olga and {Beaton}, Rachael L. and {Bovy}, Jo and {Donor}, John and {Pinsonneault}, Marc H. and {Poovelil}, Vijith Jacob and {Sobeck}, Jennifer},
        title = "{APOGEE Data and Spectral Analysis from SDSS Data Release 16: Seven Years of Observations Including First Results from APOGEE-South}",
      journal = {\aj},
         year = 2020,
        month = sep,
       volume = {160},
       number = {3},
          eid = {120},
        pages = {120},
          doi = {10.3847/1538-3881/aba592},
archivePrefix = {arXiv},
       eprint = {2007.05537},
 primaryClass = {astro-ph.GA},
       adsurl = {https://ui.adsabs.harvard.edu/abs/2020AJ....160..120J}
}

@ARTICLE{Hardegree-Ullman2019,
       author = {{Hardegree-Ullman}, Kevin K. and {Cushing}, Michael C. and {Muirhead}, Philip S. and {Christiansen}, Jessie L.},
        title = "{Kepler Planet Occurrence Rates for Mid-type M Dwarfs as a Function of Spectral Type}",
      journal = {\aj},
         year = 2019,
        month = aug,
       volume = {158},
       number = {2},
          eid = {75},
        pages = {75},
          doi = {10.3847/1538-3881/ab21d2},
archivePrefix = {arXiv},
       eprint = {1905.05900},
 primaryClass = {astro-ph.EP},
       adsurl = {https://ui.adsabs.harvard.edu/abs/2019AJ....158...75H}
}

@ARTICLE{Gizis2000,
       author = {{Gizis}, John E. and {Monet}, David G. and {Reid}, I. Neill and {Kirkpatrick}, J. Davy and {Liebert}, James and {Williams}, Rik J.},
        title = "{New Neighbors from 2MASS: Activity and Kinematics at the Bottom of the Main Sequence}",
      journal = {\aj},
         year = 2000,
        month = aug,
       volume = {120},
       number = {2},
        pages = {1085-1099},
          doi = {10.1086/301456},
archivePrefix = {arXiv},
       eprint = {astro-ph/0004361},
 primaryClass = {astro-ph},
       adsurl = {https://ui.adsabs.harvard.edu/abs/2000AJ....120.1085G}
}

@ARTICLE{Quintana2014,
       author = {{Quintana}, Elisa V. and {Barclay}, Thomas and {Raymond}, Sean N. and {Rowe}, Jason F. and {Bolmont}, Emeline and {Caldwell}, Douglas A. and {Howell}, Steve B. and {Kane}, Stephen R. and {Huber}, Daniel and {Crepp}, Justin R. and {Lissauer}, Jack J. and {Ciardi}, David R. and {Coughlin}, Jeffrey L. and {Everett}, Mark E. and {Henze}, Christopher E. and {Horch}, Elliott and {Isaacson}, Howard and {Ford}, Eric B. and {Adams}, Fred C. and {Still}, Martin and {Hunter}, Roger C. and {Quarles}, Billy and {Selsis}, Franck},
        title = "{An Earth-Sized Planet in the Habitable Zone of a Cool Star}",
      journal = {Science},
         year = 2014,
        month = apr,
       volume = {344},
       number = {6181},
        pages = {277-280},
          doi = {10.1126/science.1249403},
archivePrefix = {arXiv},
       eprint = {1404.5667},
 primaryClass = {astro-ph.EP},
       adsurl = {https://ui.adsabs.harvard.edu/abs/2014Sci...344..277Q}
}

@article{Kanodia2018,
doi = {10.3847/2515-5172/aaa4b7},
url = {https://dx.doi.org/10.3847/2515-5172/aaa4b7},
year = {2018},
month = {jan},
publisher = {The American Astronomical Society},
volume = {2},
number = {1},
pages = {4},
author = {Kanodia, Shubham and Wright, Jason},
title = {Python Leap Second Management and Implementation of Precise Barycentric Correction (barycorrpy)},
journal = {Research Notes of the AAS}
}

@article{Wright2014,
   title={Barycentric Corrections at 1cms-1 for Precise Doppler Velocities},
   volume={126},
   ISSN={1538-3873},
   url={http://dx.doi.org/10.1086/678541},
   DOI={10.1086/678541},
   number={943},
   journal={Publications of the Astronomical Society of the Pacific},
   publisher={IOP Publishing},
   author={Wright, J. T. and Eastman, J. D.},
   year={2014},
   month=sep, pages={838–852} }

@article{Astropy2022,
   title={The Astropy Project: Sustaining and Growing a Community-oriented Open-source Project and the Latest Major Release (v5.0) of the Core Package*},
   volume={935},
   ISSN={1538-4357},
   url={http://dx.doi.org/10.3847/1538-4357/ac7c74},
   DOI={10.3847/1538-4357/ac7c74},
   number={2},
   journal={The Astrophysical Journal},
   publisher={American Astronomical Society},
   author = {{Astropy Collaboration}},
   note = {Full author list available in the original publication},
   year={2022},
   month=aug, pages={167} }

@ARTICLE{Pickles1998,
       author = {{Pickles}, A.~J.},
        title = "{A Stellar Spectral Flux Library: 1150-25000 {\r{A}}}",
      journal = {\pasp},
         year = 1998,
        month = jul,
       volume = {110},
       number = {749},
        pages = {863-878},
          doi = {10.1086/316197},
       adsurl = {https://ui.adsabs.harvard.edu/abs/1998PASP..110..863P}
}

@ARTICLE{Hardegree-Ullman2025,
       author = {{Hardegree-Ullman}, Kevin K. and {Apai}, D{\'a}niel and {Haffert}, Sebastiaan Y. and {Schlecker}, Martin and {Kasper}, Markus and {Kammerer}, Jens and {Wagner}, Kevin},
        title = "{Bioverse: Giant Magellan Telescope and Extremely Large Telescope Direct Imaging and High-resolution Spectroscopy Assessment{\textemdash}Surveying Exo-Earth O$_{2}$ and Testing the Habitable Zone Oxygen Hypothesis}",
      journal = {\aj},
         year = 2025,
        month = mar,
       volume = {169},
       number = {3},
          eid = {171},
        pages = {171},
          doi = {10.3847/1538-3881/adb02f},
archivePrefix = {arXiv},
       eprint = {2405.11423},
 primaryClass = {astro-ph.EP},
       adsurl = {https://ui.adsabs.harvard.edu/abs/2025AJ....169..171H}
}

@ARTICLE{Ramirez2018,
       author = {{Ramirez}, Ramses M.},
        title = "{A More Comprehensive Habitable Zone for Finding Life on Other Planets}",
      journal = {Geosciences},
         year = 2018,
        month = jul,
       volume = {8},
       number = {8},
          eid = {280},
        pages = {280},
          doi = {10.3390/geosciences8080280},
archivePrefix = {arXiv},
       eprint = {1807.09504},
 primaryClass = {astro-ph.EP},
       adsurl = {https://ui.adsabs.harvard.edu/abs/2018Geosc...8..280R}
}

@ARTICLE{Thompson2022,
       author = {{Thompson}, Maggie A. and {Krissansen-Totton}, Joshua and {Wogan}, Nicholas and {Telus}, Myriam and {Fortney}, Jonathan J.},
        title = "{The case and context for atmospheric methane as an exoplanet biosignature}",
      journal = {Proceedings of the National Academy of Science},
         year = 2022,
        month = apr,
       volume = {119},
       number = {14},
          eid = {e2117933119},
        pages = {e2117933119},
          doi = {10.1073/pnas.2117933119},
archivePrefix = {arXiv},
       eprint = {2204.04257},
 primaryClass = {astro-ph.EP},
       adsurl = {https://ui.adsabs.harvard.edu/abs/2022PNAS..11917933T}
}

@ARTICLE{Brogi2017,
       author = {{Brogi}, M. and {Line}, M. and {Bean}, J. and {D{\'e}sert}, J. -M. and {Schwarz}, H.},
        title = "{A Framework to Combine Low- and High-resolution Spectroscopy for the Atmospheres of Transiting Exoplanets}",
      journal = {\apjl},
         year = 2017,
        month = apr,
       volume = {839},
       number = {1},
          eid = {L2},
        pages = {L2},
          doi = {10.3847/2041-8213/aa6933},
archivePrefix = {arXiv},
       eprint = {1612.07008},
 primaryClass = {astro-ph.EP},
       adsurl = {https://ui.adsabs.harvard.edu/abs/2017ApJ...839L...2B}
}

@ARTICLE{Smith2024,
       author = {{Smith}, Peter C.~B. and {Line}, Michael R. and {Bean}, Jacob L. and {Brogi}, Matteo and {August}, Prune and {Welbanks}, Luis and {Desert}, Jean-Michel and {Lunine}, Jonathan and {Sanchez}, Jorge and {Mansfield}, Megan and {Pino}, Lorenzo and {Rauscher}, Emily and {Kempton}, Eliza and {Zalesky}, Joseph and {Fowler}, Martin},
        title = "{A Combined Ground-based and JWST Atmospheric Retrieval Analysis: Both IGRINS and NIRSpec Agree that the Atmosphere of WASP-77A b Is Metal-poor}",
      journal = {\aj},
         year = 2024,
        month = mar,
       volume = {167},
       number = {3},
          eid = {110},
        pages = {110},
          doi = {10.3847/1538-3881/ad17bf},
archivePrefix = {arXiv},
       eprint = {2312.13069},
 primaryClass = {astro-ph.EP},
       adsurl = {https://ui.adsabs.harvard.edu/abs/2024AJ....167..110S}
}

@ARTICLE{Kasper2023,
       author = {{Kasper}, David and {Bean}, Jacob L. and {Line}, Michael R. and {Seifahrt}, Andreas and {Brady}, Madison T. and {Lothringer}, Joshua and {Pino}, Lorenzo and {Fu}, Guangwei and {Pelletier}, Stefan and {St{\"u}rmer}, Julian and {Benneke}, Bj{\"o}rn and {Brogi}, Matteo and {D{\'e}sert}, Jean-Michel},
        title = "{Unifying High- and Low-resolution Observations to Constrain the Dayside Atmosphere of KELT-20b/MASCARA-2b}",
      journal = {\aj},
         year = 2023,
        month = jan,
       volume = {165},
       number = {1},
          eid = {7},
        pages = {7},
          doi = {10.3847/1538-3881/ac9f40},
archivePrefix = {arXiv},
       eprint = {2208.04759},
 primaryClass = {astro-ph.EP},
       adsurl = {https://ui.adsabs.harvard.edu/abs/2023AJ....165....7K}
}

@misc{Buchner2016,
       author = {{Buchner}, Johannes},
        title = "{PyMultiNest: Python interface for MultiNest}",
 howpublished = {Astrophysics Source Code Library, record ascl:1606.005},
         year = 2016,
        month = jun,
          eid = {ascl:1606.005},
       adsurl = {https://ui.adsabs.harvard.edu/abs/2016ascl.soft06005B}
}

@article{Feroz2009,
    author = {Feroz, F. and Hobson, M. P. and Bridges, M.},
    title = {MultiNest: an efficient and robust Bayesian inference tool for cosmology and particle physics},
    journal = {Monthly Notices of the Royal Astronomical Society},
    volume = {398},
    number = {4},
    pages = {1601-1614},
    year = {2009},
    month = {09},
    issn = {0035-8711},
    doi = {10.1111/j.1365-2966.2009.14548.x},
    url = {https://doi.org/10.1111/j.1365-2966.2009.14548.x},
    eprint = {https://academic.oup.com/mnras/article-pdf/398/4/1601/3039078/mnras0398-1601.pdf},
}

@ARTICLE{Virtanen2020,
  author  = {Virtanen, Pauli and Gommers, Ralf and Oliphant, Travis E. and
            Haberland, Matt and Reddy, Tyler and Cournapeau, David and
            Burovski, Evgeni and Peterson, Pearu and Weckesser, Warren and
            Bright, Jonathan and {van der Walt}, St{\'e}fan J. and
            Brett, Matthew and Wilson, Joshua and Millman, K. Jarrod and
            Mayorov, Nikolay and Nelson, Andrew R. J. and Jones, Eric and
            Kern, Robert and Larson, Eric and Carey, C J and
            Polat, {\.I}lhan and Feng, Yu and Moore, Eric W. and
            {VanderPlas}, Jake and Laxalde, Denis and Perktold, Josef and
            Cimrman, Robert and Henriksen, Ian and Quintero, E. A. and
            Harris, Charles R. and Archibald, Anne M. and
            Ribeiro, Ant{\^o}nio H. and Pedregosa, Fabian and
            {van Mulbregt}, Paul and {SciPy 1.0 Contributors}},
  title   = {{{SciPy} 1.0: Fundamental Algorithms for Scientific
            Computing in Python}},
  journal = {Nature Methods},
  year    = {2020},
  volume  = {17},
  pages   = {261--272},
  adsurl  = {https://rdcu.be/b08Wh},
  doi     = {10.1038/s41592-019-0686-2},
}

@ARTICLE{Said2016,
       author = {{Said}, Alejandro and {Daviz{\'o}n}, Yasser A. and {Espino-Rom{\'a}n}, Piero and {Rodr{\'\i}guez-Said}, Roberto and {Hern{\'a}ndez-Santos}, Carlos},
        title = "{Automatic Frequency Identification under Sample Loss in Sinusoidal Pulse Width Modulation Signals Using an Iterative Autocorrelation Algorithm}",
      journal = {Symmetry},
         year = 2016,
        month = aug,
       volume = {8},
       number = {8},
          eid = {78},
        pages = {78},
          doi = {10.3390/sym8080078},
       adsurl = {https://ui.adsabs.harvard.edu/abs/2016Symm....8...78S}
}

@ARTICLE{Martin1932,
       author = {{Martin}, P.~E. and {Barker}, E.~F.},
        title = "{The Infrared Absorption Spectrum of Carbon Dioxide}",
      journal = {Physical Review},
         year = 1932,
        month = aug,
       volume = {41},
       number = {3},
        pages = {291-303},
          doi = {10.1103/PhysRev.41.291},
       adsurl = {https://ui.adsabs.harvard.edu/abs/1932PhRv...41..291M}
}

@ARTICLE{Kiehl1985,
       author = {{Kiehl}, J.~T. and {Yamanouchi}, T.},
        title = "{A parameterization for absorption due to the A, B, and {\ensuremath{\gamma}} oxygen bands}",
      journal = {Tellus Series B Chemical and Physical Meteorology B},
         year = 1985,
        month = feb,
       volume = {37},
       number = {1},
        pages = {1-6},
          doi = {10.3402/tellusb.v37i1.14986},
       adsurl = {https://ui.adsabs.harvard.edu/abs/1985TellB..37....1K}
}

@ARTICLE{Yurchenko2014,
       author = {{Yurchenko}, Sergei N. and {Tennyson}, Jonathan},
        title = "{ExoMol line lists - IV. The rotation-vibration spectrum of methane up to 1500 K}",
      journal = {\mnras},
         year = 2014,
        month = may,
       volume = {440},
       number = {2},
        pages = {1649-1661},
          doi = {10.1093/mnras/stu326},
archivePrefix = {arXiv},
       eprint = {1401.4852},
 primaryClass = {astro-ph.EP},
       adsurl = {https://ui.adsabs.harvard.edu/abs/2014MNRAS.440.1649Y}
}

@ARTICLE{Yurchenko2024,
       author = {{Yurchenko}, Sergei N. and {Owens}, Alec and {Kefala}, Kyriaki and {Tennyson}, Jonathan},
        title = "{ExoMol line lists - LVII. High accuracy ro-vibrational line list for methane (CH$_{4}$)}",
      journal = {\mnras},
         year = 2024,
        month = feb,
       volume = {528},
       number = {2},
        pages = {3719-3729},
          doi = {10.1093/mnras/stae148},
       adsurl = {https://ui.adsabs.harvard.edu/abs/2024MNRAS.528.3719Y}
}

@ARTICLE{Barber2006,
       author = {{Barber}, R.~J. and {Tennyson}, J. and {Harris}, G.~J. and {Tolchenov}, R.~N.},
        title = "{A high-accuracy computed water line list}",
      journal = {\mnras},
         year = 2006,
        month = may,
       volume = {368},
       number = {3},
        pages = {1087-1094},
          doi = {10.1111/j.1365-2966.2006.10184.x},
archivePrefix = {arXiv},
       eprint = {astro-ph/0601236},
 primaryClass = {astro-ph},
       adsurl = {https://ui.adsabs.harvard.edu/abs/2006MNRAS.368.1087B}
}

@ARTICLE{Tennyson2013,
       author = {{Tennyson}, Jonathan and {Bernath}, Peter F. and {Brown}, Linda R. and {Campargue}, Alain and {Cs{\'a}sz{\'a}r}, Attila G. and {Daumont}, Ludovic and {Gamache}, Robert R. and {Hodges}, Joseph T. and {Naumenko}, Olga V. and {Polyansky}, Oleg L. and {Rothman}, Laurence S. and {Vandaele}, Ann Carine and {Zobov}, Nikolai F. and {Al Derzi}, Afaf R. and {F{\'a}bri}, Csaba and {Fazliev}, Alexander Z. and {Furtenbacher}, Tibor and {Gordon}, Iouli E. and {Lodi}, Lorenzo and {Mizus}, Irina I.},
        title = "{IUPAC critical evaluation of the rotational-vibrational spectra of water vapor, Part III: Energy levels and transition wavenumbers for H$_{2}$$^{16}$O}",
      journal = {\jqsrt},
         year = 2013,
        month = mar,
       volume = {117},
        pages = {29-58},
          doi = {10.1016/j.jqsrt.2012.10.002},
       adsurl = {https://ui.adsabs.harvard.edu/abs/2013JQSRT.117...29T}
}

@ARTICLE{Bohl2025,
       author = {{Bohl}, Abigail and {Lawrence}, Lucas and {Lowry}, Gillis and {Kaltenegger}, Lisa},
        title = "{Probing the Limits of Habitability: A Catalog of Rocky Exoplanets in the Habitable Zone}",
      journal = {arXiv e-prints},
         year = 2025,
        month = jan,
          eid = {arXiv:2501.14054},
        pages = {arXiv:2501.14054},
          doi = {10.48550/arXiv.2501.14054},
archivePrefix = {arXiv},
       eprint = {2501.14054},
 primaryClass = {astro-ph.EP},
       adsurl = {https://ui.adsabs.harvard.edu/abs/2025arXiv250114054B}
}

@ARTICLE{Hill2023,
       author = {{Hill}, Michelle L. and {Bott}, Kimberly and {Dalba}, Paul A. and {Fetherolf}, Tara and {Kane}, Stephen R. and {Kopparapu}, Ravi and {Li}, Zhexing and {Ostberg}, Colby},
        title = "{A Catalog of Habitable Zone Exoplanets}",
      journal = {\aj},
         year = 2023,
        month = feb,
       volume = {165},
       number = {2},
          eid = {34},
        pages = {34},
          doi = {10.3847/1538-3881/aca1c0},
       adsurl = {https://ui.adsabs.harvard.edu/abs/2023AJ....165...34H}
}

@ARTICLE{Edlen1966,
       author = {{Edl{\'e}n}, Bengt},
        title = "{The Refractive Index of Air}",
      journal = {Metrologia},
         year = 1966,
        month = apr,
       volume = {2},
       number = {2},
        pages = {71-80},
          doi = {10.1088/0026-1394/2/2/002},
       adsurl = {https://ui.adsabs.harvard.edu/abs/1966Metro...2...71E}
}

@ARTICLE{Kasting1993,
       author = {{Kasting}, James F. and {Whitmire}, Daniel P. and {Reynolds}, Ray T.},
        title = "{Habitable Zones around Main Sequence Stars}",
      journal = {\icarus},
         year = 1993,
        month = jan,
       volume = {101},
       number = {1},
        pages = {108-128},
          doi = {10.1006/icar.1993.1010},
       adsurl = {https://ui.adsabs.harvard.edu/abs/1993Icar..101..108K}
}

@ARTICLE{Kopparapu2013,
       author = {{Kopparapu}, Ravi Kumar and {Ramirez}, Ramses and {Kasting}, James F. and {Eymet}, Vincent and {Robinson}, Tyler D. and {Mahadevan}, Suvrath and {Terrien}, Ryan C. and {Domagal-Goldman}, Shawn and {Meadows}, Victoria and {Deshpande}, Rohit},
        title = "{Habitable Zones around Main-sequence Stars: New Estimates}",
      journal = {\apj},
         year = 2013,
        month = mar,
       volume = {765},
       number = {2},
          eid = {131},
        pages = {131},
          doi = {10.1088/0004-637X/765/2/131},
archivePrefix = {arXiv},
       eprint = {1301.6674},
 primaryClass = {astro-ph.EP},
       adsurl = {https://ui.adsabs.harvard.edu/abs/2013ApJ...765..131K}
}

@ARTICLE{Robinson2014,
       author = {{Robinson}, T.~D. and {Catling}, D.~C.},
        title = "{Common 0.1 bar tropopause in thick atmospheres set by pressure-dependent infrared transparency}",
      journal = {Nature Geoscience},
         year = 2014,
        month = jan,
       volume = {7},
       number = {1},
        pages = {12-15},
          doi = {10.1038/ngeo2020},
archivePrefix = {arXiv},
       eprint = {1312.6859},
 primaryClass = {astro-ph.EP},
       adsurl = {https://ui.adsabs.harvard.edu/abs/2014NatGe...7...12R}
}

@BOOK{Pierrehumbert2010,
       author = {{Pierrehumbert}, Raymond T.},
        title = "{Principles of Planetary Climate}",
         year = 2010,
       adsurl = {https://ui.adsabs.harvard.edu/abs/2010ppc..book.....P},
      publisher = {Cambridge University Press}
}

@ARTICLE{Parmentier2014,
       author = {{Parmentier}, Vivien and {Guillot}, Tristan},
        title = "{A non-grey analytical model for irradiated atmospheres. I. Derivation}",
      journal = {\aap},
         year = 2014,
        month = feb,
       volume = {562},
          eid = {A133},
        pages = {A133},
          doi = {10.1051/0004-6361/201322342},
archivePrefix = {arXiv},
       eprint = {1311.6597},
 primaryClass = {astro-ph.EP},
       adsurl = {https://ui.adsabs.harvard.edu/abs/2014A&A...562A.133P}
}

@book{Bishop2006,
  author    = {Bishop, Christopher M.},
  title     = {Pattern Recognition and Machine Learning},
  series    = {Information Science and Statistics},
  publisher = {Springer},
  year      = {2006},
  isbn      = {9780387310732}
}

@ARTICLE{Krishnamoorthy2011,
       author = {{Krishnamoorthy}, Aravindh and {Menon}, Deepak},
        title = "{Matrix Inversion Using Cholesky Decomposition}",
      journal = {arXiv e-prints},
         year = 2011,
        month = nov,
          eid = {arXiv:1111.4144},
        pages = {arXiv:1111.4144},
          doi = {10.48550/arXiv.1111.4144},
archivePrefix = {arXiv},
       eprint = {1111.4144},
 primaryClass = {cs.MS},
       adsurl = {https://ui.adsabs.harvard.edu/abs/2011arXiv1111.4144K}
}

@book{Walck1996,
  title={Hand-book on statistical distributions for experimentalists},
  author={Christian Walck},
  year={1996},
  url={https://api.semanticscholar.org/CorpusID:122813337},
  publisher={Stockholms universitet}
}

@ARTICLE{Kipping2025,
       author = {{Kipping}, David and {Benneke}, Bj{\"o}rn},
        title = "{Exoplaneteers Keep Overestimating Sigma Significances}",
      journal = {arXiv e-prints},
         year = 2025,
        month = jun,
          eid = {arXiv:2506.05392},
        pages = {arXiv:2506.05392},
          doi = {10.48550/arXiv.2506.05392},
archivePrefix = {arXiv},
       eprint = {2506.05392},
 primaryClass = {astro-ph.IM},
       adsurl = {https://ui.adsabs.harvard.edu/abs/2025arXiv250605392K}
}

@article{Kass1995,
 ISSN = {01621459, 1537274X},
 URL = {http://www.jstor.org/stable/2291091},
 author = {Robert E. Kass and Adrian E. Raftery},
 journal = {Journal of the American Statistical Association},
 number = {430},
 pages = {773--795},
 publisher = {[American Statistical Association, Taylor & Francis, Ltd.]},
 title = {Bayes Factors},
 urldate = {2025-10-12},
 volume = {90},
 year = {1995}
}

@book{Wilks2011,
  address = {Amsterdam; Boston},
  author = {Wilks, Daniel S.},
  description = {Statistical Methods in the Atmospheric Sciences, Volume 100, Third Edition (International Geophysics) (9780123850225): Daniel S. Wilks: Books},
  isbn = {9780123850225 0123850223},
  publisher = {Elsevier Academic Press},
  refid = {762279039},
  title = {Statistical methods in the atmospheric sciences},
  year = 2011
}

@book{Hamilton1994,
 URL = {http://www.jstor.org/stable/j.ctv14jx6sm},
 author = {James D. Hamilton},
 publisher = {Princeton University Press},
 title = {Time Series Analysis},
 urldate = {2025-10-13},
 year = {1994}
}

@ARTICLE{Morley2017,
       author = {{Morley}, Caroline V. and {Kreidberg}, Laura and {Rustamkulov}, Zafar and {Robinson}, Tyler and {Fortney}, Jonathan J.},
        title = "{Observing the Atmospheres of Known Temperate Earth-sized Planets with JWST}",
      journal = {\apj},
         year = 2017,
        month = dec,
       volume = {850},
       number = {2},
          eid = {121},
        pages = {121},
          doi = {10.3847/1538-4357/aa927b},
archivePrefix = {arXiv},
       eprint = {1708.04239},
 primaryClass = {astro-ph.EP},
       adsurl = {https://ui.adsabs.harvard.edu/abs/2017ApJ...850..121M}
}

@INPROCEEDINGS{Marconi2022,
       author = {{Marconi}, A. and {Abreu}, M. and {Adibekyan}, V. and {Alberti}, V. and {Albrecht}, S. and {Alcaniz}, J. and {Aliverti}, M. and {Allende Prieto}, C. and {Alvarado G{\'o}mez}, J.~D. and {Amado}, P.~J. and {Amate}, M. and {Andersen}, M.~I. and {Artigau}, E. and {Baker}, C. and {Baldini}, V. and {Balestra}, A. and {Barnes}, S.~A. and {Baron}, F. and {Barros}, S.~C.~C. and {Bauer}, S.~M. and {Beaulieu}, M. and {Bellido-Tirado}, O. and {Benneke}, B. and {Bensby}, T. and {Bergin}, E.~A. and {Biazzo}, K. and {Bik}, A. and {Birkby}, J.~L. and {Blind}, N. and {Boisse}, I. and {Bolmont}, E. and {Bonaglia}, M. and {Bonfils}, X. and {Borsa}, F. and {Brandeker}, A. and {Brandner}, W. and {Broeg}, C.~H. and {Brogi}, M. and {Brousseau}, D. and {Brucalassi}, A. and {Brynnel}, J. and {Buchhave}, L.~A. and {Buscher}, D.~F. and {Cabral}, A. and {Calderone}, G. and {Calvo-Ortega}, R. and {Canto Martins}, B.~L. and {Cantalloube}, F. and {Carbonaro}, L. and {Chauvin}, G. and {Chazelas}, B. and {Cheffot}, A. -L. and {Cheng}, Y.~S. and {Chiavassa}, A. and {Christensen}, L. and {Cirami}, R. and {Cook}, N.~J. and {Cooke}, R.~J. and {Coretti}, I. and {Covino}, S. and {Cowan}, N. and {Cresci}, G. and {Cristiani}, S. and {Cunha Parro}, V. and {Cupani}, G. and {D'Odorico}, V. and {de Castro Le{\~a}o}, I. and {De Cia}, A. and {De Medeiros}, J.~R. and {Debras}, F. and {Debus}, M. and {Demangeon}, O. and {Dessauges-Zavadsky}, M. and {Di Marcantonio}, P. and {Dionies}, F. and {Doyon}, R. and {Dunn}, J. and {Ehrenreich}, D. and {Faria}, J.~P. and {Feruglio}, C. and {Fisher}, M. and {Fontana}, A. and {Fumagalli}, M. and {Fusco}, T. and {Fynbo}, J. and {Gabella}, O. and {Gaessler}, W. and {Gallo}, E. and {Gao}, X. and {Genolet}, L. and {Genoni}, M. and {Giacobbe}, P. and {Giro}, E. and {Gon{\c{c}}alves}, R.~S. and {Gonzalez}, O. and {Gonz{\'a}lez Hern{\'a}ndez}, J.~I. and {Gracia T{\'e}mich}, F. and {Haehnelt}, M.~G. and {Haniff}, C. and {Hatzes}, A. and {Helled}, R. and {Hoeijmakers}, H.~J. and {Huke}, P. and {J{\"a}rvinen}, S. and {J{\"a}rvinen}, A. and {Kaminski}, A. and {Korn}, A. and {Kouach}, D. and {Kowzan}, G. and {Kreidberg}, L. and {Landoni}, M. and {Lanotte}, A. and {Lavail}, A. and {Li}, J. and {Liske}, J. and {Lovis}, C. and {Lucatello}, S. and {Lunney}, D. and {MacIntosh}, M. and {Madhusudhan}, N. and {Magrini}, L. and {Maiolino}, R. and {Malo}, L. and {Man}, A. and {Marquart}, T. and {Marques}, E.~L. and {Martins}, A.~M. and {Martins}, C.~J.~A.~P. and {Maslowski}, P. and {Mason}, C. and {Mason}, E. and {McCracken}, R.~A. and {Mergo}, P. and {Micela}, G. and {Mitchell}, T. and {Molli{\`e}re}, P. and {Monteiro}, M. and {Montgomery}, D. and {Mordasini}, C. and {Morin}, J. and {Mucciarelli}, A. and {Murphy}, M.~T. and {N'Diaye}, M. and {Neichel}, B. and {Niedzielski}, A.~T. and {Niemczura}, E. and {Nortmann}, L. and {Noterdaeme}, P. and {Nunes}, N. and {Oggioni}, L. and {Oliva}, E. and {{\"O}nel}, H. and {Origlia}, L. and {{\"O}stlin}, G. and {Palle}, E. and {Papaderos}, P. and {Pariani}, G. and {Pe{\~n}ate Castro}, J. and {Pepe}, F. and {Perreault Levasseur}, L. and {Petit}, P. and {Pino}, L. and {Piqueras}, J. and {Pollo}, A. and {Poppenhaeger}, K. and {Quirrenbach}, A. and {Rauscher}, E. and {Rebolo}, R. and {Redaelli}, E.~M.~A. and {Reffert}, S. and {Reid}, D.~T. and {Reiners}, A. and {Richter}, P. and {Riva}, M. and {Rivoire}, S. and {Rodr{\'\i}guez-L{\'o}pez}, C. and {Roederer}, I.~U. and {Romano}, D. and {Rousseau}, S. and {Rowe}, J. and {Salvadori}, S. and {Santos}, N. and {Santos Diaz}, P. and {Sanz-Forcada}, J. and {Sarajlic}, M. and {Sauvage}, J. -F. and {Sch{\"a}fer}, S. and {Schiavon}, R.~P. and {Schmidt}, T.~M. and {Selmi}, C. and {Sivanandam}, S. and {Sordet}, M. and {Sordo}, R. and {Sortino}, F. and {Sosnowska}, D. and {Sousa}, S.~G. and {Stempels}, E. and {Strassmeier}, K.~G. and {Su{\'a}rez Mascare{\~n}o}, A. and {Sulich}, A.},
        title = "{ANDES, the high resolution spectrograph for the ELT: science case, baseline design and path to construction}",
    booktitle = {Ground-based and Airborne Instrumentation for Astronomy IX},
         year = 2022,
       editor = {{Evans}, Christopher J. and {Bryant}, Julia J. and {Motohara}, Kentaro},
       series = {Society of Photo-Optical Instrumentation Engineers (SPIE) Conference Series},
       volume = {12184},
        month = aug,
          eid = {1218424},
        pages = {1218424},
          doi = {10.1117/12.2628689},
       adsurl = {https://ui.adsabs.harvard.edu/abs/2022SPIE12184E..24M}
}

@ARTICLE{Fulton2017,
       author = {{Fulton}, Benjamin J. and {Petigura}, Erik A. and {Howard}, Andrew W. and {Isaacson}, Howard and {Marcy}, Geoffrey W. and {Cargile}, Phillip A. and {Hebb}, Leslie and {Weiss}, Lauren M. and {Johnson}, John Asher and {Morton}, Timothy D. and {Sinukoff}, Evan and {Crossfield}, Ian J.~M. and {Hirsch}, Lea A.},
        title = "{The California-Kepler Survey. III. A Gap in the Radius Distribution of Small Planets}",
      journal = {\aj},
         year = 2017,
        month = sep,
       volume = {154},
       number = {3},
          eid = {109},
        pages = {109},
          doi = {10.3847/1538-3881/aa80eb},
archivePrefix = {arXiv},
       eprint = {1703.10375},
 primaryClass = {astro-ph.EP},
       adsurl = {https://ui.adsabs.harvard.edu/abs/2017AJ....154..109F}
}

@misc{Prather1989,
  author    = {{NASA conference}},
  editor    = {Michael J. Prather},
  title     = {An Assessment Model for Atmospheric Composition},
  booktitle = {NASA Conference Publication No. 3023},
  year      = {1989},
  publisher = {National Aeronautics and Space Administration, Office of Management, Scientific and Technical Information Division},
  address   = {Washington, D.C.},
  url = {https://ntrs.nasa.gov/api/citations/19890011217/downloads/19890011217.pdf}
}

@ARTICLE{Oman2008,
       author = {{Oman}, Luke and {Waugh}, Darryn W. and {Pawson}, Steven and {Stolarski}, Richard S. and {Nielsen}, J. Eric},
        title = "{Understanding the Changes of Stratospheric Water Vapor in Coupled Chemistry-Climate Model Simulations}",
      journal = {Journal of the Atmospheric Sciences},
         year = 2008,
        month = jan,
       volume = {65},
       number = {10},
        pages = {3278},
          doi = {10.1175/2008JAS2696.1},
       adsurl = {https://ui.adsabs.harvard.edu/abs/2008JAtS...65.3278O}
}

@ARTICLE{vonParis2011,
       author = {{von Paris}, P. and {Cabrera}, J. and {Godolt}, M. and {Grenfell}, J.~L. and {Hedelt}, P. and {Rauer}, H. and {Schreier}, F. and {Stracke}, B.},
        title = "{Spectroscopic characterization of the atmospheres of potentially habitable planets: GL 581 d as a model case study}",
      journal = {\aap},
         year = 2011,
        month = oct,
       volume = {534},
          eid = {A26},
        pages = {A26},
          doi = {10.1051/0004-6361/201117091},
archivePrefix = {arXiv},
       eprint = {1108.3670},
 primaryClass = {astro-ph.EP},
       adsurl = {https://ui.adsabs.harvard.edu/abs/2011A&A...534A..26V}
}

@ARTICLE{vonParis2013,
       author = {{von Paris}, P. and {Hedelt}, P. and {Selsis}, F. and {Schreier}, F. and {Trautmann}, T.},
        title = "{Characterization of potentially habitable planets: Retrieval of atmospheric and planetary properties from emission spectra}",
      journal = {\aap},
         year = 2013,
        month = mar,
       volume = {551},
          eid = {A120},
        pages = {A120},
          doi = {10.1051/0004-6361/201220009},
archivePrefix = {arXiv},
       eprint = {1301.0217},
 primaryClass = {astro-ph.EP},
       adsurl = {https://ui.adsabs.harvard.edu/abs/2013A&A...551A.120V}
}

@INPROCEEDINGS{Bekki2009,
       author = {{Bekki}, S. and {Lefevre}, F.},
        title = "{Stratospheric ozone: History and concepts and interactions with climate}",
    booktitle = {European Physical Journal Web of Conferences},
         year = 2009,
       series = {European Physical Journal Web of Conferences},
       volume = {1},
        month = feb,
    publisher = {EDP},
        pages = {113},
          doi = {10.1140/epjconf/e2009-00914-y},
       adsurl = {https://ui.adsabs.harvard.edu/abs/2009EPJWC...1..113B}
}

@ARTICLE{Kuttippurath2024,
       author = {{Kuttippurath}, Jayanarayanan and {Pillai Gopikrishnan}, Gopalakrishna and {M{\"u}ller}, Rolf and {Godin-Beekmann}, Sophie and {Brioude}, Jerome},
        title = "{No Severe Ozone Depletion In The Tropical Stratosphere In Recent Decades}",
      journal = {Atmospheric Chemistry \& Physics},
         year = 2024,
        month = jun,
       volume = {24},
        pages = {6743-6756},
          doi = {10.5194/acp-24-6743-2024},
       adsurl = {https://ui.adsabs.harvard.edu/abs/2024ACP....24.6743K}
}

@misc{Lan2025,
  author       = {{Lan}, X. and {Thoning}, K.W. and {Dlugokencky}, E.J},
  title        = {Trends in CO2, CH4, N2O, SF6},
  howpublished = {\url{https://gml.noaa.gov/ccgg/trends/}},
  note         = {Accessed: 2025-10, Version: 2025-09},
  year         = {2025},
  organization = {National Oceanic and Atmospheric Administration, Earth System Research Laboratories},
}

@article{Bainbridge1966,
author = {Bainbridge, A. E. and Heidt, LEROY E.},
title = {Measurements of methane in the troposphere and lower stratosphere},
journal = {Tellus},
volume = {18},
number = {2-3},
pages = {221-225},
url = {https://onlinelibrary.wiley.com/doi/abs/10.1111/j.2153-3490.1966.tb00230.x},
eprint = {https://onlinelibrary.wiley.com/doi/pdf/10.1111/j.2153-3490.1966.tb00230.x},
year = {1966}
}

@Article{ElAmraoui2014,
AUTHOR = {El Amraoui, L. and Atti\'e, J.-L. and Ricaud, P. and Lahoz, W. A. and Piacentini, A. and Peuch, V.-H. and Warner, J. X. and Abida, R. and Barr\'e, J. and Zbinden, R.},
TITLE = {Tropospheric CO vertical profiles deduced from total columns using data assimilation:   methodology and validation},
JOURNAL = {Atmospheric Measurement Techniques},
VOLUME = {7},
YEAR = {2014},
NUMBER = {9},
PAGES = {3035--3057},
URL = {https://amt.copernicus.org/articles/7/3035/2014/},
DOI = {10.5194/amt-7-3035-2014}
}

@ARTICLE{Pan2004,
       author = {{Pan}, L.~L. and {Randel}, W.~J. and {Gary}, B.~L. and {Mahoney}, M.~J. and {Hintsa}, E.~J.},
        title = "{Definitions and sharpness of the extratropical tropopause: A trace gas perspective}",
      journal = {Journal of Geophysical Research (Atmospheres)},
         year = 2004,
        month = dec,
       volume = {109},
       number = {D23},
          eid = {D23103},
        pages = {D23103},
          doi = {10.1029/2004JD004982},
       adsurl = {https://ui.adsabs.harvard.edu/abs/2004JGRD..10923103P}
}

@INPROCEEDINGS{Marconi2024,
       author = {{Marconi}, A. and {Abreu}, M. and {Adibekyan}, V. and {Alberti}, V. and {Albrecht}, S. and {Alcaniz}, J. and {Aliverti}, M. and {Allende Prieto}, C. and {Alvarado-Gomez}, J.~D. and {Alves}, C.~S. and {Amado}, P.~J. and {Amate}, M. and {Andersen}, M.~I. and {Antoniucci}, S. and {Artigau}, E. and {Bailet}, C. and {Baker}, C. and {Baldini}, V. and {Balestra}, A. and {Barnes}, S.~A. and {Baron}, F. and {Barros}, S.~C.~C. and {Bauer}, S.~M. and {Beaulieu}, M. and {Bellido-Tirado}, O. and {Benneke}, B. and {Bensby}, T. and {Bergin}, E.~A. and {Berio}, P. and {Biazzo}, K. and {Bigot}, L. and {Bik}, A. and {Birkby}, J.~L. and {Blind}, N. and {Boebion}, O. and {Boisse}, I. and {Bolmont}, E. and {Bolton}, J.~S. and {Bonaglia}, M. and {Bonfils}, X. and {Bonhomme}, L. and {Borsa}, F. and {Bouret}, J. -C. and {Brandeker}, A. and {Brandner}, W. and {Broeg}, C.~H. and {Brogi}, M. and {Brousseau}, D. and {Brucalassi}, A. and {Brynnel}, J. and {Buchhave}, L.~A. and {Buscher}, D.~F. and {Cabona}, L. and {Cabral}, A. and {Calderone}, G. and {Calvo-Ortega}, R. and {Cantalloube}, F. and {Canto Martins}, B.~L. and {Carbonaro}, L. and {Caujolle}, Y. and {Chauvin}, G. and {Chazelas}, B. and {Cheffot}, A. -L. and {Cheng}, Y.~S. and {Chiavassa}, A. and {Christensen}, L. and {Cirami}, R. and {Cirasuolo}, M. and {Cook}, N.~J. and {Cooke}, R.~J. and {Coretti}, I. and {Covino}, S. and {Cowan}, N. and {Cresci}, G. and {Cristiani}, S. and {Cunha Parro}, V. and {Cupani}, G. and {D'Odorico}, V. and {Dadi}, K. and {de Castro Le{\~a}o}, I. and {De Cia}, A. and {De Medeiros}, J.~R. and {Debras}, F. and {Debus}, M. and {Delorme}, A. and {Demangeon}, O. and {Derie}, F. and {Dessauges-Zavadsky}, M. and {Di Marcantonio}, P. and {Di Stefano}, S. and {Dionies}, F. and {Domiciano de Souza}, A. and {Doyon}, R. and {Dunn}, J. and {Egner}, S. and {Ehrenreich}, D. and {Faria}, J.~P. and {Ferruzzi}, D. and {Feruglio}, C. and {Fisher}, M. and {Fontana}, A. and {Frank}, B.~S. and {Fuesslein}, C. and {Fumagalli}, M. and {Fusco}, T. and {Fynbo}, J. and {Gabella}, O. and {Gaessler}, W. and {Gallo}, E. and {Gao}, X. and {Genolet}, L. and {Genoni}, M. and {Giacobbe}, P. and {Giro}, E. and {Gon{\c{c}}alves}, R.~S. and {Gonzalez}, O.~A. and {Gonz{\'a}lez-Hern{\'a}ndez}, J.~I. and {Gouvret}, C. and {Gracia T{\'e}mich}, F. and {Haehnelt}, M.~G. and {Haniff}, C. and {Hatzes}, A. and {Helled}, R. and {Hoeijmakers}, H.~J. and {Hughes}, I. and {Huke}, P. and {Ivanisenko}, Y. and {J{\"a}rvinen}, A.~S. and {J{\"a}rvinen}, S.~P. and {Kaminski}, A. and {Kern}, J. and {Knoche}, J. and {Kordt}, A. and {Korhonen}, H. and {Korn}, A.~J. and {Kouach}, D. and {Kowzan}, G. and {Kreidberg}, L. and {Landoni}, M. and {Lanotte}, A.~A. and {Lavail}, A. and {Lavie}, B. and {Lee}, D. and {Lehmitz}, M. and {Li}, J. and {Li}, W. and {Liske}, J. and {Lovis}, C. and {Lucatello}, S. and {Lunney}, D. and {MacIntosh}, M.~J. and {Madhusudhan}, N. and {Magrini}, L. and {Maiolino}, R. and {Maldonado}, J. and {Malo}, L. and {Man}, A.~W.~S. and {Marquart}, T. and {Marques}, C.~M.~J. and {Marques}, E.~L. and {Martinez}, P. and {Martins}, A. and {Martins}, C.~J.~A.~P. and {Martins}, J.~H.~C. and {Maslowski}, P. and {Mason}, C. and {Mason}, E. and {McCracken}, R.~A. and {Melo e Sousa}, M.~A.~F. and {Mergo}, P. and {Micela}, G. and {Milakovi{\'c}}, D. and {Molli{\`e}re}, P. and {Monteiro}, M.~A. and {Montgomery}, D. and {Mordasini}, C. and {Morin}, J. and {Mucciarelli}, A. and {Murphy}, M.~T. and {N'Diaye}, M. and {Nardetto}, N. and {Neichel}, B. and {Neri}, N. and {Niedzielski}, A.~T. and {Niemczura}, E. and {Nisini}, B. and {Nortmann}, L. and {Noterdaeme}, P. and {Nunes}, N.~J. and {Oggioni}, L. and {Olchewsky}, F. and {Oliva}, E. and {{\"O}nel}, H. and {Origlia}, L. and {{\"O}stlin}, G. and {Ouellette}, N.~N. -Q. and {Pall{\'e}}, E. and {Papaderos}, P. and {Pariani}, G. and {Pasquini}, L.},
        title = "{ANDES, the high resolution spectrograph for the ELT: science goals, project overview, and future developments}",
    booktitle = {Ground-based and Airborne Instrumentation for Astronomy X},
         year = 2024,
       editor = {{Bryant}, Julia J. and {Motohara}, Kentaro and {Vernet}, Jo{\"e}l. R.~D.},
       series = {Society of Photo-Optical Instrumentation Engineers (SPIE) Conference Series},
       volume = {13096},
        month = jul,
          eid = {1309613},
        pages = {1309613},
          doi = {10.1117/12.3017966},
archivePrefix = {arXiv},
       eprint = {2407.14601},
 primaryClass = {astro-ph.IM},
       adsurl = {https://ui.adsabs.harvard.edu/abs/2024SPIE13096E..13M}
}

@ARTICLE{Roederer2024,
       author = {{Roederer}, Ian U. and {Alvarado-G{\'o}mez}, Juli{\'a}n D. and {Allende Prieto}, Carlos and {Adibekyan}, Vardan and {Aguado}, David S. and {Amado}, Pedro J. and {Amazo-G{\'o}mez}, Eliana M. and {Baratella}, Martina and {Barnes}, Sydney A. and {Bensby}, Thomas and {Bigot}, Lionel and {Chiavassa}, Andrea and {Domiciano de Souza}, Armando and {Gonz{\'a}lez Hern{\'a}ndez}, J.~I. and {Hansen}, Camilla Juul and {J{\"a}rvinen}, Silva P. and {Korn}, Andreas J. and {Lucatello}, Sara and {Magrini}, Laura and {Maiolino}, Roberto and {Di Marcantonio}, Paolo and {Marconi}, Alessandro and {De Medeiros}, Jos{\'e} R. and {Mucciarelli}, Alessio and {Nardetto}, Nicolas and {Origlia}, Livia and {Peroux}, Celine and {Poppenh{\"a}ger}, Katja and {Reiners}, Ansgar and {Rodr{\'\i}guez-L{\'o}pez}, Cristina and {Romano}, Donatella and {Salvadori}, Stefania and {Tisserand}, Patrick and {Venn}, Kim and {Wade}, Gregg A. and {Zanutta}, Alessio},
        title = "{The discovery space of ELT-ANDES. Stars and stellar populations}",
      journal = {Experimental Astronomy},
         year = 2024,
        month = apr,
       volume = {57},
       number = {2},
          eid = {17},
        pages = {17},
          doi = {10.1007/s10686-024-09938-8},
archivePrefix = {arXiv},
       eprint = {2311.16320},
 primaryClass = {astro-ph.IM},
       adsurl = {https://ui.adsabs.harvard.edu/abs/2024ExA....57...17R}
}

@ARTICLE{Martins2024,
       author = {{Martins}, C.~J.~A.~P. and {Cooke}, R. and {Liske}, J. and {Murphy}, M.~T. and {Noterdaeme}, P. and {Schmidt}, T.~M. and {Alcaniz}, J.~S. and {Alves}, C.~S. and {Balashev}, S. and {Cristiani}, S. and {Di Marcantonio}, P. and {G{\'e}nova Santos}, R. and {Gon{\c{c}}alves}, R.~S. and {Gonz{\'a}lez Hern{\'a}ndez}, J.~I. and {Maiolino}, R. and {Marconi}, A. and {Marques}, C.~M.~J. and {Melo e Sousa}, M.~A.~F. and {Nunes}, N.~J. and {Origlia}, L. and {P{\'e}roux}, C. and {Vinzl}, S. and {Zanutta}, A.},
        title = "{Cosmology and fundamental physics with the ELT-ANDES spectrograph}",
      journal = {Experimental Astronomy},
         year = 2024,
        month = feb,
       volume = {57},
       number = {1},
          eid = {5},
        pages = {5},
          doi = {10.1007/s10686-024-09928-w},
archivePrefix = {arXiv},
       eprint = {2311.16274},
 primaryClass = {astro-ph.CO},
       adsurl = {https://ui.adsabs.harvard.edu/abs/2024ExA....57....5M}
}

@ARTICLE{DOdorico2024,
       author = {{D'Odorico}, Valentina and {Bolton}, James S. and {Christensen}, Lise and {De Cia}, Annalisa and {Zackrisson}, Erik and {Kordt}, Aron and {Izzo}, Luca and {Li}, Jiangtao and {Maiolino}, Roberto and {Marconi}, Alessandro and {Richter}, Philipp and {Saccardi}, Andrea and {Salvadori}, Stefania and {Vanni}, Irene and {Feruglio}, Chiara and {Fumagalli}, Michele and {Fynbo}, Johan P.~U. and {Noterdaeme}, Pasquier and {Papaderos}, Polychronis and {P{\'e}roux}, C{\'e}line and {Verma}, Aprajita and {Di Marcantonio}, Paolo and {Origlia}, Livia and {Zanutta}, Alessio},
        title = "{Galaxy formation and symbiotic evolution with the inter-galactic medium in the age of ELT-ANDES}",
      journal = {Experimental Astronomy},
         year = 2024,
        month = dec,
       volume = {58},
       number = {3},
          eid = {21},
        pages = {21},
          doi = {10.1007/s10686-024-09967-3},
archivePrefix = {arXiv},
       eprint = {2311.16803},
 primaryClass = {astro-ph.GA},
       adsurl = {https://ui.adsabs.harvard.edu/abs/2024ExA....58...21D}
}

@article{Brogi2013,
doi = {10.1088/0004-637X/767/1/27},
url = {https://doi.org/10.1088/0004-637X/767/1/27},
year = {2013},
month = {mar},
publisher = {The American Astronomical Society},
volume = {767},
number = {1},
pages = {27},
author = {Brogi, M. and Snellen, I. A. G. and de Kok, R. J. and Albrecht, S. and Birkby, J. L. and de Mooij, E. J. W.},
title = {DETECTION OF MOLECULAR ABSORPTION IN THE DAYSIDE OF EXOPLANET 51 PEGASI b?},
journal = {The Astrophysical Journal}
}

@ARTICLE{deKok2013,
       author = {{de Kok}, R.~J. and {Brogi}, M. and {Snellen}, I.~A.~G. and {Birkby}, J. and {Albrecht}, S. and {de Mooij}, E.~J.~W.},
        title = "{Detection of carbon monoxide in the high-resolution day-side spectrum of the exoplanet HD 189733b}",
      journal = {\aap},
         year = 2013,
        month = jun,
       volume = {554},
          eid = {A82},
        pages = {A82},
          doi = {10.1051/0004-6361/201321381},
archivePrefix = {arXiv},
       eprint = {1304.4014},
 primaryClass = {astro-ph.EP},
       adsurl = {https://ui.adsabs.harvard.edu/abs/2013A&A...554A..82D}
}

@article{Thorngren2026,
doi = {10.3847/1538-4365/ae0e71},
url = {https://doi.org/10.3847/1538-4365/ae0e71},
year = {2026},
month = {feb},
publisher = {The American Astronomical Society},
volume = {283},
number = {1},
pages = {10},
author = {Thorngren, Daniel P. and Sing, David K. and Mukherjee, Sagnick},
title = {Bayesian Model Comparison and Significance: Widespread Errors and How to Correct Them},
journal = {The Astrophysical Journal Supplement Series}
}

@ARTICLE{Damiano2024,
       author = {{Damiano}, Mario and {Bello-Arufe}, Aaron and {Yang}, Jeehyun and {Hu}, Renyu},
        title = "{LHS 1140 b Is a Potentially Habitable Water World}",
      journal = {\apjl},
         year = 2024,
        month = jun,
       volume = {968},
       number = {2},
          eid = {L22},
        pages = {L22},
          doi = {10.3847/2041-8213/ad5204},
archivePrefix = {arXiv},
       eprint = {2403.13265},
 primaryClass = {astro-ph.EP},
       adsurl = {https://ui.adsabs.harvard.edu/abs/2024ApJ...968L..22D}
}

@ARTICLE{Kausch2015,
       author = {{Kausch}, W. and {Noll}, S. and {Smette}, A. and {Kimeswenger}, S. and {Barden}, M. and {Szyszka}, C. and {Jones}, A.~M. and {Sana}, H. and {Horst}, H. and {Kerber}, F.},
        title = "{Molecfit: A general tool for telluric absorption correction. II. Quantitative evaluation on ESO-VLT/X-Shooterspectra}",
      journal = {\aap},
         year = 2015,
        month = apr,
       volume = {576},
          eid = {A78},
        pages = {A78},
          doi = {10.1051/0004-6361/201423909},
archivePrefix = {arXiv},
       eprint = {1501.07265},
 primaryClass = {astro-ph.IM},
       adsurl = {https://ui.adsabs.harvard.edu/abs/2015A&A...576A..78K}
}

@ARTICLE{Smette2015,
       author = {{Smette}, A. and {Sana}, H. and {Noll}, S. and {Horst}, H. and {Kausch}, W. and {Kimeswenger}, S. and {Barden}, M. and {Szyszka}, C. and {Jones}, A.~M. and {Gallenne}, A. and {Vinther}, J. and {Ballester}, P. and {Taylor}, J.},
        title = "{Molecfit: A general tool for telluric absorption correction. I. Method and application to ESO instruments}",
      journal = {\aap},
         year = 2015,
        month = apr,
       volume = {576},
          eid = {A77},
        pages = {A77},
          doi = {10.1051/0004-6361/201423932},
archivePrefix = {arXiv},
       eprint = {1501.07239},
 primaryClass = {astro-ph.IM},
       adsurl = {https://ui.adsabs.harvard.edu/abs/2015A&A...576A..77S}
}

@ARTICLE{Meech2022,
       author = {{Meech}, Annabella and {Aigrain}, Suzanne and {Brogi}, Matteo and {Birkby}, Jayne L.},
        title = "{Applications of a Gaussian process framework for modelling of high-resolution exoplanet spectra}",
      journal = {\mnras},
         year = 2022,
        month = may,
       volume = {512},
       number = {2},
        pages = {2604-2617},
          doi = {10.1093/mnras/stac662},
archivePrefix = {arXiv},
       eprint = {2203.09428},
 primaryClass = {astro-ph.EP},
       adsurl = {https://ui.adsabs.harvard.edu/abs/2022MNRAS.512.2604M}
}

@ARTICLE{Kjaersgaard2023,
       author = {{Kj{\ae}rsgaard}, R.~D. and {Bello-Arufe}, A. and {Rathcke}, A.~D. and {Buchhave}, L.~A. and {Clemmensen}, L.~K.~H.},
        title = "{TAU: A neural network based telluric correction framework}",
      journal = {\aap},
         year = 2023,
        month = sep,
       volume = {677},
          eid = {A120},
        pages = {A120},
          doi = {10.1051/0004-6361/202346652},
       adsurl = {https://ui.adsabs.harvard.edu/abs/2023A&A...677A.120K}
}

@ARTICLE{Piskunov2025,
       author = {{Piskunov}, Nikolai and {Rains}, Adam D. and {Boldt-Christmas}, Linn},
        title = "{TSD: An inverse problem approach for recovering the exoplanetary atmosphere transmission spectrum from high-resolution spectroscopy}",
      journal = {arXiv e-prints},
         year = 2025,
        month = sep,
          eid = {arXiv:2509.12737},
        pages = {arXiv:2509.12737},
          doi = {10.48550/arXiv.2509.12737},
archivePrefix = {arXiv},
       eprint = {2509.12737},
 primaryClass = {astro-ph.EP},
       adsurl = {https://ui.adsabs.harvard.edu/abs/2025arXiv250912737P}
}

@ARTICLE{Gandhi2020,
       author = {{Gandhi}, Siddharth and {Brogi}, Matteo and {Webb}, Rebecca K.},
        title = "{Seeing above the clouds with high-resolution spectroscopy}",
      journal = {\mnras},
         year = 2020,
        month = oct,
       volume = {498},
       number = {1},
        pages = {194-204},
          doi = {10.1093/mnras/staa2424},
archivePrefix = {arXiv},
       eprint = {2008.11464},
 primaryClass = {astro-ph.EP},
       adsurl = {https://ui.adsabs.harvard.edu/abs/2020MNRAS.498..194G}
}

@ARTICLE{Piaulet2025,
       author = {{Piaulet-Ghorayeb}, Caroline and {Benneke}, Bj{\"o}rn and {Turbet}, Martin and {Moore}, Keavin and {Roy}, Pierre-Alexis and {Lim}, Olivia and {Doyon}, Ren{\'e} and {Fauchez}, Thomas J. and {Albert}, Lo{\"\i}c and {Radica}, Michael and {Coulombe}, Louis-Philippe and {Lafreni{\`e}re}, David and {Cowan}, Nicolas B. and {Belzile}, Danika and {Musfirat}, Kamrul and {Kaur}, Mehramat and {L'Heureux}, Alexandrine and {Johnstone}, Doug and {MacDonald}, Ryan J. and {Allart}, Romain and {Dang}, Lisa and {Kaltenegger}, Lisa and {Pelletier}, Stefan and {Rowe}, Jason F. and {Taylor}, Jake and {Turner}, Jake D.},
        title = "{Strict Limits on Potential Secondary Atmospheres on the Temperate Rocky Exo-Earth TRAPPIST-1 d}",
      journal = {\apj},
         year = 2025,
        month = aug,
       volume = {989},
       number = {2},
          eid = {181},
        pages = {181},
          doi = {10.3847/1538-4357/adf207},
archivePrefix = {arXiv},
       eprint = {2508.08416},
 primaryClass = {astro-ph.EP},
       adsurl = {https://ui.adsabs.harvard.edu/abs/2025ApJ...989..181P}
}

@ARTICLE{Espinoza2025,
       author = {{Espinoza}, N{\'e}stor and {Allen}, Natalie H. and {Glidden}, Ana and {Lewis}, Nikole K. and {Seager}, Sara and {Ca{\~n}as}, Caleb I. and {Grant}, David and {Gressier}, Am{\'e}lie and {Courreges}, Shelby and {Stevenson}, Kevin B. and {Ranjan}, Sukrit and {Col{\'o}n}, Knicole and {Morris}, Brett M. and {MacDonald}, Ryan J. and {Long}, Douglas and {Wakeford}, Hannah R. and {Valenti}, Jeff A. and {Alderson}, Lili and {Batalha}, Natasha E. and {Challener}, Ryan C. and {Huang}, Jingcheng and {Lin}, Zifan and {Louie}, Dana R. and {Mullens}, Elijah and {Valentine}, Daniel and {Mountain}, C. Matt and {Pueyo}, Laurent and {Perrin}, Marshall D. and {Bellini}, Andrea and {Kammerer}, Jens and {Libralato}, Mattia and {Rebollido}, Isabel and {Rickman}, Emily and {Sohn}, Sangmo Tony and {van der Marel}, Roeland P.},
        title = "{JWST-TST DREAMS: NIRSpec/PRISM Transmission Spectroscopy of the Habitable Zone Planet TRAPPIST-1 e}",
      journal = {\apjl},
         year = 2025,
        month = sep,
       volume = {990},
       number = {2},
          eid = {L52},
        pages = {L52},
          doi = {10.3847/2041-8213/adf42e},
archivePrefix = {arXiv},
       eprint = {2509.05414},
 primaryClass = {astro-ph.EP},
       adsurl = {https://ui.adsabs.harvard.edu/abs/2025ApJ...990L..52E}
}

@ARTICLE{Glidden2025,
       author = {{Glidden}, Ana and {Ranjan}, Sukrit and {Seager}, Sara and {Espinoza}, N{\'e}stor and {MacDonald}, Ryan J. and {Allen}, Natalie H. and {Ca{\~n}as}, Caleb I. and {Grant}, David and {Gressier}, Am{\'e}lie and {Stevenson}, Kevin B. and {Batalha}, Natasha E. and {Lewis}, Nikole K. and {Long}, Douglas and {Wakeford}, Hannah R. and {Alderson}, Lili and {Challener}, Ryan C. and {Col{\'o}n}, Knicole and {Huang}, Jingcheng and {Lin}, Zifan and {Louie}, Dana R. and {Mullens}, Elijah and {Sotzen}, Kristin S. and {Valenti}, Jeff A. and {Valentine}, Daniel and {Clampin}, Mark and {Mountain}, C. Matt and {Perrin}, Marshall and {van der Marel}, Roeland P.},
        title = "{JWST-TST DREAMS: Secondary Atmosphere Constraints for the Habitable Zone Planet TRAPPIST-1 e}",
      journal = {\apjl},
         year = 2025,
        month = sep,
       volume = {990},
       number = {2},
          eid = {L53},
        pages = {L53},
          doi = {10.3847/2041-8213/adf62e},
archivePrefix = {arXiv},
       eprint = {2509.05407},
 primaryClass = {astro-ph.EP},
       adsurl = {https://ui.adsabs.harvard.edu/abs/2025ApJ...990L..53G}
}

@ARTICLE{Suissa2020,
       author = {{Suissa}, Gabrielle and {Wolf}, Eric T. and {Kopparapu}, Ravi kumar and {Villanueva}, Geronimo L. and {Fauchez}, Thomas and {Mandell}, Avi M. and {Arney}, Giada and {Gilbert}, Emily A. and {Schlieder}, Joshua E. and {Barclay}, Thomas and {Quintana}, Elisa V. and {Lopez}, Eric and {Rodriguez}, Joseph E. and {Vanderburg}, Andrew},
        title = "{The First Habitable-zone Earth-sized Planet from TESS. III. Climate States and Characterization Prospects for TOI-700 d}",
      journal = {\aj},
         year = 2020,
        month = sep,
       volume = {160},
       number = {3},
          eid = {118},
        pages = {118},
          doi = {10.3847/1538-3881/aba4b4},
archivePrefix = {arXiv},
       eprint = {2001.00955},
 primaryClass = {astro-ph.EP},
       adsurl = {https://ui.adsabs.harvard.edu/abs/2020AJ....160..118S}
}

@ARTICLE{Brogi2019,
       author = {{Brogi}, Matteo and {Line}, Michael R.},
        title = "{Retrieving Temperatures and Abundances of Exoplanet Atmospheres with High-resolution Cross-correlation Spectroscopy}",
      journal = {\aj},
         year = 2019,
        month = mar,
       volume = {157},
       number = {3},
          eid = {114},
        pages = {114},
          doi = {10.3847/1538-3881/aaffd3},
archivePrefix = {arXiv},
       eprint = {1811.01681},
 primaryClass = {astro-ph.EP},
       adsurl = {https://ui.adsabs.harvard.edu/abs/2019AJ....157..114B}
}

@ARTICLE{Gibson2020,
       author = {{Gibson}, Neale P. and {Merritt}, Stephanie and {Nugroho}, Stevanus K. and {Cubillos}, Patricio E. and {de Mooij}, Ernst J.~W. and {Mikal-Evans}, Thomas and {Fossati}, Luca and {Lothringer}, Joshua and {Nikolov}, Nikolay and {Sing}, David K. and {Spake}, Jessica J. and {Watson}, Chris A. and {Wilson}, Jamie},
        title = "{Detection of Fe I in the atmosphere of the ultra-hot Jupiter WASP-121b, and a new likelihood-based approach for Doppler-resolved spectroscopy}",
      journal = {\mnras},
         year = 2020,
        month = apr,
       volume = {493},
       number = {2},
        pages = {2215-2228},
          doi = {10.1093/mnras/staa228},
archivePrefix = {arXiv},
       eprint = {2001.06430},
 primaryClass = {astro-ph.EP},
       adsurl = {https://ui.adsabs.harvard.edu/abs/2020MNRAS.493.2215G}
}

@ARTICLE{Fujii2018,
       author = {{Fujii}, Yuka and {Angerhausen}, Daniel and {Deitrick}, Russell and {Domagal-Goldman}, Shawn and {Grenfell}, John Lee and {Hori}, Yasunori and {Kane}, Stephen R. and {Pall{\'e}}, Enric and {Rauer}, Heike and {Siegler}, Nicholas and {Stapelfeldt}, Karl and {Stevenson}, Kevin B.},
        title = "{Exoplanet Biosignatures: Observational Prospects}",
      journal = {Astrobiology},
         year = 2018,
        month = jun,
       volume = {18},
       number = {6},
        pages = {739-778},
          doi = {10.1089/ast.2017.1733},
archivePrefix = {arXiv},
       eprint = {1705.07098},
 primaryClass = {astro-ph.EP},
       adsurl = {https://ui.adsabs.harvard.edu/abs/2018AsBio..18..739F}
}

@ARTICLE{cadieux2024b,
       author = {{Cadieux}, Charles and {Doyon}, Ren{\'e} and {MacDonald}, Ryan J. and {Turbet}, Martin and {Artigau}, {\'E}tienne and {Lim}, Olivia and {Radica}, Michael and {Fauchez}, Thomas J. and {Salhi}, Salma and {Dang}, Lisa and {Albert}, Lo{\"\i}c and {Coulombe}, Louis-Philippe and {Cowan}, Nicolas B. and {Lafreni{\`e}re}, David and {L'Heureux}, Alexandrine and {Piaulet-Ghorayeb}, Caroline and {Benneke}, Bj{\"o}rn and {Cloutier}, Ryan and {Charnay}, Benjamin and {Cook}, Neil J. and {Fournier-Tondreau}, Marylou and {Plotnykov}, Mykhaylo and {Valencia}, Diana},
        title = "{Transmission Spectroscopy of the Habitable Zone Exoplanet LHS 1140 b with JWST/NIRISS}",
      journal = {\apjl},
         year = 2024,
        month = jul,
       volume = {970},
       number = {1},
          eid = {L2},
        pages = {L2},
          doi = {10.3847/2041-8213/ad5afa},
archivePrefix = {arXiv},
       eprint = {2406.15136},
 primaryClass = {astro-ph.EP},
       adsurl = {https://ui.adsabs.harvard.edu/abs/2024ApJ...970L...2C}
}

@ARTICLE{meiervaldes2025,
       author = {{Meier Vald{\'e}s}, E.~A. and {Demory}, B. -O. and {Diamond-Lowe}, H. and {Mendon{\c{c}}a}, J.~M. and {August}, P.~C. and {Fortune}, M. and {Allen}, N.~H. and {Kitzmann}, D. and {Gressier}, A. and {Hooton}, M. and {Jones}, K.~D. and {Buchhave}, L.~A. and {Espinoza}, N. and {Fisher}, C.~E. and {Gibson}, N.~P. and {Heng}, K. and {Hoeijmakers}, J. and {Prinoth}, B. and {Rathcke}, A.~D. and {Eastman}, J.~D.},
        title = "{Hot Rocks Survey: II. The thermal emission of TOI-1468 b reveals a bare hot rock}",
      journal = {\aap},
         year = 2025,
        month = jun,
       volume = {698},
          eid = {A68},
        pages = {A68},
          doi = {10.1051/0004-6361/202453449},
archivePrefix = {arXiv},
       eprint = {2503.19772},
 primaryClass = {astro-ph.EP},
       adsurl = {https://ui.adsabs.harvard.edu/abs/2025A&A...698A..68M}
}

@ARTICLE{luque2025,
       author = {{Luque}, Rafael and {Coy}, Brandon Park and {Xue}, Qiao and {Feinstein}, Adina D. and {Ahrer}, Eva-Maria and {Changeat}, Quentin and {Zhang}, Michael and {Moran}, Sarah E. and {Bean}, Jacob L. and {Kite}, Edwin and {Weiner Mansfield}, Megan and {Pall{\'e}}, Enric},
        title = "{A Dark, Bare Rock for TOI-1685 b from a JWST NIRSpec G395H Phase Curve}",
      journal = {\aj},
         year = 2025,
        month = jul,
       volume = {170},
       number = {1},
          eid = {49},
        pages = {49},
          doi = {10.3847/1538-3881/addb40},
archivePrefix = {arXiv},
       eprint = {2412.03411},
 primaryClass = {astro-ph.EP},
       adsurl = {https://ui.adsabs.harvard.edu/abs/2025AJ....170...49L}
}

@ARTICLE{fortune2025,
       author = {{Fortune}, Mark and {Gibson}, Neale P. and {Diamond-Lowe}, Hannah and {Mendon{\c{c}}a}, Jo{\~a}o M. and {Gressier}, Am{\'e}lie and {Kitzmann}, Daniel and {Allen}, Natalie H. and {August}, Prune C. and {Ih}, Jegug and {Meier Vald{\'e}s}, Erik and {Zgraggen}, Merlin and {Buchhave}, Lars A. and {Demory}, Brice-Olivier and {Espinoza}, N{\'e}stor and {Heng}, Kevin and {Jones}, Kathryn and {Rathcke}, Alexander D.},
        title = "{Hot Rocks Survey: III. A deep eclipse for LHS 1140c and a new Gaussian process method to account for correlated noise in individual pixels}",
      journal = {\aap},
         year = 2025,
        month = sep,
       volume = {701},
          eid = {A25},
        pages = {A25},
          doi = {10.1051/0004-6361/202554198},
archivePrefix = {arXiv},
       eprint = {2505.22186},
 primaryClass = {astro-ph.EP},
       adsurl = {https://ui.adsabs.harvard.edu/abs/2025A&A...701A..25F}
}

@ARTICLE{allen2025,
       author = {{Allen}, Natalie H. and {Espinoza}, N{\'e}stor and {Diamond-Lowe}, Hannah and {Mendon{\c{c}}a}, Jo{\~a}o M. and {Demory}, Brice-Olivier and {Gressier}, Am{\'e}lie and {Ih}, Jegug and {Fortune}, Mark and {August}, Prune C. and {Holmberg}, M{\r{a}}ns and {Meier Vald{\'e}s}, Erik and {Zgraggen}, Merlin and {Buchhave}, Lars A. and {Burgasser}, Adam J. and {Fisher}, Chloe and {Gibson}, Neale P. and {Heng}, Kevin and {Hoeijmakers}, Jens and {Kitzmann}, Daniel and {Prinoth}, Bibiana and {Rathcke}, Alexander D. and {Morris}, Brett M.},
        title = "{Hot Rocks Survey. IV. Emission from LTT 3780 b Is Consistent with a Bare Rock}",
      journal = {\aj},
         year = 2025,
        month = oct,
       volume = {170},
       number = {4},
          eid = {240},
        pages = {240},
          doi = {10.3847/1538-3881/adfc51},
archivePrefix = {arXiv},
       eprint = {2508.14210},
 primaryClass = {astro-ph.EP},
       adsurl = {https://ui.adsabs.harvard.edu/abs/2025AJ....170..240A}
}

@ARTICLE{gressier2024,
       author = {{Gressier}, Am{\'e}lie and {Espinoza}, N{\'e}stor and {Allen}, Natalie H. and {Sing}, David K. and {Banerjee}, Agnibha and {Barstow}, Joanna K. and {Valenti}, Jeff A. and {Lewis}, Nikole K. and {Birkmann}, Stephan M. and {Challener}, Ryan C. and {Manjavacas}, Elena and {Alves de Oliveira}, Catarina and {Crouzet}, Nicolas and {Beck}, Tracy. L.},
        title = "{Hints of a Sulfur-rich Atmosphere around the 1.6 R $_{{\ensuremath{\oplus}}}$ Super-Earth L98-59 d from JWST NIRspec G395H Transmission Spectroscopy}",
      journal = {\apjl},
         year = 2024,
        month = nov,
       volume = {975},
       number = {1},
          eid = {L10},
        pages = {L10},
          doi = {10.3847/2041-8213/ad73d1},
archivePrefix = {arXiv},
       eprint = {2408.15855},
 primaryClass = {astro-ph.EP},
       adsurl = {https://ui.adsabs.harvard.edu/abs/2024ApJ...975L..10G}
}

@ARTICLE{august2025,
       author = {{August}, P.~C. and {Buchhave}, L.~A. and {Diamond-Lowe}, H. and {Mendon{\c{c}}a}, J.~M. and {Gressier}, A. and {Rathcke}, A.~D. and {Allen}, N.~H. and {Fortune}, M. and {Jones}, K.~D. and {Meier Vald{\'e}s}, E.~A. and {Demory}, B. -O. and {Espinoza}, N. and {Fisher}, C.~E. and {Gibson}, N.~P. and {Heng}, K. and {Hoeijmakers}, J. and {Hooton}, M.~J. and {Kitzmann}, D. and {Prinoth}, B. and {Eastman}, J.~D. and {Barnes}, R.},
        title = "{Hot Rocks Survey I: A possible shallow eclipse for LHS 1478 b}",
      journal = {\aap},
         year = 2025,
        month = mar,
       volume = {695},
          eid = {A171},
        pages = {A171},
          doi = {10.1051/0004-6361/202452611},
archivePrefix = {arXiv},
       eprint = {2410.11048},
 primaryClass = {astro-ph.EP},
       adsurl = {https://ui.adsabs.harvard.edu/abs/2025A&A...695A.171A}
}

@ARTICLE{belloarufe2025,
       author = {{Bello-Arufe}, Aaron and {Damiano}, Mario and {Bennett}, Katherine A. and {Hu}, Renyu and {Welbanks}, Luis and {MacDonald}, Ryan J. and {Seligman}, Darryl Z. and {Sing}, David K. and {Tokadjian}, Armen and {Oza}, Apurva V. and {Yang}, Jeehyun},
        title = "{Evidence for a Volcanic Atmosphere on the Sub-Earth L 98-59 b}",
      journal = {\apjl},
         year = 2025,
        month = feb,
       volume = {980},
       number = {2},
          eid = {L26},
        pages = {L26},
          doi = {10.3847/2041-8213/adaf22},
archivePrefix = {arXiv},
       eprint = {2501.18680},
 primaryClass = {astro-ph.EP},
       adsurl = {https://ui.adsabs.harvard.edu/abs/2025ApJ...980L..26B}
}

@ARTICLE{yang2013,
       author = {{Yang}, Jun and {Cowan}, Nicolas B. and {Abbot}, Dorian S.},
        title = "{Stabilizing Cloud Feedback Dramatically Expands the Habitable Zone of Tidally Locked Planets}",
      journal = {\apjl},
         year = 2013,
        month = jul,
       volume = {771},
       number = {2},
          eid = {L45},
        pages = {L45},
          doi = {10.1088/2041-8205/771/2/L45},
archivePrefix = {arXiv},
       eprint = {1307.0515},
 primaryClass = {astro-ph.EP},
       adsurl = {https://ui.adsabs.harvard.edu/abs/2013ApJ...771L..45Y}
}

@ARTICLE{rackham2018,
       author = {{Rackham}, Benjamin V. and {Apai}, D{\'a}niel and {Giampapa}, Mark S.},
        title = "{The Transit Light Source Effect: False Spectral Features and Incorrect Densities for M-dwarf Transiting Planets}",
      journal = {\apj},
         year = 2018,
        month = feb,
       volume = {853},
       number = {2},
          eid = {122},
        pages = {122},
          doi = {10.3847/1538-4357/aaa08c},
archivePrefix = {arXiv},
       eprint = {1711.05691},
 primaryClass = {astro-ph.EP},
       adsurl = {https://ui.adsabs.harvard.edu/abs/2018ApJ...853..122R}
}

@ARTICLE{rathcke2025,
       author = {{Rathcke}, Alexander D. and {Buchhave}, Lars A. and {de Wit}, Julien and {Rackham}, Benjamin V. and {August}, Prune C. and {Diamond-Lowe}, Hannah and {Mendon{\c{C}}a}, Jo{\~a}o M. and {Bello-Arufe}, Aaron and {L{\'o}pez-Morales}, Mercedes and {Kitzmann}, Daniel and {Heng}, Kevin},
        title = "{Stellar Contamination Correction Using Back-to-back Transits of TRAPPIST-1 b and c}",
      journal = {\apjl},
         year = 2025,
        month = jan,
       volume = {979},
       number = {1},
          eid = {L19},
        pages = {L19},
          doi = {10.3847/2041-8213/ada5c7},
archivePrefix = {arXiv},
       eprint = {2412.16541},
 primaryClass = {astro-ph.EP},
       adsurl = {https://ui.adsabs.harvard.edu/abs/2025ApJ...979L..19R}
}

@ARTICLE{cauley2018,
       author = {{Cauley}, P. Wilson and {Kuckein}, Christoph and {Redfield}, Seth and {Shkolnik}, Evgenya L. and {Denker}, Carsten and {Llama}, Joe and {Verma}, Meetu},
        title = "{The Effects of Stellar Activity on Optical High-resolution Exoplanet Transmission Spectra}",
      journal = {\aj},
         year = 2018,
        month = nov,
       volume = {156},
       number = {5},
          eid = {189},
        pages = {189},
          doi = {10.3847/1538-3881/aaddf9},
archivePrefix = {arXiv},
       eprint = {1808.09558},
 primaryClass = {astro-ph.EP},
       adsurl = {https://ui.adsabs.harvard.edu/abs/2018AJ....156..189C}
}

@ARTICLE{Tamuz2005,
       author = {{Tamuz}, O. and {Mazeh}, T. and {Zucker}, S.},
        title = "{Correcting systematic effects in a large set of photometric light curves}",
      journal = {\mnras},
         year = 2005,
        month = feb,
       volume = {356},
       number = {4},
        pages = {1466-1470},
          doi = {10.1111/j.1365-2966.2004.08585.x},
archivePrefix = {arXiv},
       eprint = {astro-ph/0502056},
 primaryClass = {astro-ph},
       adsurl = {https://ui.adsabs.harvard.edu/abs/2005MNRAS.356.1466T}
}

@INPROCEEDINGS{Mazeh2007,
       author = {{Mazeh}, T. and {Tamuz}, O. and {Zucker}, S.},
        title = "{The Sys-Rem Detrending Algorithm: Implementation and Testing}",
    booktitle = {Transiting Extrapolar Planets Workshop},
         year = 2007,
       editor = {{Afonso}, C. and {Weldrake}, D. and {Henning}, Th.},
       series = {Astronomical Society of the Pacific Conference Series},
       volume = {366},
        month = jul,
        pages = {119},
          doi = {10.48550/arXiv.astro-ph/0612418},
archivePrefix = {arXiv},
       eprint = {astro-ph/0612418},
 primaryClass = {astro-ph},
       adsurl = {https://ui.adsabs.harvard.edu/abs/2007ASPC..366..119M}
}

@ARTICLE{Meadows2018,
       author = {{Meadows}, Victoria S. and {Reinhard}, Christopher T. and {Arney}, Giada N. and {Parenteau}, Mary N. and {Schwieterman}, Edward W. and {Domagal-Goldman}, Shawn D. and {Lincowski}, Andrew P. and {Stapelfeldt}, Karl R. and {Rauer}, Heike and {DasSarma}, Shiladitya and {Hegde}, Siddharth and {Narita}, Norio and {Deitrick}, Russell and {Lustig-Yaeger}, Jacob and {Lyons}, Timothy W. and {Siegler}, Nicholas and {Grenfell}, J. Lee},
        title = "{Exoplanet Biosignatures: Understanding Oxygen as a Biosignature in the Context of Its Environment}",
      journal = {Astrobiology},
         year = 2018,
        month = jun,
       volume = {18},
       number = {6},
        pages = {630-662},
          doi = {10.1089/ast.2017.1727},
archivePrefix = {arXiv},
       eprint = {1705.07560},
 primaryClass = {astro-ph.EP},
       adsurl = {https://ui.adsabs.harvard.edu/abs/2018AsBio..18..630M}
}

@BOOK{Brockwell1991,
       author = {{Brockwell}, Peter J. and {Davis}, Richard A.},
        title = "{Time series: Theory and methods}",
         year = 1991,
       adsurl = {https://ui.adsabs.harvard.edu/abs/1991tstm.book.....B}
}
\bibliographystyle{aasjournal}

\end{document}